\documentstyle[preprint,aps,eqsecnum]{revtex}
\begin{document}
\draft
\tightenlines

\newcommand{\beq}{\begin{equation}}
\newcommand{\eeq}{\end{equation}}
\newcommand{\bea}{\begin{eqnarray}}
\newcommand{\eea}{\end{eqnarray}}
\newcommand{\cir}{{\buildrel \circ \over =}}

\title{Centers of Mass and Rotational Kinematics  for the  Relativistic
 N-Body Problem in the Rest-Frame Instant Form.}

\author{David Alba}

\address
{Dipartimento di Fisica\\
Universita' di Firenze\\
L.go E.Fermi 2 (Arcetri)\\
50125 Firenze, Italy\\
E-mail: ALBA@FI.INFN.IT}

\author{and}

\author{Luca Lusanna}

\address
{Sezione INFN di Firenze\\
L.go E.Fermi 2 (Arcetri)\\
50125 Firenze, Italy\\
E-mail: LUSANNA@FI.INFN.IT}

\author{and}

\author{Massimo Pauri}

\address
{Dipartimento di Fisica\\
Universita' di Parma\\
Campus Universitario, Viale delle Scienze\\
43100 Parma, Italy\\
E-mail: PAURI@PR.INFN.IT}

\maketitle

\begin{abstract}

In the Wigner-covariant rest-frame instant form of dynamics it is
possible to develop a relativistic kinematics for the N-body problem
which solves all the problems raised till now on this topic. The
Wigner hyperplanes, orthogonal to the total timelike 4-momentum of any
N-body configuration, define the intrinsic rest frame and realize the
separation of the center-of-mass motion. The point chosen as origin of
each Wigner hyperplane can be made to coincide with the covariant
non-canonical Fokker-Pryce center of inertia. This is distinct from
the canonical pseudo-vector describing the decoupled motion of the
center of mass (having the same Euclidean covariance as the quantum
Newton-Wigner 3-position operator) and the non-canonical pseudo-vector
for the M\o ller center of energy. These are the only {\it external}
notions of relativistic center of mass, definable only in terms of the
{\it external} Poincar\'e group realization. Inside the Wigner
hyperplane, an {\it internal}  unfaithful realization of the
Poincar\'e group is defined while the analogous three concepts of
center of mass weakly {\it coincide} due to the first class
constraints defining the rest frame (vanishing of the {\it internal}
3-momentum). This unique {\it internal} center of mass is consequently
a {gauge variable} which, through a gauge fixing, can be localized
atthe origin of the Wigner hyperplane. An adapted canonical basis of
relative variables is found by means of the classical counterpart of
the Gartenhaus-Schwartz transformation. The invariant mass of the
N-body configuration is the Hamiltonian for the relative motions. In
this framework we can introduce the same {\it dynamical body frames},
{\it orientation-shape} variables, {\it spin frame} and {\it canonical
spin bases} for the rotational kinematics developed for the
non-relativistic N-body problem.

\vskip 1truecm

\today

\vskip 1truecm

\end{abstract}
\vfill\eject

\vfill\eject

\section{Introduction.}

In the non-relativistic N-body problem  the separation of the {\it
absolute translational motion} of the center of mass from the relative
motions can be easily carried out, due to the Abelian nature of the
translation symmetry group. This implies that the associated Noether
constants of motion (the conserved total 3-momentum) are in
involution, so that the center-of-mass degrees of  freedom decouple.
Moreover, the fact that the non-relativistic kinetic energy of the
relative motions is a quadratic form in the relative velocities allows
the introduction of special sets of relative coordinates, the {\it
Jacobi normal relative coordinates}, that diagonalize the quadratic
form and correspond to different patterns of clustering of the centers
of mass of the particles. Each set of Jacobi  coordinates organizes
the N particles into a {\it hierarchy of clusters}, in which each
cluster of two or more particles has a mass given by an eigenvalue
(reduced masses) of the quadratic form; the Jacobi normal coordinates
join the centers of mass of cluster pairs.

On the other hand, the non-Abelian nature of the rotation symmetry
group, whose associated Noether constants of motion (the conserved
total angular momentum) are not in involution, prevents the
possibility of a global separation of {\it absolute rotations} from
the relative motions, so that there is no global definition of {\it
absolute vibrations}. This has the consequence that an {\it isolated}
deformable body can undergo rotations by changing its own shape (as
shown by the {\it falling cat} and  the {\it diver}). It was just to
deal with these problems that the theory of the {\it orientation-shape
SO(3) principal bundle} \cite{little}  has been developed in the
context of molecular physics, emphasizing the {\it gauge} nature of a
{\it static} (i.e. velocity-independent) definition of {\it body
frame} for a deformable body. As a consequence, both the laboratory
and body frame angular velocities as well as the orientational
variables of the static body frame become {\it unobservable gauge}
variables. This approach is associated with a set of point canonical
transformations, which allow to define the body frame components of
relative motions in a velocity-independent way.

In a previous paper \cite{iten2} we showed that a more general class
of non-point canonical transformations exists for $N \geq 3$, which
allows to identify a family of {\it canonical spin bases} connected to
the patterns of the possible {\it clusterings of the spins} associated
with relative motions (namely the components of the center-of-mass
angular momenta). The definition of these {\it spin bases} is
independent of the use of Jacobi normal relative coordinates, just  as
the patterns of spin clustering are independent of the patterns of
center-of-mass Jacobi clustering.

There exist two basic frames associated to each spin basis: the {\it
spin frame} and the {\it dynamical body frame}. Their construction is
guaranteed by the fact that, besides the existence on the relative
phase space of a  Hamiltonian symmetry {\it left} action of
SO(3)\footnote{We adhere to the definitions used in Ref.\cite{little};
in the mathematical literature our {\it left} action is a {\it right}
action.} \footnote{The generators are the center-of-mass angular
momentum, Noether constants of motion.} on the relative phase space,
it is possible to define as many Hamiltonian non-symmetry  {\it right}
actions of SO(3)
\footnote{The generators are not constants of motion.} as the possible patterns of
spin clustering. While for N=3 the unique canonical spin basis
coincides with a special class of global cross sections of the trivial
orientation-shape SO(3) principal bundle, for $N \geq 4$ the existing
{\it spin bases} and {\it dynamical body frames} turn out to be
unrelated  to the local cross sections of the {\it static} non-trivial
orientation-shape SO(3) principal bundle, and {\it evolve} in a
dynamical way dictated by the equations of motion. Both the
orientation variables and the angular velocities become {\it
measurable} quantities in each canonical spin basis.

In this way we get for each N a finite number of physically
well-defined separations between {\it rotational} and {\it
vibrational} degrees of freedom. The unique {\it body frame} of rigid
bodies is replaced by a discrete number of {\it evolving dynamical
body frames} and of {\it spin canonical bases}, both of which are
grounded on patterns of spin couplings, direct analogues of the
coupling of quantum angular momenta. These results might be useful in
non-relativistic nuclear and molecular physics.

Besides translations and rotations, every isolated non-relativistic
system admits the internal energy, the total mass and the Galilei
boosts (which amounts essentially to the definition of the center of
mass) as constants of the motion. Altogether, there are 11 constants
of motion (one of them is a central charge) with which one gets a
realization of the kinematical extended Galilei algebra
\cite{levy,pauri3}.

The problem we want to tackle is what happens when we replace Galilean
spacetime with Minkowski spacetime. Precisely what can be said in this
case about the separation of the center of mass from the relative
motions (the Abelian translation symmetry) and about the treatment of
rotations (the non-Abelian rotational symmetry) already for the
simplest system of N free  scalar positive-energy particles?

The first immediate issue is how to describe a relativistic scalar
particle. Among the various possibilities (see Refs.\cite{luu} for a
review of the various options) we will choose to start from the
manifestly Lorentz covariant approach using Dirac's first class
constraints to identify free particles\footnote{We shall use $c=1$
everywhere and the convention $\eta^{\mu\nu}=\epsilon (+---)$ for the
Minkowski metric (with $\epsilon = \pm 1$ according to the either
particle physics or general relativity convention).}

\beq
p^2_i-\epsilon m_i^2 \approx 0.
\label{I1}
\eeq

\noindent  The associated Lagrangian description starts from the
4-vector positions $x^{\mu}_i(\tau )$ and the action $S=
\int d\tau \Big( -\epsilon \sum_im_i \sqrt{\epsilon {\dot x}_i^2(\tau )}
\Big)$, where $\tau$ is a Lorentz scalar mathematical time parameter
\footnote{An affine parameter for the particle timelike worldlines.}.
Therefore Lorentz covariance implies the use of singular Lagrangians
and of the associated Dirac's theory of constraints for the
Hamiltonian description. The time variables $x^o_i(\tau )$ are the
{\it gauge variables} associated to the mass-shell constraints, which
have the two topologically disjoint solutions $p^o_i \approx \pm
\sqrt{m^2_i+{\vec p}_i^2}$. As discussed in Ref.\cite{pons,lus} this
implies that:

i) a combination of the time variables may be identified with the
clock of one arbitrary observer labelling the evolution of the
isolated system;

ii) the $N-1$ relative times are connected with the observer freedom
of looking at the N particles either at the same time or with any
prescribed delay among them.

The introduction of interactions in this picture without destroying
the first class nature of the constraints \footnote{See Ref.\cite{luu}
for the models with second class constraints corresponding to gauge
fixings of the relative times.} is a difficult problem, which  gave
origin, in the two-particle case, to the DrozVincent-Komar-Todorov
model\cite{dvkt}. On the other hand, its extension to N particles was
never given in closed form.

When the particle is charged and interacts with a dynamical
(non-external) electromagnetic field a problem of covariance
reappears. The standard description of a charged scalar particle
interacting with the electromagnetic field is based on the action

\beq
S=-\epsilon m \int d\tau \sqrt{\epsilon {\dot x}^2(\tau )} -e \int
d\tau \int d^4z \delta^4(z-x(\tau )) {\dot x}^{\mu}(\tau ) A_{\mu}(z)
-{1\over 4}\int d^4z F^{\mu\nu}(z)F_{\mu\nu}(z).
\label{I2}
\eeq

\noindent If we evaluate the canonical momenta of the isolated system {\it
charged particle plus electromagnetic field}, we find two primary
constraints:

\bea
&&\chi (\tau ) = \Big( p-eA(x(\tau ))\Big)^2-\epsilon m^2\approx
0,\nonumber  \\
 &&\pi^o(z^o,\vec z)\approx 0.
 \label{I3}
 \eea

One realizes immediately that it is impossible to evaluate the Poisson
bracket of the two constraints, because there is no concept of {\it
equal time}. Also the Dirac Hamiltonian, which should be
$H_D=H_c+\lambda (\tau ) \chi (\tau) +\int d^3z
\lambda^o(z^o,\vec z)\pi^o(z^o,\vec z)$ \footnote{With $H_c$ the canonical
Hamiltonian and with $\lambda (\tau )$, $\lambda^o(z^o,\vec z)$
Dirac's multipliers.}, does not make sense for the same reason. This
problem is also present at the level of the Euler-Lagrange equations:
precisely the formulation of a Cauchy problem for a system of coupled
equations some of which are ordinary differential equations in the
affine parameter $\tau$ along the particle worldline, while the others
are partial differential equations depending on  Minkowski coordinates
$z^{\mu}$. Since the problem is due the absence of a covariant concept
of {\it equal time} between the field and particle variables, a new
formulation of the problem is needed.

In Ref.\cite{lus}, after a discussion of the many time formalism, a
solution of the previous covariance problem was found in a way
suggested in the context of a description able to incorporate the
gravitational field. There one considers an arbitrary 3+1 splitting of
Minkowski spacetime with spacelike hypersurfaces which is equivalent
to a congruence of timelike accelerated observers. This is essentially
Dirac's reformulation\cite{dirac} of classical field theory (suitably
extended to particles) on arbitrary spacelike hypersurfaces ({\it
equal time} surfaces): it is also the classical basis of the
Tomonaga-Schwinger formulation of quantum field theory. In this way,
for each isolated system (containing any combination of particles,
strings and fields) one gets its reformulation as a parametrized
Minkowski theory\cite{lus}, with the extra bonus of having the theory
already prepared to the coupling to gravity in its ADM formulation,
but with the price that the functions $z^{\mu}(\tau ,\vec \sigma )$
describing the embedding of the spacelike hypersurface in Minkowski
spacetime become {\it additional configuration variables} in the
action principle. Since the action is invariant under separate
$\tau$-reparametrizations and space-diffeomorphisms, there are first
class constraints ensuring the independence of the description from
the choice of the 3+1 splitting: the embedding configuration variables
$z^{\mu}(\tau ,\vec \sigma )$ are the {\it gauge} variables associated
with this kind of general covariance.

Let us remark that, since the intersection of a timelike worldline
with a spacelike hypersurface corresponding to a value $\tau$ of the
time parameter is identified by 3 numbers $\vec \sigma =\vec \eta
(\tau )$ and {\it not by four}, in parametrized Minkowski theories
each particle must have a well defined sign of the energy: therefore
we cannot simultaneously describe the two topologically disjoint
branches of the mass hyperboloid as in the standard manifestly
Lorentz-covariant theories. As a consequence, there are no more
mass-shell constraints. Each particle with a definite sign of the
energy is described by the canonical coordinates ${\vec
\eta}_i(\tau )$, ${\vec \kappa}_i(\tau )$ with the derived 4-position
of the particles given by $x^{\mu}_i(\tau )=z^{\mu}(\tau ,{\vec
\eta}_i(\tau ))$. The derived 4-momenta $p^{\mu}_i(\tau )$ are
${\vec \kappa}_i$-dependent solutions of $p^2_i-\epsilon m^2_i =0$
with the chosen sign of the energy.

In Minkowski spacetime, due to the independence of  parametrized
theories from the 3+1 splitting, we can restrict the foliation to have
spacelike hyperplanes as leaves. In particular, for each configuration
of the isolated system with timelike 4-momentum, we can restrict to
the special foliation whose leaves are the hyperplanes orthogonal to
the conserved total 4-momentum ({\it Wigner hyperplanes}). This
special foliation is intrinsically determined by the configuration of
the isolated system only. In this way\cite{lus} it is possible to
define the {\it Wigner-covariant rest-frame instant form of dynamics}
for every isolated system whose configurations have well defined and
finite Poincar\'e generators with timelike total 4-momentum
\footnote{See Ref.\cite{dir} for the traditional forms of dynamics.}
\footnote{See Appendix A for a review of parametrized Minkowski
theories and of the rest-frame instant form of dynamics.}.

This formulation provides a clarification of the roles of the various
relativistic centers of mass. This is a long standing problem which
arose just after the foundation of special relativity in the first
decade of the last century. In the next ninty years it became clear
that the definition of a relativistic center of mass is highly
non-trivial: no definition can enjoy all the properties of the
non-relativistic center of mass. See
Refs.\cite{pau,com1,mol,com2,com3,com4} for a partial bibliography of
all the existing attempts and Ref.\cite{re} for reviews.

As shown in Appendix A, in the rest-frame instant form on Wigner
hyperplanes only four first class constraints survive and the original
configuration variables $z^{\mu}(\tau ,\vec \sigma )$, ${\vec
\eta}_i(\tau )$ and their conjugate momenta $\rho_{\mu}(\tau ,\vec
\sigma )$, ${\vec \kappa}_i(\tau )$ are reduced to:

i) a decoupled particle ${\tilde x}^{\mu}_s(\tau )$, $p^{\mu}_s$ (the
only remnant of the spacelike hypersurface) with a positive mass
$\epsilon_s=\sqrt{\epsilon p^2_s}$ determined by the first class
constraint $\epsilon_s-M_{sys} \approx 0$ \footnote{$M_{sys}$ being
the invariant mass of the isolated system.} and with its rest-frame
Lorentz scalar time $T_s={{{\tilde x}_s\cdot p_s}\over {\epsilon_s}}$
put equal to the mathematical time as the gauge fixing $T_s-\tau
\approx 0$ to the previous constraint. Here, ${\tilde x}^{\mu}_s(\tau
)$ is a {\it non-covariant canonical} variable for the {\it external
4-center of mass}. After the elimination of $T_s$ and $\epsilon_s$
with the previous pair of second class constraints, one remains with a
decoupled free point ({\it point particle clock}) of mass $M_{sys}$
and canonical 3-coordinates ${\vec z}_s=\epsilon_s ({\vec {\tilde
x}}_s-{{{\vec p}_s}\over {p^o_s}} {\tilde x}^o)$, ${\vec k}_s={{{\vec
p}_s}\over {\epsilon_s}}$ \footnote{${\vec z}_s/\epsilon_s$ is the
classical analogue of the Newton-Wigner 3-position operator\cite{com1}
and, like it, is only covariant under the Euclidean subgroup of the
Poincar\'e group only.}. The non-covariant canonical ${\tilde
x}^{\mu}_s(\tau )$ must not be confused with the 4-vector
$x^{\mu}_s(\tau )=z^{\mu}(\tau ,\vec
\sigma =0)$ identifying the origin of the 3-coordinates $\vec \sigma$ inside
the Wigner hyperplanes. The worldline $x^{\mu}_s(\tau )$ is arbitrary
because it depends on $x^{\mu}_s(0)$ and its 4-velocity ${\dot
x}^{\mu}_s(\tau )$ depends on the Dirac multipliers associated with
the 4 first class constraints \footnote{Therefore this arbitrary
worldline may be considered as an arbitrary {\it centroid} for the
isolated system.}, as it will be shown in the next Section. The unit
timelike 4-vector $u^{\mu}(p_s)=p_s^{\mu}/\epsilon_s$ is orthogonal to
the Wigner hyperplanes and describes their orientation in the chosen
inertial frame.

ii) the particle canonical variables ${\vec \eta}_i(\tau )$, ${\vec
\kappa}_i(\tau )$ \footnote{They are Wigner spin-1 3-vectors, like the coordinates
$\vec \sigma$.} inside the Wigner hyperplanes. They are restricted by
the three first class constraints (the {\it rest-frame conditions})
${\vec \kappa}_{+}=\sum_{i=1}^N {\vec \kappa}_i \approx 0$.

Therefore, we need a doubling of the concepts:

1) there is the {\it external} viewpoint of an arbitrary inertial
Lorentz observer, who describes the Wigner hyperplanes, as leaves of a
foliation of Minkowski spacetime, determined by the timelike
configurations of the isolated system. A change of inertial observer
by means of a Lorentz transformation rotates the Wigner hyperplanes
and induces a Wigner rotation of the 3-vectors inside each Wigner
hyperplane. Every such hyperplane inherits an induced {\it internal
Euclidean structure} while an {\it external} realization of the
Poincar\'e group induces the {\it internal} Euclidean action.

As said above, an arbitrary worldline $x^{\mu}_s(\tau )$ is chosen as
origin of the {\it internal} 3-coordinates on the Wigner hyperplanes;
its velocity ${\dot x}^{\mu}_s(\tau )$ is determined only after the
introduction of four gauge fixings for the four first class
constraints (one of them is $T_s-\tau \approx 0$).

Three {\it external} concepts of 4-center of mass can be defined (each
one of which there has an {\it internal} 3-location inside the Wigner
hyperplanes):\hfill\break
\hfill\break
a) the {\it external} non-covariant canonical {\it 4-center of mass}
(also named {\it center of spin}\cite{com3}) ${\tilde x}^{\mu}_s$
(with 3-location ${\vec {\tilde \sigma}}$),\hfill\break
\hfill\break
b) the {\it external} non-covariant non-canonical M\o ller {\it
4-center of energy}\cite{mol} $R^{\mu}_s$ (with 3-location ${\vec
\sigma}_R$),\hfill\break
\hfill\break
c) the {\it external} covariant non-canonical Fokker-Pryce {\it
4-center of inertia}\cite{com2,com3} $Y^{\mu}_s$ (with 3-location
${\vec
\sigma}_Y$).

Only the canonical non-covariant center of mass ${\tilde
x}^{\mu}_s(\tau )$ is relevant in the Hamiltonian treatment with Dirac
constraints, while only the Fokker-Pryce $Y^{\mu}_s$ is a 4-vector by
construction

2) there is the {\it internal} viewpoint inside the Wigner hyperplanes
associated to a unfaithful {\it internal} realization of the
Poincar\'e algebra: the {\it internal} 3-momentum ${\vec
\kappa}_{+}$ vanishes due to the rest-frame conditions. The {\it internal}
energy and angular momentum are the invariant mass $M_{sys}$ and the
spin (the angular momentum with respect to ${\tilde x}^{\mu}_s(\tau
)$) of the isolated system, respectively.   Three {\it internal}
3-centers of mass: the {\it internal} canonical 3-center of mass can
be correspondingly defined, the {\it internal} M\o ller 3-center of
energy and the {\it internal} Fokker-Pryce 3-center of inertia. But,
due to the rest-frame conditions, they {\it coincide} and become
essentially the {it gauge} variable conjugate to ${\vec \kappa}_{+}$.
As a natural gauge fixing for the rest-frame conditions ${\vec
\kappa}_{+}\approx 0$, we can add the vanishing of the {\it internal}
Lorentz boosts: this is equivalent to locate the internal canonical
3-center of mass ${\vec q}_{+}$ in $\vec
\sigma =0$, i.e. in the origin $x^{\mu}_s(\tau )=z^{\mu}(\tau ,\vec 0)$. With
these gauge fixings and with $T_s-\tau \approx 0$, the worldline
$x^{\mu}_s(\tau )$ becomes uniquely determined except for the
arbitrariness in the choice of $x^{\mu}_s(0)$ [$\,
u^{\mu}(p_s)=p^{\mu}_s/\epsilon_s$]

\beq
x^{\mu}_s(\tau )=x^{\mu}_s(0) + u^{\mu}(p_s) T_s,
\label{I4}
\eeq

\noindent   and coincides with the {\it
external} covariant non-canonical Fokker-Pryce 4-center of inertia,
$x^{\mu}_s(\tau ) = x^{\mu}_s(0) + Y^{\mu}_s$.

This doubling of concepts replaces the separation of the
non-relativistic 3-center of mass due to the Abelian translation
symmetry. The non-relativistic conserved 3-momentum is replaced by the
{\it external} ${\vec p}_s=\epsilon_s {\vec k}_s$, while the {\it
internal} 3-momentum vanishes, ${\vec
\kappa}_{+}\approx 0$, as a definition of rest frame.

In the final gauge we have $\epsilon_s \equiv M_{sys}$, $T_s \equiv
\tau $ and the canonical basis ${\vec z}_s$, ${\vec k}_s$, ${\vec
\eta}_i$, ${\vec
\kappa}_i$ restricted by the three pairs of second class constraints
${\vec \kappa}_{+}=\sum_{i=1}^N{\vec \kappa}_i \approx 0$, ${\vec
q}_{+} \approx 0$, so that 6N canonical variables describe the N
particles like in the non-relativistic case. We still need a canonical
transformation ${\vec \eta}_i$, ${\vec \kappa}_i$ $\,\, \mapsto \,\,$
${\vec q}_{+} [\approx 0]$, ${\vec \kappa}_{+} [\approx 0]$, ${\vec
\rho}_a$, ${\vec \pi}_a$ ($a=1,..,N-1$) in order to identify a set of relative
canonical variables. The final 6N-dimensional canonical basis is
${\vec z}_s$, ${\vec k}_s$, ${\vec \rho}_a$, ${\vec \pi}_a$. To get
this result  we need a highly non-linear canonical transformation,
which can be obtained by exploiting the Gartenhaus-Schwartz singular
transformation \cite{garten}.

In the end, we obtain the Hamiltonian for relative motions as a sum of
N square roots, each one containing a squared mass and a quadratic
form in the relative momenta. This Hamiltonian goes into its
non-relativistic counterpart in the limit $c\,
\rightarrow \infty$. This fact has the following implications:

a) if one tries to make the inverse Legendre transformation to find
the associated Lagrangian, it turns out that, due to the presence of
square roots, the Lagrangian is a hyperelliptic function of ${\dot
{\vec
\rho}}_a$ already in the free case. A closed form exists only for N=2,
$m_1=m_2=m$: $L=-\epsilon m \sqrt{4-{\dot {\vec \rho}}^2}$. This
exceptional case already shows that the existence of the limiting
velocity  $c$ (i.e. of the light-cone) forbids the traditional linear
relation between the spin and the angular velocity.

b) the N quadratic forms in the relative momenta appearing in the
relative Hamiltonian cannot be diagonalized simultaneously. In any
case, the Hamiltonian is a sum of square roots, so that concepts like
{\it reduced masses}, {\it Jacobi normal relative coordinates} and
{\it tensor of inertia} cannot be extended to special relativity. As a
consequence, the orientation-shape SO(3) principal bundle of
Ref.\cite{little} can be defined only by using unspecified relative
coordinates.

c) the best way of studying rotational kinematics is by using the {\it
canonical spin bases} of Ref.\cite{iten2} with their {\it spin frames}
and {\it dynamical body frames}: they can be build in the same way as
in the non-relativistic case starting from the canonical basis ${\vec
\rho}_a$, ${\vec \pi}_a$.

Once these points are understood in the free case, the introduction of
mutual action-at-a-distance interactions among the particles can be
done without extra complications.

The paper is organized as follows. In Section II we  review the
rest-frame instant form on the Wigner hyperplane of N positive energy
free scalar particles. In Section III we discuss the {\it internal}
realization of the Poincar\'e algebra and we define the {\it internal
center-of-mass} concepts. In Section IV we discuss the {\it external}
realization of the Poincar\'e algebra and we  define the {\it external
center-of-mass} concepts. In Section V we construct the canonical
relative variables associated with the canonical {\it internal} center
of mass. In Section VI we analyze the relativistic rotational
kinematics of relative motions inside the Wigner hyperplane using the
same Hamiltonian methods for the construction of the spin bases of
Ref.\cite{iten2}. Some final comments on open problems are given in
the Conclusions.

Appendix A contains a review of parametrized Minkowski theories and of
the rest-frame instant form of dynamics. Some notations on spacelike
hypersurfaces are listed in Appendix B. The results of Section V are
extended to spinning particles in Appendix C. Some formulas  for the
Euler angles are reported in Appendix D. Finally, the treatment of the
3-body case is explicitly given in Appendix E.

\vfill\eject

\section{The Rest-Frame Instant Form of N Free Scalar Relativistic Particles}

Let us consider a  system of N free scalar positive-energy particles
in the framework of parametrized Minkowski theory (see Appendices A
and B).

The configuration variables are a 3-vector ${\vec
\eta}_i(\tau )$ for each particle [$x^{\mu }
_i(\tau )=z^{\mu }(\tau ,{\vec \eta}_i(\tau ))$] \cite{lus,albad}.
One has to choose the sign of the energy of each particle, because
there are not mass-shell constraints (like $\epsilon
p_i^2-m^2_i\approx 0$) at our disposal, due to the fact that one has
only three degrees of freedom for particle, determining the
intersection of a timelike trajectory and of the spacelike
hypersurface $\Sigma_{\tau}$. For each choice of the sign of the
energy of the N particles, one describes only one of the $2^N$
branches of the mass spectrum of the manifestly covariant approach
based on the coordinates $x^{\mu }_i(\tau )$, $p^{\mu }_i(\tau )$,
i=1,..,N, and on the constraints $\epsilon p^2_i-m^2_i\approx 0$ (in
the free case). In this way, one gets a description of relativistic
particles with a given sign of the energy with consistent couplings to
fields
\footnote{This is true for scalar positive-energy particles \cite{crater}. For
spinning positive-energy particles one has to add \cite{horace}
non-minimally some coupling of the spin to the electric field which
would be missed in the projection from the Lorentz covariant theory
(with $2^N$ branches of the mass spectrum) to the theory describing
only the branch, in which all the particles have positive energy. This
can be done by performing an approximate Foldy-Wouthuysen
transformation of the Lorentz covariant theory in  presence of
electromagnetic fields, because the electric field is the source of
possible crossings of the deformed $2^N$ branches (classical
counterpart of pair production). The additional spin-electric field
coupling is the source of the spin-orbit term in the quantum electron
Hamiltonian.}.

The system of N free scalar and positive energy particles is described
by the action\cite{lus,albad,crater}

\begin{eqnarray}
S&=& \int d\tau d^3\sigma \, {\cal L}(\tau ,\vec
\sigma )=\int d\tau L(\tau ),\nonumber \\
 &&{\cal L}(\tau ,\vec \sigma )=-\sum_{i=1}^N\delta^3(\vec \sigma -{\vec \eta}_i
(\tau ))m_i\sqrt{ g_{\tau\tau}(\tau ,\vec \sigma )+2g_{\tau {\check
r}} (\tau ,\vec \sigma ){\dot \eta}^{\check r}_i(\tau )+g_{{\check
r}{\check s}} (\tau ,\vec \sigma ){\dot \eta}_i^{\check r}(\tau ){\dot
\eta}_i^{\check s} (\tau )  },\nonumber \\
 &&L(\tau
)=-\sum_{i=1}^Nm_i\sqrt{ g_{\tau\tau}(\tau ,{\vec \eta}_i (\tau
))+2g_{\tau {\check r}}(\tau ,{\vec \eta}_i(\tau )){\dot \eta}^{\check
r}
_i(\tau )+g_{{\check r}{\check s}}(\tau ,{\vec \eta}_i(\tau )){\dot \eta}_i
^{\check r}(\tau ){\dot \eta}_i^{\check s}(\tau )  },
\label{II1}
\end{eqnarray}

\noindent where the configuration variables are $z^{\mu}(\tau ,\vec \sigma )$
and ${\vec \eta}_i(\tau )$, i=1,..,N. The action is invariant under separate
$\tau$- and $\vec \sigma$-reparametrizations.

The canonical momenta are

\begin{eqnarray}
\rho_{\mu}(\tau ,\vec \sigma )&=&-{ {\partial {\cal L}(\tau ,\vec \sigma )}
\over {\partial z^{\mu}_{\tau}(\tau ,\vec \sigma )} }=\sum_{i=1}^N\delta^3
(\vec \sigma -{\vec \eta}_i(\tau ))m_i\nonumber \\
 &&{{z_{\tau\mu}(\tau ,\vec \sigma )+z_{{\check r}\mu}(\tau ,\vec \sigma )
{\dot \eta}_i^{\check r}(\tau )}\over {\sqrt{g_{\tau\tau}(\tau ,\vec
\sigma )+ 2g_{\tau {\check r}}(\tau ,\vec \sigma ){\dot
\eta}_i^{\check r}(\tau )+ g_{{\check r}{\check s}}(\tau ,\vec \sigma
){\dot \eta}_i^{\check r}(\tau ){\dot
\eta}_i^{\check s}(\tau ) }} }=\nonumber \\
&=&[(\rho_{\nu}l^{\nu})l_{\mu}+(\rho_{\nu}z^{\nu}_{\check r})\gamma^{{\check r}
{\check s}}z_{{\check s}\mu}](\tau ,\vec \sigma ),\nonumber \\
 &&{}\nonumber \\
\kappa_{i{\check r}}(\tau )&=&-{ {\partial L(\tau )}\over {\partial {\dot
\eta}_i^{\check r}(\tau )} }=\nonumber \\
&=&m_i{ {g_{\tau {\check r}}(\tau ,{\vec \eta}_i(\tau ))+g_{{\check r}
{\check s}}(\tau ,{\vec \eta}_i(\tau )){\dot
\eta}_i^{\check s}(\tau )}\over { \sqrt{g_{\tau\tau}(\tau ,{\vec
\eta}_i(\tau ))+ 2g_{\tau {\check r}}(\tau ,{\vec \eta}_i(\tau )){\dot
\eta}_i^{\check r}(\tau )+ g_{{\check r}{\check s}}(\tau ,{\vec
\eta}_i(\tau )){\dot \eta}_i^{\check r} (\tau ){\dot \eta}_i^{\check
s}(\tau ) }} },\nonumber \\
 &&{}\nonumber \\
&&\lbrace z^{\mu}(\tau ,\vec \sigma ),\rho_{\nu}(\tau ,{\vec
\sigma}^{'}\rbrace
=-\eta^{\mu}_{\nu}\delta^3(\vec \sigma -{\vec \sigma}^{'}),\nonumber \\
&&\lbrace \eta^{\check r}_i(\tau ),\kappa_{j{\check s}}(\tau )\rbrace =
-\delta_{ij}\delta^{\check r}_{\check s}.
\label{II2}
\end{eqnarray}

The canonical Hamiltonian $H_{c}$ is zero,  the Dirac Hamiltonian is
given by Eq.(\ref{a3})[there are no other system-dependent primary
constraints] and Eqs.(\ref{a2}) become

\begin{eqnarray}
{\cal H}_{\mu}(\tau ,\vec \sigma )&=& \rho_{\mu}(\tau ,\vec \sigma )-l_{\mu}
(\tau ,\vec \sigma )\sum_{i=1}^N\delta^3(\vec \sigma -{\vec \eta}_i(\tau ))
\sqrt{ m^2_i-\gamma^{{\check r}{\check s}}(\tau ,\vec \sigma )
\kappa_{i{\check r}}(\tau )\kappa_{i{\check s}}(\tau ) }-\nonumber \\
&-&z_{{\check r}\mu}
(\tau ,\vec \sigma )\gamma^{{\check r}{\check s}}(\tau ,\vec \sigma )
\sum_{i=1}^N\delta^3(\vec \sigma -{\vec \eta}_i(\tau ))\kappa_{i{\check s}}
\approx 0.
\label{II3}
\end{eqnarray}

The conserved Poincar\'e generators are (the suffix ``s" denotes the
hypersurface $\Sigma_{\tau}$)

\begin{eqnarray}
&&p^{\mu}_s=\int d^3\sigma \rho^{\mu}(\tau ,\vec \sigma ),\nonumber \\
&&J_s^{\mu\nu}=\int d^3\sigma [z^{\mu}(\tau ,\vec \sigma )\rho^{\nu}(\tau ,
\vec \sigma )-z^{\nu}(\tau ,\vec \sigma )\rho^{\mu}(\tau ,\vec \sigma )].
\label{II4}
\end{eqnarray}

After the restriction to spacelike hyperplanes, the Dirac Hamiltonian
is reduced to Eq.(\ref{a8}) with the surviving ten constraints given
by

\begin{eqnarray}
{\tilde {\cal H}}^{\mu}(\tau )&=&\int d^3\sigma {\cal H}^{\mu}(\tau ,\vec
\sigma )=\, p^{\mu}_s-l^{\mu}\sum_{i=1}^N\sqrt{m^2_i+{\vec \kappa}^2_i
(\tau )}+b^{\mu}_{\check r}(\tau )\sum_{i=1}^N\kappa_{i{\check r}}(\tau )
\approx 0,\nonumber \\
{\tilde {\cal H}}^{\mu\nu}(\tau )&=&b^{\mu}_{\check r}(\tau )\int d^3\sigma
\sigma^{\check r}\, {\cal H}^{\nu}(\tau ,\vec \sigma )-b^{\nu}_{\check r}(\tau )
\int d^3\sigma \sigma^{\check r}\, {\cal H}^{\mu}(\tau ,\vec \sigma )=
\nonumber \\
&=&S_s^{\mu\nu}(\tau )-[b^{\mu}_{\check r}(\tau
)b^{\nu}_{\tau}-b^{\nu}_{\check r}(\tau
)b^{\mu}_{\tau}]\sum_{i=1}^N\eta_i^{\check r}(\tau )
\sqrt{m^2_i+{\vec \kappa}^2_i(\tau )}-\nonumber \\
&-&[b^{\mu}_{\check r}(\tau )b^{\nu}_{\check s}(\tau )-b^{\nu}_{\check r}(\tau )
b^{\mu}_{\check s}(\tau )]\sum_{i=1}^N\eta_i^{\check r}(\tau )\kappa_i^{\check
s}(\tau )\approx 0.
\label{II5}
\end{eqnarray}

Here $S^{\mu\nu}_s$ is the spin part of the Lorentz generators

\begin{eqnarray}
J^{\mu\nu}_s&=&x^{\mu}_sp^{\nu}_s-x^{\nu}_sp^{\mu}_s+S^{\mu\nu}_s,\nonumber \\
&&S^{\mu\nu}_s=b^{\mu}_{\check r}(\tau )\int d^3\sigma \sigma^{\check r}
\rho^{\nu}(\tau ,\vec \sigma )-b^{\nu}_{\check r}(\tau )\int d^3\sigma
\sigma^{\check r}\rho^{\mu}(\tau ,\vec \sigma ).
\label{II6}
\end{eqnarray}

On the Wigner hyperplane we have the following constraints and Dirac
Hamiltonian\cite{lus,crater}

\begin{eqnarray}
{\tilde {\cal H}}^{\mu}(\tau )&=&p^{\mu}_s-u^{\mu}(p_s) \sum_{i=1}^N
\sqrt{m_i
^2+{\vec \kappa}_i^2} +\epsilon^{\mu}_r(u(p_s)) \sum_{i=1}^N\kappa_{ir}=
\nonumber \\
&=&u^{\mu}(p_s) [\epsilon_s-\sum_{i=1}^N\sqrt{m_i^2+ {\vec
\kappa}_i^2}] +\epsilon^{\mu}_r(u(p_s)) \sum_{i=1}^N \kappa_{ir}
\approx 0,\nonumber \\
&&{}\nonumber \\ &&or\nonumber \\
 &&{}\nonumber \\
 &&\epsilon_s-M_{sys}\approx 0,\quad\quad
M_{sys}=\sum_{i=1}^N\sqrt{m_i^2+{\vec \kappa}_i^2},\nonumber \\
 &&{\vec p}_{sys} = {\vec
\kappa}_{+}=\sum_{i=1}^N {\vec \kappa}_i \approx 0,\nonumber \\
 &&{}\nonumber \\
 H_D&=&
\lambda^{\mu}(\tau ) {\tilde {\cal H}}_{\mu}(\tau )=
\lambda (\tau ) [\epsilon_s-M_{sys}]-\vec \lambda (\tau ) \sum_{i=1}^N {\vec \kappa}_i,
\nonumber \\
&&{}\nonumber \\
&&\lambda (\tau ) \approx -{\dot x}_{s \mu}(\tau )u^{\mu}(p_s),\nonumber \\
&&\lambda_r(\tau )\approx -{\dot x}_{s \mu}(\tau )\epsilon^{\mu}_r(u(p_s)),
\nonumber \\
&&{}\nonumber \\
 {\dot {\tilde x}}^{\mu}_s(\tau ) &=&-\lambda (\tau )
u^{\mu}(p_s),\nonumber \\
 {\dot x}_s^{\mu}(\tau ) &\approx& -
{\tilde \lambda}^{\mu}(\tau )=-\lambda (\tau )
u^{\mu}(p_s)+\epsilon^{\mu}_r(u(p_s)) \lambda_r(\tau ).
\label{II7}
\end{eqnarray}

While the Dirac multiplier $\lambda (\tau )$  is determined by the
gauge fixing $T_s-\tau \approx 0$, the 3 Dirac's multipliers $\vec
\lambda (\tau )$ describe the classical zitterbewegung of the origin
of the coordinates on the Wigner hyperplane: each gauge-fixing $\vec
\chi (\tau )\approx 0$ to the 3 first class constraints ${\vec
\kappa}_{+}\approx 0$ (defining the {\it internal rest-frame}) gives a
different determination of the multipliers $\vec
\lambda (\tau )$ and therefore  identifies a different worldline for the
covariant non-canonical origin $x^{(\vec \chi )\mu}_s(\tau )$ which
induces the  definition $\vec \chi$ of the {\it internal 3-center of
mass} conjugate to ${\vec \kappa}_{+}$
\footnote{Naturally each choice $\vec \chi$ leads to a different set of
relative canonical conjugate variables.}.

The embedding describing Wigner hyperplanes is $z^{\mu}(\tau ,\vec
\sigma ) = x^{\mu}_s(\tau ) + \epsilon^{\mu}_r(u(p_s)) \sigma^r$, with the
$\epsilon^{\mu}_r(u(p_s))$ defined in Eqs.(\ref{b13}).

The various spin tensors and vectors are \cite{lus}

\begin{eqnarray}
J^{\mu\nu}_s&=&x^{\mu}_s p^{\nu}_s- x^{\nu}_s p^{\mu}_s+ S^{\mu\nu}_s=
{\tilde x}^{\mu}_s p^{\nu}_s - {\tilde x}^{\nu}_s p^{\mu}_s +{\tilde S}
^{\mu\nu}_s,\nonumber \\
&&{}\nonumber \\
S^{\mu\nu}_s&=&[u^{\mu}(p_s)\epsilon^{\nu}(u(p_s))-u^{\nu}(p_s)\epsilon^{\mu}
(u(p_s))] {\bar S}^{\tau r}_s+\epsilon^{\mu}(u(p_s))\epsilon^{\nu}(u(p_s))
{\bar S}^{rs}_s\equiv \nonumber \\
&\equiv& \Big[ \epsilon^{\mu}_r(u(p_s)) u^{\nu}(p_s)-\epsilon^{\nu}
(u(p_s)) u^{\mu}(p_s)\Big] \sum_{i=1}^N \eta^r_i \sqrt{m^2_ic^2+{\vec \kappa}
_i^2}+\nonumber \\
&+&\Big[ \epsilon^{\mu}_r(u(p_s)) \epsilon^{\nu}_s(u(p_s))-\epsilon^{\nu}_r
(u(p_s)) \epsilon^{\mu}_r(u(p_s))\Big] \sum_{i=1}^N \eta^r_i\kappa^s_i,
\nonumber \\
&&{}\nonumber \\
{\bar S}^{AB}_s&=&\epsilon^A_{\mu}(u(p_s)) \epsilon^B_{\nu}(u(p_s)) S^{\mu\nu}_s
,\nonumber \\
&&{\bar S}^{rs}_s\equiv \sum_{i=1}^N(\eta^r_i\kappa^s_i-
\eta^s_i\kappa^r_i),\quad {\bar S}^{\tau r}_s\equiv -
\sum_{i=1}^N\eta^r_i\sqrt{m^2_ic^2+{\vec \kappa}_i^2},\nonumber \\
&&{}\nonumber \\ {\tilde S}^{\mu\nu}_s&=&S^{\mu\nu}_s+{1\over
{\sqrt{\epsilon p^2_s}(p^o_s+
\sqrt{\epsilon p^2_s})}}\Big[ p_{s\beta}(S^{\beta\mu}_s p^{\nu}_s-S^{\beta\nu}_s
p^{\mu}_s)+\sqrt{p^2_s}(S^{o\mu}_s p^{\nu}_s-S^{o\nu}_s p^{\mu}_s)\Big],
\nonumber \\
&&{\tilde S}^{ij}_s=\delta^{ir}\delta^{js} {\bar S}_s^{rs},\quad\quad
{\tilde S}^{oi}_s=-{{\delta^{ir} {\bar S}^{rs}_s\, p^s_s}\over {p^o_s+
\sqrt{\epsilon p^2_s}}},\nonumber \\
&&{}\nonumber \\ {\vec  {\bar S}} &\equiv & {\vec {\bar
S}}=\sum_{i+1}^N{\vec
\eta}_i\times {\vec
\kappa}_i\approx
\sum_{i=1}^N {\vec \eta}_i\times {\vec \kappa}_i-{\vec \eta}_{+}\times
{\vec \kappa}_{+} = \sum_{a=1}^{N-1} {\vec \rho}_a\times {\vec \pi}_a.
\label{II8}
\end{eqnarray}

Let us remark that while
$L^{\mu\nu}_s=x^{\mu}_sp^{\nu}_s-x^{\nu}_sp^{\mu}_s$ and
$S^{\mu\nu}_s$ are not constants of the motion due to the classical
zitterbewegung, both ${\tilde L}_s^{\mu\nu}={\tilde
x}^{\mu}_sp_s^{\nu}-{\tilde x}^{\nu}_sp^{\mu}_s$ and ${\tilde
S}^{\mu\nu}_s$ are conserved.

The only remaining canonical variables describing the Wigner
hyperplane in the final Dirac brackets are the non-covariant canonical
coordinate ${\tilde x}^{\mu}_s(\tau )$ and $p^{\mu}_s$. The point with
coordinates ${\tilde x}^{\mu}_s(\tau )$ is the decoupled canonical
{\it external 4-center of mass}, playing the role of a kinematical
external 4-center of mass and of a decoupled observer with his
parametrized clock ({\it point particle clock}). Its velocity ${\dot
{\tilde x}}^{\mu}_s(\tau )$ is parallel to $p^{\mu}_s$, so that it has
no classical zitterbewegung.

The connection between $x^{\mu}_s(\tau )$ and ${\tilde x}_s^{\mu}(\tau
)$ is given in Eq.(\ref{IV1}) in Section IV. Let us remark that the
constant $x^{\mu}_s(0)$ [and ${\tilde x}
^{\mu}_s(0)$] is arbitrary, reflecting the arbitrariness in the absolute
location of the origin of the {\it internal} coordinates on each
hyperplane in Minkowski spacetime.

After the separation of the relativistic canonical non-covariant {\it
external} 4-center of mass ${\tilde x}_s^{\mu}(\tau )$, on the Wigner
hyperplane the N particles are described by the 6N Wigner spin-1
3-vectors ${\vec \eta}_i(\tau )$, ${\vec \kappa}_i (\tau )$ restricted
by the rest-frame condition ${\vec \kappa}_{+}=\sum^N
_{i=1} {\vec \kappa}_i \approx 0$.

The canonical variables ${\tilde x}^{\mu}_s$, $p^{\mu}_s$ for the {\it
external} 4-center of mass, may be replaced by the canonical
pairs\cite{ll}\footnote{They make explicit the interpretation of
${\tilde x}^{\mu}_s$ as a point particle clock.}

\begin{eqnarray}
T_s&=& {{p_s\cdot {\tilde x}_s}\over {\epsilon_s}}={{p_s
 \cdot x_s}\over {\epsilon_s}},\nonumber \\
 \epsilon_s&=&\pm \sqrt{\epsilon p^2_s},\nonumber \\
 {\vec z}_s&=&\epsilon_s ({\vec {\tilde
x}}_s- {{ {\vec p}_s}\over {p^o_s}} {\tilde x}^o_s), \nonumber \\
 {\vec k}_s&=&{{ {\vec p}_s}\over {\epsilon_s}},
\label{II9}
\end{eqnarray}

\noindent with the inverse transformation

\begin{eqnarray}
{\tilde x}^o_s&=&\sqrt{1+{\vec k}
_s^2}(T_s+{{{\vec k}_s\cdot {\vec z}_s}\over {\epsilon_s}}), \nonumber \\
 {\vec {\tilde x}}_s&=&{{{\vec z}_s}\over {\epsilon_s}}+(T_s+{{{\vec
k}_s\cdot {\vec z}_s}\over {\epsilon_s}}){\vec k}_s, \nonumber \\
 p^o_s&=&\epsilon_s \sqrt{1+{\vec k}_s^2}, \nonumber \\
 {\vec p}_s&=&\epsilon_s {\vec k}_s.
\label{II10}
\end{eqnarray}

This non-point canonical transformation in the rest-frame instant form
can be summarized as   [$\epsilon_s - M_{sys} \approx 0$, ${\vec
\kappa}_{+}=\sum_{i=1}^N {\vec \kappa}_i \approx 0$]

\begin{equation}
\begin{minipage}[t]{5cm}
\begin{tabular}{|l|l|} \hline
${\tilde x}_s^{\mu}$& ${\vec \eta}_i$ \\  \hline
 $p^{\mu}_s$& ${\vec \kappa}_i$ \\ \hline
\end{tabular}
\end{minipage} \ {\longrightarrow \hspace{1cm}} \
\begin{minipage}[t]{5 cm}
\begin{tabular}{|l|l|l|} \hline
$\epsilon_s$& ${\vec z}_s$   & ${\vec \eta}_i$   \\ \hline
 $T_s$ & ${\vec k}_s$& ${\vec \kappa}_i$ \\ \hline
\end{tabular}
\end{minipage}
\label{II11}
\end{equation}

The invariant mass $M_{sys}$ of the system, which is also the {\it
internal} energy of the isolated system, replaces the non-relativistic
Hamiltonian $H_{rel}$ for the relative degrees of freedom, after the
addition of the gauge-fixing $T_s-\tau
\approx 0$ \footnote{Implying $\lambda (\tau )=-\epsilon$ and
identifying the time parameter $\tau$, that labels the leaves of the
foliation  with the Lorentz scalar time of the center of mass in the
rest frame, $T_s=p_s\cdot {\tilde x}_s/M_{sys}$; $M_{sys}$  generates
the evolution in this time.}: this reminds of the frozen
Hamilton-Jacobi theory, in which the time evolution can be
reintroduced by using the energy generator of the Poincar\'e group as
Hamiltonian \footnote{See Refs.\cite{pon} for a different derivation
of this result.}.

After the gauge fixings $T_s-\tau \approx 0$, the final Hamiltonian
and the embedding of the Wigner hyperplane into Minkowski spacetime
become

\bea
 H_D&=& M_{sys} -\vec \lambda (\tau ) \cdot {\vec \kappa}_{+} ,\nonumber \\
 &&{}\nonumber \\
z^{\mu}(\tau ,\vec \sigma ) &=& x^{\mu}_s(\tau ) + \epsilon^{\mu
}_r(u(p_s)) \sigma^r = x^{\mu}_s(0) + u^{\mu}(p_s) \tau +
\epsilon^{\mu}_r(u(p_s)) \sigma^r,\nonumber \\
 &&{}\nonumber \\
 &&with \nonumber \\
&&{}\nonumber \\
 {\dot x}^{\mu}_s(\tau )\, &{\buildrel \circ \over =}& {{d\, x^{\mu}_s(\tau )}\over
 {d \tau}}+ \{ x^{\mu}_s(\tau ), H_D \} = u^{\mu}(p_s)+\epsilon^{\mu}_r(u(p_s))
\lambda_r(\tau ),
\label{II12}
\eea

\noindent where $x^{\mu}_s(0)$ is an arbitrary point and $\epsilon^{\mu
}_r(u(p_s)) =L^{\mu}{}_r(p_s,{\buildrel \circ \over p}_s)$.  This
equation visualizes the role of the Dirac multipliers as sources of
the classical zittebewegung.

After the gauge fixings $T_s-\tau \approx 0$, ${\vec q}_{+}\approx 0$,
the embedding $z^{\mu}(\tau ,\vec \sigma ) = x^{\mu}_s(\tau )
+\epsilon^{\mu}_r(u(p_s)) \sigma^r$ describing Wigner hyperplanes
becomes $z^{\mu}(\tau ,\vec \sigma ) = x^{\mu}_s(0)+ u^{\mu}(p_s) \tau
+ \epsilon^{\mu}_r(u(p_s)) \sigma^r$.

The particles' worldlines in Minkowski spacetime and the associated
momenta are

\begin{eqnarray}
x^{\mu}_i(\tau )&=&z^{\mu}(\tau ,{\vec \eta}_i(\tau ))=x^{\mu}_s(\tau
)+\epsilon^{\mu}_r(u(p_s)) \eta^r_i(\tau ),\nonumber \\
 p^{\mu}_i(\tau )&=&\sqrt{m^2_i+{\vec \kappa}^2_i(\tau )} u^{\mu}(p_s) +
\epsilon^{\mu}_r(u(p_s)) \kappa_{ir}(\tau )\,\, \Rightarrow \epsilon p^2_i=m_i^2.
\label{II13}
\end{eqnarray}

Inside the  Wigner hyperplane three  degrees of freedom of the
isolated system \footnote{They describe an {\it internal}
center-of-mass 3-variable ${\vec \sigma}_{com}$ defined inside the
Wigner hyperplane and conjugate to ${\vec \kappa}_{+}$; when the
${\vec \sigma}_{com}$ are canonical variables they are denoted ${\vec
q}_{+}$.} become gauge variables. To eliminate the three first class
constraints ${\vec \kappa}_{+} \approx 0$ the natural gauge fixing  is
$\vec \chi ={\vec q}_{+}\approx 0$ implying $\lambda_{\check r}(\tau
)=0$: in this way the {\it internal} 3-center of mass gets located in
the origin $x^{\mu}_s(\tau )=z^{\mu}(\tau ,\vec
\sigma =0)$ of the Wigner hyperplane. The determination of ${\vec q}_{+}$
for the N particle system will be done with the group theoretical
methods of Ref.\cite{pauri} in the next Section.

The same problem arises when one considers the rest-frame description
of fields. A basis with a {\it center of phase} has already been found
for a real Klein-Gordon field both in the covariant approach\cite{lon}
and on spacelike hypersurfaces \cite{mate}\footnote{In this paper
there is a first treatment of the topics which will be treated in
Sections III and IV}. In this case also the {\it internal} center of
mass has been found, but not yet a canonical basis containing it.

It turns out that the Wigner hyperplane is the natural setting for the
study of the Dixon multipoles of extended relativistic
systems\cite{dixon} \footnote{In a next paper we will study Dixon's
multipoles for the N-body problem \cite{iten3}.} and for defining
their canonical relative variables with respect to the center of mass.
Also, the Wigner hyperplane with its natural Euclidean metric
structure offers a natural solution to the problem of boost for
lattice gauge theories and realizes explicitly the Machian view of
dynamics according to which  only relative motions are relevant.

The {\it external} rest-frame instant form realization of the
Poincar\'e generators \footnote{There are  four independent
Hamiltonians $p^o_s$ and $J^{oi}_s$ functions of the system invariant
mass $M_{sys}$; we give also the expression in the basis $T_s$,
$\epsilon_s$, ${\vec z}_s$, ${\vec k}_s$.} with non-fixed invariants
$\epsilon p^2_s = \epsilon_s^2 \approx M^2_{sys}$, $-\epsilon p^2_s
{\vec {\bar S}}_s^2 \approx -\epsilon M^2_{sys} {\vec {\bar S}}^2$, is
obtained from Eq.(\ref{II8}):

\begin{eqnarray}
p^{\mu}_s,&&\nonumber  \\
 J^{\mu\nu}_s&=&{\tilde x}^{\mu}_sp^{\nu}_s-{\tilde x}^{\nu}_s p^{\mu}_s
 + {\tilde S}^{\mu\nu}_s,\nonumber \\
 &&{}\nonumber \\
 p^o_s&=& \sqrt{\epsilon_s^2+{\vec p}_s^2}= \epsilon_s \sqrt{1+ {\vec
k}_s^2}\approx  \sqrt{M^2_{sys}+{\vec p}^2_s}=M_{sys} \sqrt{1+{\vec
k}_s^2},\nonumber \\
 {\vec p}_s&=& \epsilon_s{\vec k}_s\approx M_{sys} {\vec k}_s,
\nonumber \\
 J^{ij}_s&=&{\tilde x}^i_sp^j_s-{\tilde x}^j_sp^i_s +
\delta^{ir}\delta^{js}\sum_{i=1}^N(\eta^r_i\kappa^s_i-\eta^s_i\kappa^r_i)=
z^i_sk^j_s-z^j_sk^i_s+\delta^{ir}\delta^{js} \epsilon^{rsu}{\bar
S}^u_s,\nonumber \\
 K^i_s&=&J^{oi}_s= {\tilde x}^o_sp^i_s-{\tilde x}^i_s
\sqrt{\epsilon_s^2+{\vec p}_s^2}-{1\over {\epsilon_s+\sqrt{\epsilon^2_s+
{\vec p}_s^2}}} \delta^{ir} p^s_s \sum_{i=1}^N(\eta^r_i\kappa^s_i-
\eta^s_i\kappa^r_i)=\nonumber \\
 &=&-\sqrt{1+{\vec k}_s^2} z^i_s-{{\delta^{ir} k^s_s\epsilon
^{rsu}{\bar S}^u_s}\over {1+\sqrt{1+{\vec k}_s^2} }}\approx {\tilde x}^o_sp^i_s
-{\tilde x}^i_s\sqrt{M^2_{sys}+{\vec p}^2_s}-{{\delta^{ir}p^s_s\epsilon^{rsu}{\bar
S}^u_s}\over {M_{sys}+\sqrt{M_{sys}^2+{\vec p}^2_s}}}.
\label{II14}
\end{eqnarray}

On the other hand, the {\it internal} realization of the Poincar\'e
algebra is built inside the Wigner hyperplane by using the expression
of ${\bar S}_s^{AB}$ given by Eq.(\ref{II8}) \footnote{This {\it
internal Poincar\'e algebra realization} must not be confused with the
previous {\it external} one based on ${\tilde S}^{\mu\nu}_s$; $\Pi$
and $W^2$ are the two non-fixed invariants of this realization.}

\begin{eqnarray}
&&M_{sys}=H_M=\sum_{i=1}^N  \sqrt{m^2_i+{\vec
\kappa}_i^2},\nonumber \\
 &&{\vec \kappa}_{+}=\sum_{i=1}^N {\vec
\kappa}_i\, (\approx 0),\nonumber \\
 &&\vec J=\sum_{i=1}^N {\vec \eta}_i\times {\vec
\kappa}_i,\quad\quad J^r={\bar S}^r={1\over 2}\epsilon^{ruv}{\bar
S}^{uv} \equiv {\bar S}^r_s,\nonumber \\
 &&\vec K=- \sum_{i=1}^N
\sqrt{m^2_i+{\vec \kappa}_i^2}\,\, {\vec
\eta}_i,\quad\quad K^r=J^{or}={\bar S}_s^{\tau r}, \nonumber \\
&&{}\nonumber \\
 &&\Pi = M^2_{sys}-{\vec \kappa}_{+}^2 \approx M^2_{sys} > 0,\nonumber \\
 &&W^2=-\epsilon (M^2_{sys}-{\vec \kappa}^2_{+}) {\vec {\bar
S}}^2_s \approx -\epsilon M^2_{sys} {\vec {\bar S}}^2_s.
\label{II15}
\end{eqnarray}

The meaning of the constraints $\epsilon_s-M_{sys}\approx 0$, ${\vec
\kappa}_{+}\approx 0$ is: i) the constraint $\epsilon_s-M_{sys}\approx 0$
is the bridge connecting the {\it external} and {\it internal}
realizations \footnote{The external spin coincides with the internal
angular momentum due to Eqs.(\ref{a11}).}; ii) the constraints ${\vec
\kappa}_{+}\approx 0$, together with $\vec K\approx 0$ \footnote{As we shall
see in the next Section $\vec K\approx 0$ is implied by the natural
gauge fixing ${\vec q}_{+}\approx 0$.}, imply a unfaithful {\it
internal} realization in which the only non-zero generators are the
conserved energy and the spin of an isolated system.

For isolated systems the constraint manifold\cite{india} is a
stratified manifold with each stratum corresponding to a type of
Poincar\'e orbit. The main stratum (dense in the constraint manifold)
corresponds to all configurations of the isolated system belonging to
timelike Poincar\'e orbits with $\epsilon p^2_s\approx \epsilon
M^2_{sys} > 0$. As said in Ref.\cite{lucenti}, this implies that the
center-of-mass coordinates have been adapted to the co-adjoint orbits
of the Poincar\'e group. But this canonical basis does not yet
correspond to a typical form of the Poincar\'e group\cite{pauri2} in
its canonical action on the phase space of the isolated system,
because the second Poincar\'e invariant \footnote{The Pauli-Lubanski
invariant ${\vec W}_s^2=-p_s^2{\vec S}_s^2$.} does not appear among
the canonical variables. In Ref.\cite{lucenti} a canonical basis
including both Poincar\'e invariants was found (all the coordinates
are adapted to the co-adjoint action of the Poincar\'e group). As a
consequence the new relative variables are adapted to the SO(3) group.

In Ref.\cite{lus} a naive {\it internal} center-of-mass variable
${\vec \eta}_{+} ={1\over N}\sum_{i=1}^N {\vec \eta}_i$ was introduced
and there was the definition of relative variables ${\vec \rho}_a$,
${\vec \pi}_a$ with the following point canonical transformation

\begin{eqnarray}
&&\begin{minipage}[t]{3cm}
\begin{tabular}{|l|} \hline
${\vec \eta}_i$ \\  \hline
 ${\vec \kappa}_i$ \\ \hline
\end{tabular}
\end{minipage} \ {\longrightarrow \hspace{.2cm}} \
\begin{minipage}[t]{2 cm}
\begin{tabular}{|l|l|} \hline
${\vec \eta}_{+}$   & ${\vec \rho}_a$   \\ \hline
 ${\vec \kappa}_{+}$&${\vec \pi}_a$ \\ \hline
\end{tabular}
\end{minipage}  , \quad\quad a=1,..,N-1\nonumber \\
&&{}\nonumber \\
 {\vec \eta}_i&=&{\vec \eta}_{+}+{1\over {\sqrt{N}}}
\sum_{a=1}^{N-1}
\gamma_{ai}{\vec \rho}_a,\nonumber \\
{\vec \kappa}_i&=&{1\over N} {\vec \kappa}_{+}+\sqrt{N} \sum_{a=1}^{N-1}
\gamma_{ai} {\vec \pi}_a,\nonumber \\
&&{}\nonumber \\
{\vec \eta}_{+}&=&{1\over N} \sum_{i=1}^N {\vec \eta}_i,\nonumber \\
{\vec \kappa}_{+}&=&\sum_{i=1}^N {\vec \kappa}_i\approx 0,\nonumber \\
{\vec \rho}_a&=&\sqrt{N} \sum_{i=1}^N \gamma_{ai} {\vec \eta}_i,\nonumber \\
{\vec \pi}_a&=&{1\over {\sqrt{N}}} \sum_{i=1}^N \gamma_{ai} {\vec
\kappa}_i,\nonumber \\
&&{}\nonumber \\
&&\{ \eta_i^r,\kappa^s_j\} =\delta_{ij}\delta^{rs},\quad\quad
\{ \eta^r_{+},\kappa^s_{+} \} =\delta^{rs},\quad\quad \{ \rho^r_a,\pi^s_b \} =
\delta_{ab}\delta^{rs},\nonumber \\
&&{}\nonumber \\
&&\sum_{i=1}^N\gamma_{ai}=0,\quad\quad \sum_{a=1}^{N-1}\gamma_{ai}
\gamma_{aj}=\delta_{ij}-{1\over N},\quad\quad
\sum_{i=1}^N\gamma_{ai}\gamma_{bi}=\delta_{ab},
\label{II16}
\end{eqnarray}

This is a family of canonical transformations depending on ${1\over
2}(N-1)(N-2)$ free parameters (the independent parameters in the
$\gamma_{ai}$ \cite{luu}).

Let us see whether we can take ${\vec \sigma}_{sys}={\vec \eta}_{+}$.

In the gauge\footnote{It entails $0\approx {\dot T}_s-1 ={\dot x}_s\cdot u(p_s)-1=
-\lambda (\tau )-1$; after going to Dirac brackets we get $T_s \equiv \tau$ and
 $\epsilon_s\equiv \pm M_{sys}$.}

\begin{equation}
T_s - \tau \approx 0,
\label{II17}
\end{equation}

\noindent
 the Hamiltonian and
the rest-frame constraints are

\begin{eqnarray}
H_D&=&M_{sys}-\vec \lambda (\tau )\cdot {\vec \kappa}_{+},\nonumber \\
 {\vec \kappa}_{+} &\approx& 0.
\label{II18}
\end{eqnarray}

\noindent with the invariant mass given by

\begin{eqnarray}
M_{sys} &=& H_M=\nonumber \\
 &=&\sum_{i=1}^N  \sqrt{m_i^2+{\vec \kappa}_i^2}=\sum_{i=1}^N
\sqrt{m^2_i+[{1\over N}{\vec \kappa}_{+}+\sqrt{N}\sum_{a=1}^{N-1}\gamma_{ai}
{\vec \pi}_a]^2}\approx \nonumber \\
 &\approx& \sum_{i=1}^N
\sqrt{m^2_i+N \sum_{a,b}^{1..N-1}\gamma_{ai}\gamma
_{bi} {\vec \pi}_a\cdot {\vec \pi}_b}.
\label{II19}
\end{eqnarray}

For the origin of coordinates we get

\begin{equation}
x^{\mu}_s(T_s)=x^{\mu}_s(0)+u^{\mu}(p_s) T_s +\epsilon^{\mu}_r(u(p_s))
\int_0^{T_s} d\tau \lambda_r(\tau ).
\label{II20}
\end{equation}

The Hamilton equations are [$\tau \equiv T_s$]

\begin{eqnarray}
{\dot {\vec \eta}}_i(\tau )\, &{\buildrel \circ \over =}\,& {{ {\vec
\kappa}_i(\tau )}\over {\sqrt{m^2_i+{\vec \kappa}_i^2(\tau )}}} - \vec \lambda
(\tau ),\nonumber \\
 {\dot {\vec \kappa}}_i(\tau )\, &{\buildrel \circ \over
=}\,& 0,\nonumber \\
\Rightarrow && {\vec \kappa}_i(\tau )\, {\buildrel \circ \over =}\, m_i
{{ [{\dot {\vec \eta}}_i+\vec \lambda ](\tau )}\over {\sqrt{1- ({\dot
{\vec \eta}}_i(\tau )+\vec \lambda (\tau ))^2} }},\nonumber \\
 &&\sqrt{m^2_i+{\vec \kappa}_i^2(\tau )}\, {\buildrel \circ \over =}\,
{{m_i}\over {\sqrt{1-({\dot {\vec \eta}}_i(\tau )+\vec
\lambda (\tau ))^2} }},\nonumber \\
\Rightarrow && L_D=\sum_{i=1}^N{\vec \kappa}_i\cdot {\dot {\vec
\eta}}_i-H_D=\sum_{i=1}^N{\vec \kappa}_i\cdot ({\dot {\vec \eta}}_i+\vec
\lambda )-H_M=\nonumber \\
&&=-\sum_{i=1}^N m_i \sqrt{1-({\dot {\vec
\eta}}_i+\vec \lambda )^2}=\nonumber \\
&&=-\sum_{i=1}^Nm_i \sqrt{1- {\Big[ {\dot {\vec
\eta}}_{+}+
\vec \lambda +{1\over {\sqrt{N}}}\sum_{a=1}^{N-1}\gamma_{ai}{\dot {\vec \rho}}
_a \Big] }^2 }.
\label{II21}
\end{eqnarray}

\noindent with Euler-Lagrange equations for ${\vec \eta}_i(\tau )$ and $\vec \lambda (\tau )$

\begin{eqnarray}
&&{d\over {d\tau}} \Big[ m_i {{{\dot {\vec \eta}}_i(\tau )+\vec
\lambda (\tau )}\over { \sqrt{1-({\dot {\vec \eta}}_i(\tau )+\vec
\lambda (\tau ))^2} }} \Big]
\, {\buildrel \circ \over =}\, 0,\nonumber \\
&&\sum_{i=1}^N m_i {{{\dot {\vec \eta}}_i(\tau )+\vec \lambda (\tau
)}\over {\sqrt{1-({\dot {\vec
\eta}}_i(\tau )+\vec \lambda (\tau ))^2} }} \, {\buildrel \circ \over =}\, 0.
\label{II22}
\end{eqnarray}

If $\vec \lambda (\tau )={\dot {\vec g}}(\tau )$, the solutions of the Hamilton
and Euler-Lagrange equations are

\begin{eqnarray}
&&{\vec \kappa}_i(\tau )\, {\buildrel \circ \over =}\, {\vec \beta}_i,\quad
with\quad {\vec \kappa}_{+}\, {\buildrel \circ \over =}\, \sum_{i=1}^N{\vec
\beta}_i =0,\nonumber \\
&&{\vec \eta}_i(\tau)+\vec g(\tau)\, {\buildrel \circ \over =}\,
  {\vec \alpha}_i
+ \tau { {{\vec \beta}_i} \over
         { \sqrt{m^2_i+ {\vec \beta}_i^2} } },\nonumber \\
&&\Rightarrow \quad {\vec \eta}_{+}(\tau )+\vec g(\tau )\,{\buildrel \circ
\over =}\, {1\over N}\sum_{i=1}^N {\vec \alpha}_i+{{\tau}\over N}\sum_{i=1}^N
{{{\vec \beta}_i}\over {\sqrt{m^2_i+{\vec \beta}_i^2}}},\nonumber
\\ &&given\quad {\vec \eta}_{+}\quad \Rightarrow \vec g,\nonumber \\
&&{\vec \rho}_a(\tau )\, {\buildrel \circ \over =}\, \sqrt{N}
\sum_{i=1}^N\gamma_{ai}{\vec \alpha}_i+\tau \sqrt{N}\sum_{i=1}^N{{\gamma_{ai}
{\vec \beta}_i}\over {\sqrt{m^2_i+{\vec \beta}_i^2}}},\nonumber \\
&&{\vec \pi}_a(\tau )\, {\buildrel \circ \over =}\, {1\over
{\sqrt{N}}}\sum_{i=1}^N\gamma_{ai}{\vec \beta}_i.
\label{II23}
\end{eqnarray}

Let us add the
gauge fixings ${\vec \eta}_{+} \approx 0$: their time constancy imply

\begin{eqnarray}
{\dot {\vec \eta}}_{+}\, &{\buildrel \circ \over =}\,& \{ {\vec \eta}_{+},H_D \}
={1\over N} \sum_{i=1}^N {{ {\vec \kappa}_i}\over {\sqrt{m_i^2+{\vec \kappa}
_i^2} }} -\vec \lambda (\tau ) \approx 0,\nonumber \\
\Rightarrow && \vec \lambda (\tau ) \approx
{1\over N} \sum_{i=1}^N {{ {\vec \kappa}_i}\over {\sqrt{m_i^2+{\vec
\kappa}_i^2} }} \not= 0,\nonumber \\
\Rightarrow&& L_D{|}_{{\vec \eta}_{+}=0} =-\sum_{i=1}^Nm_i \sqrt{1-
{\Big[ \vec \lambda +{1\over
{\sqrt{N}}}\sum_{a=1}^{N-1}\gamma_{ai}{\dot {\vec \rho}}_a\Big] }^2}
\not= \nonumber \\
 &&\not= - \sum_{i=1}^Nm_i \sqrt{1-{1\over
{N}}\sum_{ab}^{1..N-1}\gamma_{ai}\gamma_{bi} {\dot {\vec
\rho}}_a\cdot {\dot {\vec \rho}}_b}.
\label{II24}
\end{eqnarray}

Therefore ${\vec \eta}_{+}\approx 0$ is not the searched natural gauge
fixing ${\vec q}_{+}\approx 0$ for the separation of the
center-of-mass motion. The origin in this gauge is

\begin{equation}
x^{({\vec \eta}_{+})\mu}_s(T_s)=x^{\mu}_s(0)+[ u^{\mu}(p_s)+{1\over N}
\sum_{i=1}^N {{\kappa_{ir}}\over {\sqrt{m^2_i+{\vec \kappa}_i^2}}}
\epsilon^{\mu}_r(u(p_s))] T_s.
\label{II25}
\end{equation}

If we go to Dirac brackets with respect to ${\vec \eta}_{+}\approx 0$,
${\vec \kappa}_{+}\approx 0$, we get the following Hamiltonian for the
relative variables

\begin{equation}
H_M=M_{sys}=\sum_{i=1}^N  \sqrt{m^2_i+N\sum_{a,b}^{1..N-1}
\gamma_{ai}
\gamma_{bi}{\vec \pi}_a\cdot {\vec \pi}_b}.
\label{II26}
\end{equation}

However, it is practically impossible to get the associated Lagrangian
$L_R({\vec \rho}_a,{\dot {\vec \rho}}_a)$ for the relative motions.

For ${\vec \eta}_{+}=0$ one gets $\vec g\, {\buildrel \circ \over =}\,
{1\over N}\sum_{i=1}^N{\vec \alpha}_i+{{\tau}\over
N}\sum_{i=1}^N{{{\vec \beta}_i}
\over {\sqrt{m^2_i+{\vec \beta}_i^2}}}$, ${\vec \eta}_i\, {\buildrel \circ
\over =}\, {\vec \alpha}_i-{1\over N}\sum_{k=1}^N{\vec \alpha}_k+\tau ({{
{\vec \beta}_i}\over {\sqrt{m^2_i+{\vec \beta}_i^2}}}-{1\over N}
\sum_{k=1}^N {{ {\vec \beta}_k}\over {\sqrt{m_k^2+{\vec
\beta}_k^2}}})$.

In Section III we will find the natural canonical {\it internal}
3-center-of-mass variable ${\vec q}_{+}$ (replacing the naive ${\vec
\eta}_{+}$) whose vanishing implies $\vec \lambda (\tau )=0$. It will
be seen that, unlike in the non-relativistic theory, ${\vec q}_{+}$ is
not a linear combination of the ${\vec \eta}_i$'s with coefficients
depending on the masses, but it is connected to the M\o ller {\it
internal} 3-center of energy, in which the masses are replaced by the
particle energies.

\vfill\eject

\section{The {\it Internal} Relativistic Center-of-Mass Variables
on the Wigner Hyperplane}

In this Section we will study the {\it internal} center of mass
variables, while Section IV will be devoted to the {\it external}
ones.

In the relativistic case of N free scalar particles with positive
energy the Hamiltonian kinetic energy is not a quadratic form in the
momenta and the Lagrangian form is unknown. The first problem is to
separate the global translations: this is the old problem of the
definition of  a relativistic center of mass \footnote{No definition
can retain all the properties of the non-relativistic center of
mass.}. As we have seen in Sections I and II, the rest-frame instant
form of dynamics allows to clarify the problem provided one splits of
the concept of relativistic center of mass into an {\it external} one
(a pseudo-4-vector) and an {\it internal} one (a Wigner spin 1
3-vector).

The determination of the {\it internal} 3-center of mass  can be done
using the group theoretical methods of Ref.\cite{pauri} (see also
Ref.\cite{pau}): given a realization on the phase space of the ten
Poincar\'e generators one can build three 3-position variables only in
terms of them, which are:\hfill\break
 i) a canonical {\it internal} center of mass  ${\vec q}_{+}$
 \footnote{Or {\it center of spin}: the classical analogue\cite{com3,com4}
 of the Newton-Wigner position operator\cite{com1}.};
\hfill\break
ii) a non-canonical {\it internal} M\o ller {\it center of energy}
${\vec R}_{+}$ \cite{mol};\hfill\break
 iii) a non-canonical {\it internal} Fokker-Pryce
{\it center of inertia} ${\vec y}_{+}$ \cite{com2,com3}. \hfill\break
 On Wigner hyperplanes, due to ${\vec \kappa}_{+}\approx 0$, we will see that they all
 coincide: ${\vec q}_{+} \approx {\vec R}_{+} \approx {\vec y}_{+}$. \hfill\break

Following Ref.\cite{pauri} we will determine of ${\vec R}_{+}$, ${\vec
q}_{+}$, ${\vec y}_{+}$  starting from the {\it internal} realization
(\ref{II15}) of the Poincar\'e algebra. We get the following Wigner
spin 1  3-vectors:

i) The {\it internal} Moller 3-center of energy and the associated
spin vector

\begin{eqnarray}
{\vec R}_{+}&=& - {1\over {M_{sys}}} \vec K =
{{\sum_{i=1}^N\sqrt{m^2_i+ {\vec \kappa}^2_i}\,\, {\vec
\eta}_i}\over {\sum_{k=1}^N\sqrt{m_k^2+{\vec
\kappa}_k^2}}},\nonumber \\
{\vec S}_R &=& \vec J -{\vec R}_{+}\times {\vec
\kappa}_{+},\nonumber \\
 &&\{ R^r_{+},\kappa^s_{+} \} =\delta^{rs},\quad\quad \{ R^r_{+},M_{sys}
\} = {{\kappa^r_{+}}\over {M_{sys}}},\nonumber \\
 &&\{ R^r_{+},R^s_{+} \} =-{1\over {M_{sys}^2}} \epsilon^{rsu}S_R^u,
\nonumber \\
 &&\{ S_R^r,S_R^s \} =\epsilon^{rsu}(S_R^u-{1\over {M_{sys}^2}}
{\vec S_R} \cdot {\vec \kappa}_{+}\,\, \kappa_{+}^u),\quad\quad \{
S_R^r ,M_{sys} \} =0.
\label{III1}
\end{eqnarray}

Let us notice that with the gauge fixing ${\vec R}_{+} \approx  0$ we
have

\begin{eqnarray}
{\vec R}_{+}\approx 0 & \Rightarrow &
 {\dot {\vec R}}_{+}\, {\buildrel \circ \over =} \,
 \{ {\vec R}_{+},H_D \} = \nonumber \\
&=& { {{\vec \kappa}_{+}} \over
      {\sum_{k=1}^N\sqrt{m^2_k+{\vec \kappa}_k^2}} }
   -\vec \lambda (\tau )
      { {\sum_{i=1}^N\sqrt{m^2_i+{\vec \kappa}_i^2}} \over
        {\sum_{k=1}^N\sqrt{m^2_k+{\vec \kappa}_k^2} }}
  \approx -\vec \lambda (\tau ) \approx 0.
\label{III2}
\end{eqnarray}

Moreover the {\it internal} boost generator of Eq.(\ref{II15}) may be
rewritten as $\vec K=-M_{sys} {\vec R}_{+}$, so that ${\vec
R}_{+}\approx 0$ implies $\vec K \approx 0$.

ii) The canonical {\it internal} 3-center of mass \footnote{It
satisfies $\{ q^r_{+},q^s_{+}
\} =0$, $\{ q^r_{+},\kappa^s_{+} \} =
\delta^{rs}$, $\{ J^r,q^s_{+} \} =\epsilon^{rsu}q^u_{+}$, $I_s {\vec q}_{+}=-
{\vec q}_{+}$, $I_t^{*}{\vec q}_{+}={\vec q}_{+}$.} and the associated
spin vector

\begin{eqnarray}
{\vec q}_{+}&=& {\vec R}_{+}- {{ \vec J\times \vec \Omega}\over
{\sqrt{M^2_{sys}-{\vec \kappa}^2_{+}}(M_{sys}+\sqrt{M^2_{sys}-{\vec
\kappa}^2_{+}})}}=\nonumber \\
&=&-{{\vec K}\over {\sqrt{M^2_{sys}-{\vec \kappa}^2_{+}}}}+{{ \vec
J\times {\vec \kappa}_{+}}\over {\sqrt{M_{sys}^2-{\vec
\kappa}^2_{+}}(M_{sys}+\sqrt{M^2_{sys}-{\vec \kappa}^2_{+}})}}+\nonumber \\
 &+&{{\vec K\cdot {\vec \kappa}_{+}\,\, {\vec
\kappa}_{+}}\over {M_{sys}\sqrt{M^2_{sys}-{\vec \kappa}^2_{+}}
\Big( M_{sys}+\sqrt{M^2_{sys}
-{\vec \kappa}^2_{+}}\Big) }},\nonumber \\
 &&\approx {\vec
R}_{+}\quad for\quad {\vec \kappa}_{+}\approx 0;\quad\quad
\{ {\vec q}_{+},M_{sys} \} ={{{\vec \kappa}_{+}}\over {M_{sys}}},\nonumber \\
&&{}\nonumber \\
 {\vec S}_q &=&\vec J-{\vec q}_{+}\times {\vec
\kappa}_{+}= {{M_{sys}\vec J}\over {\sqrt{M^2_{sys}-{\vec
\kappa}^2_{+}}}}+\nonumber \\
 &+&{{ \vec K\times {\vec \kappa}_{+}}\over
{\sqrt{M^2_{sys}-{\vec \kappa}^2_{+}}}}-{{\vec J\cdot {\vec
\kappa}_{+}\,\, {\vec \kappa}_{+}}\over {\sqrt{M^2_{sys}-{\vec
\kappa}^2_{+}}\Big( M_{sys}+\sqrt{M^2_{sys}-{\vec \kappa}^2_{+}}\Big) }}
\approx {\vec {\bar S}} =\vec J,\nonumber \\
&&\{ {\vec S}_q,{\vec \kappa}_{+} \} =\{ {\vec S}_q,{\vec q}_{+} \} =0,
\quad\quad \{ S_q^r,S_q^s \} =\epsilon^{rsu}S_q^u.
\label{III3}
\end{eqnarray}

Let us remark that the {\it scheme A} for the {\it internal}
realization of the Poincar\'e group\cite{pauri} contains the canonical
pairs ${\vec
\kappa}_{+}$, ${\vec q}_{+}$, $S^3_q$, $arctg\, {{S^2_q}\over
{S_q^1}}$, and the two Casimirs invariants $|{\vec
S}_q|=\sqrt{-W^2/\Pi^2}$, $\sqrt{\Pi}$ (see Eq.(\ref{III5})).

In terms of the original variables one has

\begin{eqnarray}
{\vec S}_R &=& \sum_{i=1}^N {\vec \eta}_i\times {\vec \kappa}_i- {{
\sum_{i=1}^N\sqrt{m^2_i+{\vec \kappa}_i^2}\, {\vec \eta}_i\times
{\vec \kappa}_{+} }\over {\sum_{k=1}^N\sqrt{m_k^2+{\vec
\kappa}^2_k} }}=\nonumber \\
 &=&\sum_{i=1}^N {\vec \eta}_i\times \Big[
{\vec \kappa}_i -{{\sqrt{m^2_i+ {\vec \kappa}_i^2} }\over
{\sum_{k=1}^N\sqrt{m_k^2+{\vec \kappa}_k^2} }} {\vec \kappa}_{+}
\Big] \approx {\vec {\bar S}}=\vec J,\nonumber \\
 {\vec q}_{+} &=& {\vec R}_{+}-\nonumber \\
 &-& {{ \sum_{i=1}^N\Big[ {\vec \eta}_i\, {\vec
\kappa}_{+}\cdot ({\vec \kappa}
_i-{{\sqrt{m^2_i+{\vec \kappa}_i^2}}\over {\sum_{k=1}^N\sqrt{m_k^2+
{\vec \kappa}_k^2}}} {\vec \kappa}_{+})-{\vec \kappa}_{+}\cdot {\vec
\eta}_i ({\vec \kappa}_i-{{\sqrt{m^2_i+{\vec \kappa}_i^2}}\over
{\sum_{k=1}^N
\sqrt{m_k^2+{\vec \kappa}_k^2}}} {\vec \kappa}_{+} )\Big]}\over
{ \sqrt{(\sum_{k=1}^N\sqrt{m_k^2+{\vec \kappa}_{+}^2})^2-{\vec
\kappa}^2_{+}}
\Big( \sum_{k=1}^N\sqrt{m^2_k+{\vec \kappa}_k^2}+\sqrt{(\sum_{k=1}^N
\sqrt{m_k^2+{\vec \kappa}_k^2})^2-{\vec \kappa}^2_{+}}\Big) }}\approx \nonumber \\
 &\approx& {\vec R}_{+},\nonumber \\
{\vec S}_q&=&\vec J-{\vec q}_{+}\times {\vec \kappa}_{+}= \sum_{i=1}^N {\vec
\eta}_i\times {\vec \kappa}_i-\nonumber \\
&-&{{ \sum_{i=1}^N\Big[ {\vec \eta}_i\times {\vec \kappa}_{+}\, {\vec \kappa}
_{+}\cdot ({\vec \kappa}_i- {{\sqrt{m^2_i+{\vec \kappa}_i^2}}\over
{\sum_{k=1}^N\sqrt{m_k^2+{\vec \kappa}_k^2}}} {\vec
\kappa}_{+})-{\vec
\kappa}_{+}\cdot {\vec \eta}_i\, {\vec \kappa}_i\times {\vec \kappa}_{+} \Big]}
\over { \sqrt{(\sum_{k=1}^N\sqrt{m_k^2+{\vec \kappa}_{+}^2})^2-{\vec \kappa}^2_{+}}
\Big( \sum_{k=1}^N\sqrt{m^2_k+{\vec \kappa}_k^2}+\sqrt{(\sum_{k=1}^N
\sqrt{m_k^2+{\vec \kappa}_k^2})^2-{\vec \kappa}^2_{+}}\Big) }}\nonumber \\
&\approx& {\vec {\bar S}}=\vec J,\nonumber \\
 &&{}\nonumber \\
 M_{sys}&=& H_M=
\sum_{i=1}^N \sqrt{m^2_i+{\vec \kappa}_i^2},\nonumber \\
&&\sqrt{M^2_{sys}-{\vec \kappa}_{+}^2}=
\sqrt{(\sum_{k=1}^N\sqrt{m^2_k+{\vec
\kappa}_k^2})^2-{\vec \kappa}_{+}^2} .
\label{III4}
\end{eqnarray}

iii) Besides the {\it internal} canonical 3-center of mass ${\vec q}
_{+}$ and the {\it internal} non-canonical M\o ller 3-center of energy, we can define
an {\it internal} non-canonical Fokker-Pryce center of inertia ${\vec
y}_{+}$

\begin{eqnarray}
{\vec y}_{+}&=& {\vec q}_{+}+{{ {\vec S}_q\times {\vec
\kappa}_{+}}\over {\sqrt{M^2_{sys}-{\vec \kappa}_{+}^2}
(M_{sys}+\sqrt{M^2_{sys}-{\vec \kappa}_{+}^2})}}
={\vec R}_{+}+{{ {\vec S}_q\times {\vec \kappa}_{+}}\over {M_{sys}\sqrt{M^2_{sys}
-{\vec \kappa}_{+}^2}}},\nonumber \\
 {\vec q}_{+}&=&{\vec
R}_{+}+{{ {\vec S}_q\times {\vec \kappa}_{+}}\over
{M_{sys}(M_{sys}+\sqrt{M^2_{sys}-{\vec \kappa}_{+}^2})}} =
{{M_{sys}{\vec R}_{+}+
\sqrt{M^2_{sys}-{\vec \kappa}_{+}^2} {\vec y}_{+}}\over {M_{sys}+\sqrt{M^2_{sys}
-{\vec \kappa}_{+}^2}}},\nonumber \\
 &&\{ y^r_{+},y^s_{+} \}
={1\over {M_{sys}\sqrt{M^2_{sys}-{\vec \kappa}_{+}^2} }}
\epsilon^{rsu}\Big[ S^u_q+{{ {\vec S}_q\cdot {\vec \kappa}_{+}\, \kappa^u_{+}}
\over {\sqrt{M^2_{sys}-{\vec \kappa}_{+}^2}(M_{sys}+\sqrt{M^2_{sys}-{\vec \kappa}
_{+}^2})}}\Big] ,\nonumber \\
&&{}\nonumber \\
{\vec \kappa}_{+}\approx 0 &\Rightarrow& {\vec q}_{+}\approx {\vec R}_{+}
\approx {\vec y}_{+}.
\label{III5}
\end{eqnarray}

Therefore the gauge fixings ${\vec q}_{+}\approx {\vec R}_{+}\approx {\vec y}
_{+}\approx 0$ imply $\vec \lambda (\tau )\approx 0$ and force the three
{\it internal} collective variables to coincide with the origin of the
coordinates, which now becomes

\beq
x^{({\vec q}_{+})\mu}_s(T_s)=x_s^{\mu}(0) +u^{\mu}(p_s) T_s.
\label{III6}
\eeq

By adding the gauge fixings $\vec \chi ={\vec q}_{+}\approx {\vec
R}_{+}\approx {\vec y}_{+} \approx 0$,  it can be shown that the
origin $x_s^{(\mu )}(\tau )$ becomes  simultaneously the Dixon center
of mass of an extended object\cite{dixon1} and both the
Pirani\cite{pi} and Tulczyjew\cite{tu} centroids (the Dixon multipoles
for the N-body problem on the Wigner hyperplane will be studied in
Ref.\cite{iten3}).

We need now a canonical transformation bringing from the basis ${\vec
\eta}_i$, ${\vec \kappa}_i$, to a new canonical basis ${\vec q}_{+}$,
${\vec \kappa}_{+} (\approx 0)$, ${\vec \rho}_{q,a}$, ${\vec
\pi}_{q,a}$, in which ${\vec S}_q=
\sum_{a=1}^{N-1} {\vec \rho}_{q,a}\times {\vec \pi}_{q,a}$:

\begin{equation}
\begin{minipage}[t]{1cm}
\begin{tabular}{|l|} \hline
${\vec \eta}_i$ \\  \hline
 ${\vec \kappa}_i$ \\ \hline
\end{tabular}
\end{minipage} \ {\longrightarrow \hspace{.2cm}} \
\begin{minipage}[t]{2 cm}
\begin{tabular}{|l|l|} \hline
${\vec q}_{+}$   & ${\vec \rho}_{qa}$   \\ \hline
 ${\vec \kappa}_{+}$&${\vec \pi}_{qa}$ \\ \hline
\end{tabular}
\end{minipage}
\label{III7}
\end{equation}

Let us remark that this is not a point transformation: the
relativistic internal center of mass ${\vec q}_{+}$, realizing the
effective separation of the center of mass from the relative motions
in the kinetic energy, is momentum dependent.

The canonical transformation (\ref{III6}) will be studied in Section V
by using the method of Gartenhaus-Schwartz\cite{garten} as delineated
in Ref.\cite{osborne} (see Refs.\cite{pauri,osborne,garten} for the
N=2 case).

Let us finally consider the non-relativistic limit of the Lagrangian
of Eq.(\ref{II21}) \footnote{Here $\tau =t$, with $t$ the absolute
Newton time; for the sake of simplicity we shall use the same notation
for functions of $\tau$ and $t$; having $c=1$ the non-relativistic
limit $c \rightarrow \infty$ is done by considering $velocities << 1$
and $momenta << m$.}

\begin{eqnarray}
L_D&=&-\sum_{i=1}^N m_i \sqrt{1- ({\dot {\vec
\eta}}_i(\tau )+{\vec
\lambda (\tau )})^2} {\mapsto}_{c\rightarrow \infty} \sum_{i=1}^Nm_i+ L_{Dnr},
\nonumber \\
L_{Dnr}(t)&=&\sum_{i=1}^N{{m_i}\over 2} ({\dot {\vec \eta}}_i+\vec
\lambda)^2(t),\quad\quad S_{Dnr}=\int dt L_{Dnr}(t),\nonumber \\
 &&{}\nonumber \\
 {\vec \kappa}_i(\tau )\,
&{\mapsto}_{c\rightarrow \infty}& {\vec k}_i^{'}(t)+O(c^{-2}),
\nonumber \\
 {\vec \kappa}_i^{'}(t)&=&{{\partial L_{Dnr}(t)}\over {\partial {\dot
{\vec \eta}}_i(t)}}=m_i({\dot {\vec \eta}}_i+\vec
\lambda )(t),\nonumber \\
 {\vec \pi}_{\lambda}(t)&=&{{\partial
L_{Dnr}(t)}\over {\partial {\dot {\vec
\lambda}}(t) }} =0,\nonumber \\
 &&{}\nonumber \\
 H_{c,nr}&=&{\vec \pi}_{\lambda}\cdot {\dot {\vec \lambda}}+\sum_{i=1}^N{\vec
\kappa}_i^{'}\cdot {\dot {\vec \eta}}_i-L_{Dnr}=\sum_{i=1}^N {{ {\vec \kappa}
_i^{{'}\, 2}}\over {2m_i}}-\vec \lambda \cdot {\vec \kappa}^{'}_{+},\quad
{\vec \kappa}^{'}_{+}=\sum_{i=1}^N{\vec \kappa}^{'}_i,\nonumber \\
H_{Dnr}&=&\sum_{i=1}^N {{ {\vec \kappa}_i^{{'}\, 2}(t)}\over {2m_i}}-
\vec \lambda (t)\cdot {\vec \kappa}^{'}_{+}(t)+\vec \mu (t)\cdot
{\vec \pi}_{\lambda}(t),\nonumber \\
 {\dot {\vec \pi}}_{\lambda}(t)\, &{\buildrel \circ \over =}\,&
{\vec \kappa}^{'}_{+}\approx 0,\quad (non-relativistic\,\, rest\,\,
frame).
\label{III8}
\end{eqnarray}

The Lagrangian $L_{Dnr}$ has been used in Ref.\cite{iten2} [see its
Eq.(2.1)] to describe the relative motions in the non-relativistic
rest frame. In the non-relativistic limit ${\vec q}_{+}$ tends the the
non-relativistic center of mass ${\vec q}_{nr}= {{\sum_{i=1}^Nm_i{\vec
\eta}_i}\over {\sum_{i=1}^Nm_i}}$.

In conclusion, the non-relativistic Abelian translation symmetry
generating the non-relativistic Noether constants ${\vec P}=const.$ is
splitted at the relativistic level into the two following symmetries:
i) the {\it external} Abelian translation symmetry whose Noether
constants of motion are ${\vec p}_s=\epsilon_s {\vec k}_s\approx
M_{sys}{\vec k}_s = const.$ (its conjugate variable is the {\it
external} 3-center of mass ${\vec z}_s$); ii) the {\it internal}
Abelian gauge symmetry generating the three first class constraints
${\vec \kappa}_{+}\approx 0$ (the rest-frame conditions) inside the
Wigner hyperplane (its conjugate {\it gauge} variable is the {\it
internal} 3-center of mass ${\vec q}_{+}\approx {\vec R}_{+}\approx
{\vec y}_{+}$), whose non-relativistic counterpart would be the
non-relativistic rest-frame conditions ${\vec P}\approx 0$.

\vfill\eject

\section{The {\it External} Center-of-Mass Variables on the Wigner Hyperplane.}

Let us study now the localization on the Wigner hyperplane of the {\it
external} center-of-mass variables. Let us remember\cite{lus} that the
{\it external} canonical non-covariant point of coordinates

\begin{equation}
{\tilde x}^{\mu}_s(\tau )= ({\tilde x}^o_s(\tau ); {\vec {\tilde
x}}_s(\tau ) )=z^{\mu}(\tau ,{\tilde {\vec \sigma }})=x^{\mu}_s(\tau
)-{1\over {\epsilon_s(p^o_s+\epsilon_s)}}\Big[
p_{s\nu}S_s^{\nu\mu}+\epsilon_s(S^{o\mu}_s-S^{o\nu}_s{{p_{s\nu}p_s^{\mu}}\over
{\epsilon^2_s}}) \Big],
\label{IV1}
\end{equation}

 \noindent
lies in the Wigner hyperplane $z^{\mu}(\tau ,\vec
\sigma )=x^{\mu}_s(\tau )+\epsilon^{\mu}_r(u(p_s)) \sigma^r$ at some 3-position
${\tilde \sigma}^r$, like the true coordinate origin $x^{\mu}_s(\tau
)=z^{\mu}(\tau ,\vec 0)$, because $p_s\cdot {\tilde x}_s=p_s\cdot
x_s$, see Ref.\cite{lus}.

Like in Eqs.(\ref{III1}), (\ref{III3}) and (\ref{III5}) one can
build\cite{pauri} three {\it external} 3-variables, the canonical
${\vec q}_s$, the Moller ${\vec R}_s$ and the Fokker-Pryce ${\vec Y}
_s$ by using the rest-frame realization of the Poincar\'e algebra
given in Eqs.(\ref{II14})

\begin{eqnarray}
{\vec R}_s&=& -{{1}\over {p^o_s}}{\vec K}_s=({\vec {\tilde
x}}_s-{{{\vec p}_s}\over {p^o_s}} {\tilde x}^o_s)-{{{\vec {\bar
S}}_s\times {\vec p}_s}
\over {p^o_s(p^o_s+\epsilon_s)}},\nonumber \\
{\vec q}_s&=&{\vec {\tilde x}}_s-{{{\vec p}_s}\over {p^o_s}}{\tilde
x}^o_s= {{{\vec z}_s}\over {\epsilon_s}}= {\vec R}_s+{{ {\vec {\bar
S}}_s\times {\vec p}_s}\over {p^o_s(p^o_s+
\epsilon_s)}}={{p^o_s {\vec R}_s+\epsilon_s {\vec Y}_s}\over {p^o_s+\epsilon_s}}
,\nonumber \\ {\vec Y}_s&=&{\vec q}_s+{{ {\vec {\bar S}}_s\times {\vec
p}_s}\over {\epsilon_s (p^o_s+\epsilon_s)}}={\vec R}_s+{{ {\vec {\bar
S}}_s\times {\vec p}_s}\over {p^o_s\epsilon_s}},\nonumber \\
 &&{}\nonumber \\
 &&\{ R^r_s,R^s_s \} =-{{1}\over
{(p^o_s)^2}}\epsilon^{rsu}\Omega^u_s,
\quad\quad {\vec \Omega}_s={\vec J}_s-{\vec R}_s\times {\vec p}_s,\nonumber \\
 &&{}\nonumber \\
 &&\{ q^r_s, q^s_s \} =0,\nonumber \\
 &&{}\nonumber \\
 &&\{ Y^r_s,Y^s_s \} ={1\over {\epsilon_sp^o_s}}\epsilon^{rsu}\Big[
{\bar S}^u_s +{{ {\vec {\bar S}}_s\cdot {\vec p}_s\, p^u_s}\over
{\epsilon_s(p^o_s+
\epsilon_s)}}\Big] ,\nonumber \\
&&{}\nonumber \\
 {\vec p}_s\cdot {\vec q}_s&=&{\vec p}_s\cdot {\vec
 R}_s={\vec p}_s\cdot {\vec Y}_s={\vec k}_s\cdot {\vec z}_s,\nonumber \\
 &&{}\nonumber \\
{\vec p}_s=0 &\Rightarrow& {\vec q}_s={\vec Y}_s={\vec R}_s,
\label{IV2}
\end{eqnarray}

\noindent with the same velocity and coinciding in the Lorentz rest frame where
${\buildrel \circ \over p}^{\mu}_s=\epsilon_s (1;\vec 0)$

We can now try to construct the following three {\it external}
4-positions (all located on the Wigner hyperplane): \hfill\break i)
the {\it external} canonical non-covariant 4-center of mass ${\tilde
x}_s^{\mu }$;
\hfill\break ii) the {\it external} non-canonical and non-covariant M\o
ller 4-center of energy $R^{\mu }_s$; \hfill\break
 iii) the {\it external} covariant non-canonical Fokker-Pryce 4-center of inertia
$Y^{\mu }_s$ \footnote{When there are the gauge fixings ${\vec
q}_{+}\approx 0$ it will be shown that it also coincides with the
origin $x^{\mu }_s$.}.

In Ref.\cite{pauri} in a one-time framework without constraints and at
a fixed time, it is shown that the 3-vector ${\vec Y}_s$ (but not
${\vec q}_s$ and ${\vec R}_s$) satisfies the condition $\{ K^r_s,Y^s_s
\} ={1\over {c^2}} Y^r_s\, \{ Y^s_s,p^o_s \}$ for being the space
component of a 4-vector $Y^{\mu}_s$. In the enlarged canonical
treatment including time variables, it is not clear which are the time
components to be added to ${\vec q}_s$, ${\vec R}
_s$, ${\vec Y}_s$, to rebuild 4-dimensional quantities ${\tilde x}^{\mu}_s$,
$R^{\mu}_s$, $Y^{\mu}_s$, in an arbitrary Lorentz frame $\Gamma$, in which the
origin of the Wigner hyperplane is the 4-vector $x^{\mu}_s = (x^o_s; {\vec x}
_s)$. We have from Eq.(\ref{II10})

\begin{eqnarray}
{\tilde x}^{\mu}_s(\tau )&=&
({\tilde x}^o_s(\tau ); {\vec {\tilde x}}_s(\tau ) )=
x^{\mu}_s-{1\over {\epsilon_s(p^o_s+\epsilon_s)}}\Big[
p_{s\nu}S_s^{\nu\mu}+\epsilon_s(S^{o\mu}_s-S^{o\nu}_s{{p_{s\nu}p_s^{\mu}}\over
{\epsilon^2_s}}) \Big],\quad\quad p^{\mu}_s,\nonumber \\
{\tilde x}^o_s&=&\sqrt{1+{\vec k}_s^2} (T_s+{{{\vec k}_s\cdot {\vec z}_s}\over
{\epsilon_s}})=\sqrt{1+{\vec k}^2_s}(T_s+{\vec k}_s\cdot {\vec q}_s)\not=
x^0_s,\quad\quad p^o_s=\epsilon_s\sqrt{1+{\vec k}_s^2},\nonumber \\
{\vec {\tilde x}}_s&=&{{ {\vec z}_s}\over {\epsilon_s}}+(T_s+{{{\vec k}_s\cdot
{\vec z}_s}\over {\epsilon_s}}) {\vec k}_s={\vec q}_s+(T_s+{\vec k}_s\cdot
{\vec q}_s){\vec k}_s,\quad\quad {\vec p}_s=\epsilon_s {\vec k}_s.
\label{IV3}
\end{eqnarray}

\noindent for the non-covariant (frame-dependent) canonical 4-center of mass and
its conjugate momentum.

Each Wigner hyperplane intersects the worldline of the arbitrary origin
4-vector $x^{\mu}_s(\tau )=z^{\mu}(\tau ,\vec 0)$ in $\vec \sigma =0$, the
pseudo worldline of ${\tilde x}^{\mu}_s(\tau )=z^{\mu}(\tau ,{\tilde {\vec
\sigma}})$ in some ${\vec {\tilde \sigma}}$ and the worldline of the
 Fokker-Pryce 4-vector $Y^{\mu}_s(\tau )=z^{\mu}(\tau ,{\vec \sigma}_Y)$ in
some ${\vec \sigma}_Y$; one also has $R^{\mu}_s=z^{\mu}(\tau ,{\vec
\sigma}_R)$. Since we have $T_s=u(p_s)\cdot x_s=u(p_s)\cdot {\tilde
x}_s\equiv \tau$ on the Wigner hyperplane labelled by $\tau$, we
require that also $Y^{\mu}_s$, $R^{\mu}_s$ have time components such
that they too satisfy $u(p_s)\cdot Y_s=u(p_s)\cdot R_s=T_s\equiv
\tau$. Therefore, it is reasonable to assume that ${\tilde
x}^{\mu}_s$, $Y^{\mu}_s$ and $R^{\mu}_s$ satisfy the following
equations consistently with Eqs.(\ref{IV2}), (\ref{IV3})

\begin{eqnarray}
{\tilde x}^{\mu}_s&=&( {\tilde x}^o_s; {\vec {\tilde x}}_s)= ({\tilde
x}^o_s; {\vec q}_s+{{{\vec p}_s}\over {p^o_s}} {\tilde
x}^o_s)=\nonumber \\
 &=&({\tilde x}^o_s; {{{\vec z}_s}\over {\epsilon_s}}+(T_s+{{{\vec
k}_s\cdot {\vec z}_s}\over {\epsilon_s}}){\vec k}_s )
=x^{\mu}_s+\epsilon^{\mu}_u(u(p_s)) {\tilde \sigma}^u,
\nonumber \\
 &&{}\nonumber \\
 Y^{\mu}_s&=&({\tilde x}^o_s; {\vec Y}_s)=\nonumber \\
  &=&({\tilde x}^o_s;\, {1\over {\epsilon_s}}[{\vec z}_s+{{{\vec
{\bar S}}_s\times {\vec p}_s}\over {\epsilon_s[1+u^o(p_s)]}}]+(T_s+
{{{\vec k}_s\cdot {\vec z}_s}\over {\epsilon_s}}){\vec k}_s\,
)=\nonumber \\
 &=&{\tilde x}^{\mu}_s+\eta^{\mu}_r{{({\vec {\bar
S}}_s\times {\vec p}_s)^r}\over {\epsilon_s[1+u^o(p_s)]}}=\nonumber \\
 &=&x^{\mu}_s+\epsilon^{\mu}_u(u(p_s)) \sigma^u_Y,\nonumber \\
 &&{}\nonumber \\
 R^{\mu}_s&=&({\tilde x}^o_s; {\vec R}_s)=\nonumber \\
  &=&( {\tilde x}^o_s;\, {1\over {\epsilon_s}}[{\vec z}_s-
{{{\vec {\bar S}}_s\times {\vec p}_s}\over {\epsilon_s u^o(p_s)
[1+u^o(p_s)]}}]+(T_s+ {{{\vec k}_s\cdot {\vec z}_s}\over
{\epsilon_s}}){\vec k}_s\, )=\nonumber \\
 &=&{\tilde
x}^{\mu}_s-\eta^{\mu}_r{{({\vec {\bar S}}_s\times {\vec p}_s)^r}
\over {\epsilon_su^o(p_s)[1+u^o(p_s)]}}=\nonumber \\
&=&x^{\mu}_s+\epsilon^{\mu}_u(u(p_s)) \sigma^u_R,\nonumber \\
&&{}\nonumber \\
T_s&=&u(p_s)\cdot x_s=u(p_s)\cdot {\tilde x}_s=u(p_s)\cdot Y_s=u(p_s)\cdot
R_s,\nonumber \\
&&{}\nonumber \\
{\tilde \sigma}^r&=&\epsilon_{r\mu}(u(p_s))[x^{\mu}_s-{\tilde x}^{\mu}_s]=
{{ \epsilon_{r\mu}(u(p_s)) [u_{\nu}(p_s)S^{\nu\mu}_s+S^{o\mu}_s]}\over
{[1+u^o(p_s)]}}=\nonumber \\
&=&-{\bar S}_s^{\tau r}+{{{\bar S}_s^{rs}p^s_s}\over {\epsilon_s[1+u^o(p_s)]}}
=\epsilon_s R^r_{+}+{{{\bar S}_s^{rs}u^s(p_s)}\over {1+u^o(p_s)}}
\approx \nonumber \\
&{\buildrel {{\vec q}_{+}\approx 0} \over \approx}& \epsilon_s
q^r_{+}+{{{\bar S}_s^{rs}u^s(p_s)}\over {1+ u^o(p_s)}}\approx {{{\bar
S}_s^{rs}u^s(p_s)}\over {1+u^o(p_s)}} ,\nonumber \\
\sigma^r_Y&=&\epsilon_{r\mu}(u(p_s))[x^{\mu}_s-Y^{\mu}_s]={\tilde \sigma}^r-
\epsilon_{ru}(u(p_s)){{({\vec
{\bar S}}_s\times {\vec p}_s)^u}\over {\epsilon_s[1+u^o(p_s)]}}=\nonumber \\
&=&{\tilde \sigma}^r+{{{\bar S}^{rs}_su^s(p_s)}\over {1+u^o(p_s)}}=
\epsilon_s R^r_{+} \approx \epsilon_s q^r_{+}
{\buildrel {{\vec q}_{+}\approx 0} \over \approx} 0,\nonumber \\
 &&{}\nonumber \\
  &\Rightarrow&
x^{({\vec q}_{+})\mu}_s(\tau ) = Y^{\mu}_s,\, when\, {\vec
q}_{+}\approx 0,\nonumber \\
 &&{}\nonumber \\
\sigma^r_R&=&\epsilon_{r\mu}(u(p_s))[x^{\mu}_s-R^{\mu}_s]={\tilde \sigma}^r+
\epsilon_{ru}(u(p_s)) {{({\vec {\bar S}}_s\times {\vec p}_s)^u}
\over {\epsilon_su^o(p_s)[1+u^o(p_s)]}}=\nonumber \\
&=&{\tilde \sigma}^r-{{{\bar S}_s^{rs}u^s(p_s)}\over {u^o(p_s)[1+
u^o(p_s)]}}=\epsilon_sR^r_{+}+{{[1-u^o(p_s)]{\bar
S}^{rs}_su^s(p_s)}\over {u^o(p_s)[1+u^o(p_s)]}}\approx \nonumber \\
 &{\buildrel {{\vec q}_{+}\approx 0} \over \approx}&
{{[1-u^o(p_s)]{\bar S}^{rs}_su^s(p_s)}\over {u^o(p_s)[1+u^o(p_s)]}}.
\label{IV4}
\end{eqnarray}

\noindent Therefore, {\it the external Fokker-Pryce non-canonical center of inertia
coincides with the origin $x^{({\vec q}_{+})\mu}_s(\tau )$ carrying
the internal center of mass}.

Let us remember also that the origin $x^{\mu}_s(\tau )$ corresponds to
the unique special relativistic center-of-mass-like worldline of
Refs.\cite{beig}.

In each Lorentz frame one has different pseudo-worldlines describing
$R^{\mu}_s$ and ${\tilde x}^{\mu}_s$: the 4-canonical center of mass
${\tilde x}^{\mu}_s$ lies in between $Y^{\mu}_s$ and $R^{\mu}_s$ in
every frame. If, in an arbitrary Lorentz frame, we consider the
worldline $Y^{\mu}_s$ of the covariant non-canonical Fokker-Pryce
4-center of inertia, the representation in this frame of all the
pseudo-worldlines associated with ${\tilde x}^{\mu}_s$ and $R^{\mu}_s$
fill a worldtube \cite{mol} around  $Y_s^{\mu}$ whose {\it invariant
radius} is $\rho =\sqrt{-\epsilon W^2}/p^2=|\vec S|/\sqrt{\epsilon
p^2}$ ($W^2=-\epsilon p^2{\vec S}^2$ is the Pauli-Lubanski Casimir
when $\epsilon p^2 > 0$). This is the classical intrinsic radius of
the worldtube, in which   the non-covariance effects (the
pseudo-worldlines) of the canonical 4-center of mass ${\tilde
x}_s^{\mu}$ are located. See Ref.\cite{india} for a discussion of the
properties of the {\it M$\o$ller radius}. At the quantum level $\rho$
becomes the Compton wavelength of the isolated system multiplied its
spin eigenvalue $\sqrt{s(s+1)}$ , $\rho \mapsto \hat \rho =
\sqrt{s(s+1)} \hbar /M=\sqrt{s(s+1)} \lambda_M$ with $M=\sqrt{\epsilon
p^2}$ the invariant mass and $\lambda_M=\hbar /M$ its Compton
wavelength. Therefore, the criticism to classical relativistic
physics, based on quantum pair production, concerns the testing of
distances where, due to the Lorentz signature of spacetime, one has
intrinsic classical covariance problems: it is impossible to localize
the canonical 4-center of mass ${\tilde x}_s^{\mu}$ adapted to the
first class constraints of the system (also named Pryce center of mass
and having the same covariance of the Newton-Wigner position operator)
in a frame independent way.

Let us remember \cite{lus} that $\rho$ is also a remnant of the energy
conditions of general relativity in flat Minkowski spacetime: since
the M$\o$ller non-canonical, non-covariant 4-center of energy $R^{\mu
}$ has its non-covariance  (its pseudo-worldlines) localized inside
the same worldtube with radius $\rho$ (actually  the latter was
discovered in this way) \cite{mol}, it turns out that for an extended
relativistic system with the material radius smaller of its intrinsic
radius $\rho$ one has: i) its peripheral rotation velocity can exceed
the velocity of light; ii) its classical energy density cannot be
positive definite everywhere in every frame.

\vfill\eject

\section{The Internal Relative Variables from the Gartenhaus-Schwartz
Transformation.}

Given ${\vec \eta}_i$, ${\vec \kappa}_i$, we must now find the
canonical basis ${\vec q}_{+}$, ${\vec \kappa}_{+}$, ${\vec
\rho}_{qa}$, ${\vec \pi}_{qa}$ of Eq.(\ref{III7}).

We shall use the classical analog of the
Gartenhaus-Schwartz singular transformation \cite{garten} following
the scheme used in Ref.\cite{osborne} to find the center-of-mass
subspace of phase space defined by ${\vec
\kappa}_{+}=0$ \footnote{The singular limit
$\alpha \rightarrow \infty$ is very similar to a contraction of a Lie
algebra.}

\begin{eqnarray}
U(\alpha )&=& e^{\alpha \{ .,{\vec q}_{+}\cdot {\vec \kappa}_{+} \} },
\nonumber \\
{\vec q}_{+}\cdot {\vec \kappa}_{+}&=&- {{|{\vec \kappa}_{+}|}\over
{\sum_{k=1}
^N \sqrt{m_k^2+{\vec \kappa}_k^2}}} {\vec n}_{+}\cdot \vec K,\quad\quad
{\vec n}_{+}={{{\vec \kappa}_{+}}\over {|{\vec \kappa}_{+}|}},\quad
\vec K=-\sum_{i=1}^N\sqrt{m_i^2+{\vec \kappa}_i^2} {\vec \eta}_i,
\nonumber \\
&&{}\nonumber \\
{\vec \kappa}_{+}(\alpha )&=& U(\alpha ) {\vec \kappa}_{+} = e^{-\alpha} {\vec
\kappa}_{+}\, {\rightarrow}_{\alpha \rightarrow \infty}\, 0,\nonumber \\
{\vec q}_{+}(\alpha )&=& U(\alpha ) {\vec q}_{+} = e^{\alpha} {\vec q}_{+}\,
{\rightarrow}_{\alpha \rightarrow \infty}\, \infty ,\nonumber \\
&&U(-\alpha ) {\vec q}_{+} =e^{-\alpha} {\vec q}_{+}\, {\rightarrow}_{\alpha
\rightarrow \infty}\, 0,\nonumber \\
&&{}\nonumber \\
\Rightarrow && {\vec \kappa}_{+}(\alpha ) \cdot {\vec q}_{+}(\alpha )={\vec
\kappa}_{+} \cdot {\vec q}_{+},\quad\quad {\vec n}_{+}(\alpha )={\vec n}_{+}.
\label{V1}
\end{eqnarray}

Therefore, $lim_{\alpha \rightarrow \infty} U(\alpha )$ can only be applied to
the set of functions on phase space which have vanishing Poisson bracket with
${\vec \kappa}_{+}$, namely to ${\vec \kappa}_i$ [or ${\vec \pi}_a={1\over
{\sqrt{N}}} \sum_{i=1}^N\gamma_{ai}{\vec \kappa}_i$] and to ${\vec \rho}_a=
\sqrt{N} \sum_{i=1}^N\gamma_{ai}{\vec \eta}_i$ of Eq.(\ref{II16}).

Since, for finite $\alpha$, $U(\alpha )$ is a canonical
transformation, the Poisson brackets are preserved [$\{ f(\alpha
),g(\alpha ) \} =U(\alpha ) \{ f,g \}$] even in the limit $\alpha
\rightarrow \infty$.

Let $f=f({\vec \eta}_i,{\vec \kappa}_i)$ have zero Poisson bracket with ${\vec
\kappa}_{+}$, $\{ f,{\vec \kappa}_{+} \} =0$, and let be $f(\alpha )=
U(\alpha ) f$. Then, we have

\beq
\{ {\vec \kappa}_{+},f(\alpha ) \} =e^{\alpha}
\{ {\vec \kappa}_{+}(\alpha ),f(\alpha ) \} = e^{\alpha} \Big( U(\alpha )
\{ {\vec \kappa}_{+},f \} \Big) =0.
\label{V2}
\eeq

Moreover, since the Jacobi identity
$\{ {\vec \kappa}_{+}, \{ {\vec q}_{+},f \} \} + \{ f, \{ {\vec \kappa}_{+},
{\vec q}_{+} \} \} + \{ {\vec q}_{+}, \{ f,{\vec \kappa}_{+} \} \}
\equiv 0$ implies $\{ {\vec \kappa}_{+}, \{ {\vec q}_{+},f \} \} \equiv 0$
[namely also $\{ {\vec q}_{+},f \}$ has zero Poisson bracket with ${\vec
\kappa}_{+}$ if $\{ f,{\vec \kappa}_{+} \} =0$, so that $U(\alpha ) \{ {\vec
q}_{+},f \}$ has a well defined limit for $\alpha \rightarrow \infty$]
one also has

\beq
\{ {\vec q}_{+},f(\alpha ) \} =e^{-\alpha} \{ {\vec q}_{+}(\alpha ),
f(\alpha ) \} =e^{-\alpha} \Big( U(\alpha ) \{ {\vec q}_{+},f \}
\Big)\, {\rightarrow}_{\alpha \rightarrow \infty}\, 0.
\label{V3}
\eeq

Moreover, we have

\begin{equation}
{{df(\alpha )}\over {d\alpha}} = \{ f(\alpha ),{\vec \kappa}_{+}\cdot {\vec q}
_{+} \} = \{ f(\alpha ), {\vec \kappa}_{+}(\alpha )\cdot {\vec q}_{+}(\alpha )
\} .
\label{V4}
\end{equation}

Therefore, the relative variables ${\vec \pi}_a={1\over {\sqrt{N}}} \sum_{i=1}
^N\gamma_{ai}{\vec \kappa}_i$ and  ${\vec \rho}_a=\sqrt{N} \sum_{i=1}^N\gamma
_{ai}{\vec \eta}_i$, which commute with ${\vec \kappa}_{+}$ [see Eq.(\ref{II16})],
satisfy

\begin{eqnarray}
 {\vec \pi}_a(\alpha )&=&U(\alpha ) {\vec \pi}_a\, {\rightarrow}_{\alpha
\rightarrow \infty}\, {\vec \pi}_a(\infty )\, {\buildrel {def}\over =}\,
{\vec \pi}_{qa},\nonumber \\
 {\vec \rho}_a (\alpha )&=&U(\alpha ) {\vec \rho}_a\, {\rightarrow}_{\alpha \rightarrow
\infty}\, {\vec \rho}_a(\infty )\, {\buildrel {def} \over =}\, {\vec \rho}_{qa},
\label{V5}
\end{eqnarray}

\noindent
with ${\vec \rho}_{qa}$, ${\vec \pi}_{qa}$, pairs of
canonical variables having zero Poisson bracket with ${\vec q}_{+}$, ${\vec
\kappa}_{+}$.

In this way one gets the searched  canonical transformation
(\ref{III7}).

Let us first evaluate ${\vec \pi}_{qa}$ following the scheme of Ref.
\cite{osborne}. From Eq.(\ref{V4}) we get

\begin{equation}
{{d{\vec \kappa}_i(\alpha )}\over {d\alpha}}= \{ {\vec \kappa}_i(\alpha ),
{\vec \kappa}_{+}(\alpha )\cdot {\vec q}_{+}(\alpha ) \} =-{{{\vec \kappa}
_{+}(\alpha )}\over {H_M(\alpha )}} H_i(\alpha ),
\label{V6}
\end{equation}

\noindent with the notations

\begin{eqnarray}
M_{sys}&=&H_M= \sum_{i=1}^N H_i,\quad\quad H_M(\alpha )=\sum_{i=1}^N
H_i(\alpha )\, {\rightarrow}_{\alpha \rightarrow \infty}\, H_M(\infty
 )\, {\buildrel {def}\over =}\, H_{(rel)},\nonumber \\
 H_i&=& \sqrt{m_i^2+{\vec
\kappa}_i^2},\quad\quad H_i(\alpha )=\sqrt{m_i^2
+{\vec \kappa}_i^2(\alpha )}\, {\rightarrow}_{\alpha \rightarrow
\infty}\, H_i(\infty )\, {\buildrel {def}\over =}\, H_{(rel) i},\nonumber \\
 \Pi&=& H^2_M-{\vec \kappa}_{+}^2\approx H_M^2.
\label{V7}
\end{eqnarray}

From $m^2_i= H^2_i(\alpha )-{\vec \kappa}_i^2(\alpha )$, we get

\begin{eqnarray}
{{dH_i(\alpha )}\over {d\alpha}} H_i(\alpha ) &=& {{d{\vec \kappa}_i(\alpha )}
\over {d\alpha }} \cdot {\vec \kappa}_i(\alpha ),\nonumber \\
\Rightarrow && {{dH_i(\alpha )}\over {d\alpha}} =-{\vec \kappa}_i(\alpha )\cdot
{{{\vec \kappa}_i(\alpha )}\over {H_M(\alpha )}},\nonumber \\
\Rightarrow && {{dH_M(\alpha )}\over {d\alpha}}=\sum_{i=1}^N{{dH_i(\alpha )}
\over {d\alpha}} =- {{{\vec \kappa}_{+}^2(\alpha )}\over {H_M(\alpha )}},
\nonumber \\
\Rightarrow && \Pi =H^2_M-{\vec \kappa}^2_{+}=H^2_M(\alpha )-{\vec \kappa}_{+}
^2(\alpha ) {\rightarrow}_{\alpha \rightarrow \infty}\, H^2_M(\infty )
 = H^2_{(rel)},\nonumber \\
 &&or\quad {{d\Pi}\over {d\alpha}}=0.
\label{V8}
\end{eqnarray}

Let us now introduce $\theta (\alpha )$ such that [$ch^2\, \theta (\alpha )-
sh^2\, \theta (\alpha ) =1$ also for $\alpha \rightarrow \infty$]

\begin{eqnarray}
sh\, \theta (\alpha )&=&{{|{\vec \kappa}_{+}| H_M(\alpha )- |{\vec \kappa}
_{+}(\alpha )| H_M}\over {\Pi}}\, {\rightarrow}_{\alpha \rightarrow \infty}\,
{{|{\vec \kappa}_{+}|}\over {\sqrt{\Pi}}},\nonumber \\
ch\, \theta (\alpha )&=& {{H_M H_M(\alpha ) -|{\vec \kappa}_{+}| |{\vec \kappa}
_{+}(\alpha )|}\over {\Pi}}\, {\rightarrow}_{\alpha \rightarrow \infty}\,
{{H_M}\over {\sqrt{\Pi}}},\nonumber \\
&&{}\nonumber \\
\theta (\alpha )&=&tanh^{-1}\, {{|{\vec \kappa}_{+}|}\over {H_M}}-tanh^{-1}\,
{{|{\vec \kappa}_{+}(\alpha )|}\over {H_M(\alpha )}}\, {\rightarrow}
_{\alpha \rightarrow 0}\, 0,\quad\quad {\rightarrow}_{\alpha \rightarrow \infty}\,
tanh^{-1}\, {{|{\vec \kappa}_{+}|}\over {H_M}}.
\label{V9}
\end{eqnarray}

Since we have

\begin{equation}
{{d\theta (\alpha )}\over {d\alpha}} = {{|{\vec \kappa}_{+}(\alpha
)|}\over {H_M(\alpha )}},\quad\quad {{d{\vec n}_{+}(\alpha )}\over
{d\alpha}} =0\quad \Rightarrow \quad {\vec n}_{+}(\alpha )={\vec
n}_{+},
\label{V10}
\end{equation}

\noindent we arrive at the coupled equations

\begin{eqnarray}
{{d{\vec \kappa}_i(\alpha )}\over {d\theta}}&=& -H_i(\alpha ) {\vec
n}_{+},
\nonumber \\
{{dH_i(\alpha )}\over {d\theta}} &=& -{\vec \kappa}_i(\alpha )\cdot {\vec n}
_{+},
\label{V11}
\end{eqnarray}

\noindent whose integration gives

\begin{eqnarray}
{\vec \kappa}_i(\alpha ) &=& {\vec \kappa}_i +\Big( [ch\, \theta (\alpha ) -1]
{\vec n}_{+}\cdot {\vec \kappa}_i -sh\, \theta (\alpha ) H_i\Big) {\vec n}_{+}
\nonumber \\
{\rightarrow}_{\alpha \rightarrow \infty}&& {\vec \kappa}_i(\infty )={\vec
\kappa}_i +\Big[ ({{H_M}\over {\sqrt{\Pi}}}-1) {\vec n}_{+}\cdot {\vec \kappa}
_i-{{|{\vec \kappa}_{+}|}\over {\sqrt{\Pi}}} H_i\Big] {\vec n}_{+}
\approx {\vec \kappa}_i,\nonumber \\
 &&{}\nonumber \\
 H_i(\alpha ) &=&\sqrt{m_i^2+{\vec
\kappa}_i^2(\alpha )}= ch\, \theta (\alpha ) H_i - sh\, \theta (\alpha
) {\vec n}_{+}\cdot {\vec \kappa}_i
\nonumber \\
{\rightarrow}_{\alpha \rightarrow \infty}&& H_i(\infty )=
\sqrt{m^2_i+{\vec \kappa}_i^2(\infty )}={1\over
{\sqrt{\Pi}}} ( H_M H_i-{\vec \kappa}_{+}\cdot {\vec
\kappa}_i)\approx H_i,\nonumber \\
 &&{}\nonumber \\
 \Rightarrow &&H_i=\sqrt{m^2_i+{\vec
\kappa}_i^2}=H_i(\alpha ) ch\, \theta (\alpha ) +{\vec n}_{+}\cdot
{\vec \kappa}_i(\alpha ) sh\, \theta (\alpha )=\nonumber \\
 &&={1\over
{\sqrt{\Pi}}} [ H_i(\infty ) H_M+{\vec n}_{+}\cdot {\vec \kappa}
_i(\infty ) |{\vec \kappa}_{+}|\approx H_i(\infty ),\nonumber \\
&&{}\nonumber \\
 &&with\quad \sum_{i=1}^N H_i(\infty )=H_M(\infty
) = \sqrt{\Pi}\, {\buildrel {def} \over =}\, H_{(rel)},\nonumber \\
 &&{}\nonumber \\
 {\vec \kappa}_i&=& {\vec \kappa}_i(\alpha ) +[ch\, \theta (\alpha ) -1] {\vec n}_{+}
 \cdot {\vec \kappa}_i(\alpha ) {\vec n}_{+} +sh\, \theta (\alpha ) H_i(\alpha ).
\label{V12}
\end{eqnarray}

Therefore, we get

\begin{eqnarray}
{\vec \pi}_a(\alpha ) &=& {1\over {\sqrt{N}}} \sum_{i=1}^N \gamma_{ai} {\vec
\kappa}_i(\alpha ),\nonumber \\
&&{}\nonumber \\ {\vec \pi}_{qa}&{\buildrel {def}\over =}&{\vec
\pi}_a(\infty ) ={1\over {\sqrt{N}}} \sum_{i=1}^N {\vec
\kappa}_i(\infty )=\nonumber \\
&=&{\vec \pi}_a+ {{{\vec n}_{+}}\over {\sqrt{\Pi}}} [(H_M-\sqrt{\Pi}) {\vec n}
_{+}\cdot {\vec \pi}_a-|{\vec \kappa}_{+}| H_a]=\nonumber \\
&=& {\vec \pi}_a -{{{\vec \kappa}_{+}}\over {\sqrt{H^2_M-{\vec \kappa}_{+}^2}}}
[H_a-{{H_M-\sqrt{H_M^2-{\vec \kappa}^2_{+}}}\over {{\vec \kappa}_{+}^2}} {\vec
\kappa}_{+}\cdot {\vec \pi}_a] \approx {\vec \pi}_a,\nonumber \\
 &&{}\nonumber \\
 H_a&=& {1\over {\sqrt{N}}} \sum_{i=1}^N \gamma_{ai} H_i,\nonumber \\
 &&{}\nonumber \\
 {\vec \kappa}_i(\infty )&=& \sqrt{N} \sum_{a=1}^{N-1} \gamma_{ai} {\vec \pi}_{qa},\nonumber \\
 H_{(rel) i} &=& H_i(\infty ) =  \sqrt{m_i^2+N
\sum_{ab}^{1..N-1}\gamma_{ai}\gamma_{bi}{\vec \pi}_{qa}\cdot {\vec \pi}_{qb}},
\nonumber \\
 M_{sys}&=& H_M=\sum_{i=1}^NH_i =\sqrt{\Pi +{\vec \kappa}_{+}^2}
 \approx  H_{(rel)}=H_M(\infty ) = \sqrt{\Pi} =\nonumber \\
 &=& \sum_{i=1}^N H_i(\infty ) = \sum_{i=1}^N \sqrt{m_i^2+N
\sum_{ab}^{1..N-1}\gamma_{ai}\gamma_{bi}{\vec \pi}_{qa}\cdot {\vec \pi}_{qb}}.
\label{V13}
\end{eqnarray}

Let us now evaluate ${\vec \rho}_{qa}$.

Let us first remark that the following two quantities are invariant
under the canonical transformation $U(\alpha )$:

\begin{eqnarray}
I^{(1)}_i&=& H_M H_i -{\vec \kappa}_{+}\cdot {\vec \kappa}_i =H_M(\alpha )
H_i(\alpha )-{\vec \kappa}_{+}(\alpha )\cdot {\vec \kappa}_i(\alpha ),
\nonumber \\
&&\Rightarrow \quad {{dI^{(1)}_i}\over {d\alpha}}=0,\nonumber \\
I^{(2)}&=&{{{\vec \kappa}_{+}\cdot \vec K}\over {H_M}}=-\sum_{i=1}^N |{\vec
\kappa}_{+}| ({\vec n}_{+} \cdot {{{\vec \eta}_i H_i}\over {H_M}})=
{{{\vec \kappa}_{+}(\alpha )\cdot \vec K(\alpha )}\over {H_M(\alpha )}},
\nonumber \\
&&\Rightarrow {{dI^{(2)}}\over {d\alpha}} =0,
\label{V14}
\end{eqnarray}

\noindent and that we have

\begin{eqnarray}
{1\over {H_M(\alpha ) |{\vec \kappa}_{+}(\alpha )|}}&=&{{dJ^{(1)}(\alpha )}\over
{d\alpha}},\quad with\quad J^{(1)}(\alpha )= {{sh\, \theta (\alpha )}\over
{|{\vec \kappa}_{+}(\alpha )| |{\vec \kappa}_{+}|}},\nonumber \\
&&{}\nonumber \\
{{{\vec \kappa}_i(\alpha )}\over {H^2_i(\alpha )}}&=&{{d{\vec J}^{(2)}
_i(\alpha )}\over {d\theta (\alpha )}},\quad with\quad {\vec J}^{(2)}_i(\alpha )
={{{\vec \kappa}_i(\alpha ) sh\, \theta (\alpha )}\over {H_i H_i(\alpha )}}+
(ch\, \theta (\alpha )-1) {{{\vec n}_{+}}\over {H_i}}.
\label{V15}
\end{eqnarray}

Since we have also

\begin{equation}
n^r_{+} {{\partial}\over {\partial \kappa^r_i}} n^s_{+}={{n^r_{+}}\over
{|{\vec \kappa}_{+}|}} (\delta^{rs}-n^r_{+}n^s_{+})=0,
\label{V16}
\end{equation}

\noindent we get preliminarly,  for ${\vec n}_{+}\cdot {\vec \eta}_i(\alpha )$

\begin{eqnarray}
{d\over {d\alpha}} {\vec n}_{+}\cdot {\vec \eta}_i(\alpha )
&=&\{ {\vec n}_{+}\cdot
{\vec \eta}_i(\alpha ),{\vec \kappa}_{+}(\alpha )\cdot {\vec q}_{+}(\alpha )\}
=\nonumber \\
&=&-n^r_{+} {{\partial}\over {\partial k^r_i(\alpha )}} {{{\vec n}_{+}\cdot
\vec K(\alpha ) |{\vec \kappa}_{+}(\alpha )|}\over {H_M(\alpha )}}=- n^r_{+}
n^s_{+} {{\partial}\over {\partial k^r_i(\alpha )}} {{|{\vec \kappa}
_{+}(\alpha )| K^s(\alpha )}\over {H_M(\alpha )}}.
\label{V17}
\end{eqnarray}

Then, since

\begin{eqnarray}
&&n^r_{+} {{\partial}\over {\partial k^r_i(\alpha )}} {{|{\vec \kappa}
_{+}(\alpha )|}\over {H_M(\alpha )}}= {{I^{(1)}_i}\over {H^2_M(\alpha )
H_i(\alpha )}},\nonumber \\
&&{{\partial}\over {\partial k^r_i(\alpha )}} K^s(\alpha )=- {{k^r_i(\alpha )
\eta^s_i(\alpha )}\over {H_i(\alpha )}},
\label{V18}
\end{eqnarray}

\noindent we get

\begin{eqnarray}
{d\over {d\alpha}} {\vec n}_{+}\cdot {\vec \eta}_i(\alpha )&=&- {{I^{(2)}
I^{(1)}_i}\over {H_M(\alpha )H_i(\alpha ) |{\vec \kappa}_{+}(\alpha )|}}+
{{{\vec n}_{+}\cdot {\vec \eta}_i(\alpha ) {\vec n}_{+}\cdot {\vec \kappa}
_i(\alpha )}\over {H_i(\alpha )}} {{|{\vec \kappa}_{+}(\alpha )|}\over
{H_M(\alpha )}}=\nonumber \\
&=&-{{{\vec n}_{+}\cdot {\vec \eta}_i(\alpha )}\over {H_i(\alpha )}}
{{dH_i(\alpha )}\over {d\alpha}}- {{I^{(1)}_i I^{(2)}}\over {H_i(\alpha )}}
{{dJ^{(1)}(\alpha )}\over {d\alpha}}.
\label{V19}
\end{eqnarray}

These equations have the solution

\begin{eqnarray}
{\vec n}_{+}\cdot {\vec \eta}_i(\alpha )&=&{{H_i}\over {H_i(\alpha )}} {\vec n}
_{+}\cdot {\vec \eta}_i- {{I^{(1)}_i I^{(2)}}\over {H_i(\alpha )}} {{sh\,
\theta (\alpha )}\over {|{\vec \kappa}_{+}(\alpha )| |{\vec \kappa}_{+}|}}=
\nonumber \\
&=& {{H_i}\over {H_i(\alpha )}} {\vec n}_{+}\cdot {\vec \eta}_i -{{I^{(2)}}
\over {|{\vec \kappa}_{+}|}} ( e^{\alpha}- {{H_i}\over {H_i(\alpha )}}).
\label{V20}
\end{eqnarray}

For ${\vec \eta}_i(\alpha )$ we have

\begin{eqnarray}
{{d\eta^r_i(\alpha )}\over {d\alpha}}&=& \{ \eta^r_i(\alpha ), {\vec \kappa}
_{+}(\alpha )\cdot {\vec q}_{+}(\alpha ) \} =-n^s_{+} {{\partial}\over
{\partial k^r_i(\alpha )}} {{|{\vec \kappa}_{+}(\alpha )| K^s(\alpha )}\over
{H_M(\alpha )}}=\nonumber \\
&=& {\vec n}_{+}\cdot {\vec \eta}_i(\alpha ) {{|{\vec \kappa}_{+}(\alpha )|
k^r_i(\alpha )}\over {H_i(\alpha ) H_M(\alpha )}}+\nonumber \\
&+& {{ \sum_{j=1}^N H_j(\alpha ) {\vec n}_{+}\cdot {\vec \eta}_j(\alpha )}\over
{H_M(\alpha )}}
\Big[ {{|{\vec \kappa}_{+}(\alpha )| k^r_i(\alpha )}\over {H_i(\alpha )
H_M(\alpha )}}-n^r_{+}\Big].
\label{V21}
\end{eqnarray}

By putting Eqs.(\ref{V20}) in Eq.(\ref{V21}) we get the equations
determining ${\vec \eta}_i(\alpha )$.

Instead of integrating these equations, let us study the equations for
${\vec \rho}_a(\alpha )=\sqrt{N} \sum_{i=1}^N \gamma_{ai} {\vec \eta}
_i(\alpha )$, since for interactions depending on ${\vec \eta}_i-{\vec \eta}
_j$ we have ${\vec \eta}_i(\alpha )-{\vec \eta}_j(\alpha )={1\over {\sqrt{N}}}
\sum_{a=1}^{N-1} (\gamma_{ai}-\gamma_{aj}) {\vec \rho}_a(\alpha )$.
Eqs.(\ref{V21}) and (\ref{V20}) imply

\begin{eqnarray}
{{d{\vec \rho}_a(\alpha )}\over {d\alpha}}&=&- \sum_{i,j=1}^N\sum_{b=1}^{N-1}
{\vec n}_{+}\cdot {\vec \rho}_b {{H_iH_j}\over {H_M}} \gamma_{aj}(\gamma_{bi}-
\gamma_{bj}) {{{\vec k}_j(\alpha ) |{\vec \kappa}_{+}(\alpha )|}\over {H^2
_j(\alpha ) H_M(\alpha )}},\nonumber \\
&&\Downarrow \nonumber \\
{{d{\vec \rho}_a(\alpha )}\over {d\theta (\alpha )}}&=&-\sum_{i,j=1}^N \sum
_{b=1}^{N-1} {\vec n}_{+}\cdot {\vec \rho}_b {{H_iH_j}\over {H_M}} \gamma_{aj}
(\gamma_{bi}-\gamma_{bj}) {{d{\vec J}^{(2)}_j(\alpha )}\over {d\theta
(\alpha )}},
\label{V22}
\end{eqnarray}

\noindent whose solution is

\begin{equation}
{\vec \rho}_a(\alpha )={\vec \rho}_a -\sum_{i,j=1}^N \sum_{b=1}^{N-1}{\vec n}
_{+}\cdot {\vec \rho}_b {{H_iH_j}\over {H_M}} \gamma_{aj}(\gamma_{bi}-\gamma
_{bj}) {\vec J}^{(2)}_j(\alpha ).
\label{V23}
\end{equation}

Foe $\alpha \rightarrow \infty$ we get

\begin{eqnarray}
{\vec \rho}_{qa}&{\buildrel {def}\over =}& {\vec \rho}_a(\infty
)={\vec \rho}_a-\nonumber \\
 &-&\sum_{i,j=1}^N\sum_{b=1}^{N-1}
\gamma_{aj}(\gamma_{bi}-\gamma_{bj}) {{H_i}\over {H_M}} \Big[ {{|{\vec
\kappa}_{+}| {\vec \kappa}_j(\infty )}\over {H_j(\infty )
\sqrt{\Pi}}}+({{H_M}\over {\sqrt{\Pi}}}-1) {\vec n}_{+} \Big] {\vec
n}_{+}\cdot {\vec \rho}_b=\nonumber \\
 &=& {\vec \rho}_a- \sum_{i,j=1}^N\sum_{b=1}^{N-1}
\gamma_{aj}(\gamma_{bi}-\gamma_{bj}) {{H_i}\over {H_M}}
 {{ {\vec \kappa}_j(\infty )}\over {H_j(\infty )
\sqrt{\Pi}}} {\vec \kappa}_{+}\cdot {\vec \rho}_b\, \approx {\vec \rho}_a.
\label{V24}
\end{eqnarray}

One can check that for N=2 and $\gamma_1=-\gamma_2=1/\sqrt{2}$ one
reobtains the results of Ref.\cite{osborne}.

Let us now consider the spin vector ${\vec S}_q={\vec {\bar S}}_s-{\vec q}_{+}
\times {\vec \kappa}_{+}=[{\vec \eta}_{+}-{\vec q}_{+}]\times {\vec \kappa}_{+}
+\sum_{a=1}^{N-1}{\vec \rho}_a\times {\vec \pi}_a$. For arbitrary $\alpha$ we
have ${\vec S}_q(\alpha )=\sum_{a=1}^{N-1}{\vec \rho}_a(\alpha )\times {\vec
\pi}_a(\alpha )+[{\vec \eta}_{+}(\alpha )-{\vec q}_{+}(\alpha )]\times {\vec
\kappa}_{+}(\alpha )$ and, since ${\vec q}_{+}(\alpha )\cdot {\vec \kappa}
_{+}(\alpha )$ is a scalar, $\{ {\vec S}_q(\alpha ),{\vec q}_{+}(\alpha )\cdot
{\vec \kappa}_{+}(\alpha ) \} =0$. Since $lim_{\alpha \rightarrow \infty}\,
{\vec \kappa}_{+}(\alpha )=0$, we get

\begin{equation}
{\vec S}_q(\alpha )\, \rightarrow_{\alpha \rightarrow \infty}\,\, {\vec S}_q
=\sum_{a=1}^{N-1} {\vec \rho}_{qa}\times {\vec \pi}_{qa},
\label{V25}
\end{equation}

\noindent if we can show that
${\vec \eta}_{+}(\alpha )-{\vec q}_{+}(\alpha )\,
{\rightarrow}_{\alpha \rightarrow \infty}\,\, finite\,\, value$. But, since the
boost generator may be written as

\begin{eqnarray}
\vec K(\alpha )&=&-\sum_{i=1}^N{\vec \eta}_i(\alpha )H_i(\alpha )=-{\vec \eta}
_{+}(\alpha )H_M(\alpha )+\sum_{a=1}^{N-1}{\vec \rho}_a(\alpha )H_a(\alpha ),
\nonumber \\
&&H_a(\alpha )={1\over {\sqrt{N}}}\sum_{i=1}^N\gamma_{ai}H_i(\alpha ),
\label{V26}
\end{eqnarray}

\noindent we get

\begin{eqnarray}
{\vec \eta}_{+}(\alpha )-{\vec q}_{+}(\alpha )&=& {{\sum_{a=1}^N
 {\vec \rho}_a(\alpha )H_a(\alpha )}\over {H_M(\alpha )}}+ {{ {\vec \kappa}
_{+}(\alpha )\times \Big( \sum_{a=1}^{N-1}{\vec \rho}_a(\alpha )\times {\vec
\pi}_a(\alpha )\Big)}\over {\sqrt{\Pi} (\sqrt{\Pi}+H_M)}}+\nonumber \\
&+&{{ {\vec \kappa}_{+}(\alpha )\times \Big( {\vec \kappa}_{+}(\alpha )\times
\sum_{a=1}^{N-1}{\vec \rho}_a(\alpha )H_a(\alpha )\Big)}\over {H_M(\alpha )
\sqrt{\Pi} (\sqrt{\Pi} +H_M)}}\, \rightarrow_{\alpha \rightarrow \infty}\,\,
\nonumber \\
 && \rightarrow_{\alpha \rightarrow \infty}\, {1\over {\sqrt{\Pi}}} \sum_{a=1}^N
 {\vec \rho}_a(\infty ) H_a(\infty )=\nonumber \\
 &=& {1\over {\sqrt{\Pi}}} \sum_{a=1}^{N-1} {\vec \rho}_{qa}\, {1\over {\sqrt{N}}}
 \sum_{i=1}^N \gamma_{ai} \sqrt{m_i^2+N
\sum_{ab}^{1..N-1}\gamma_{ai}\gamma_{bi}{\vec \pi}_{qa}\cdot {\vec \pi}_{qb}},
\nonumber \\
 &&\Downarrow \nonumber \\
 {\vec q}_{+}&=& {\vec \eta}_{+}- {1\over {\sqrt{N}}} {{\sum_{a=1}^{N-1} {\vec \rho}_a}\over
 {\sqrt{ (\sum_{i=1}^N\sqrt{m^2_i+{\vec \kappa}_i^2})^2-{\vec \kappa}_{+}^2} }}
 \sum_{i=1}^N \gamma_{ai} \sqrt{m_i^2+{\vec \kappa}_i^2} +\nonumber \\
 &+&(terms{\rightarrow}_{\alpha \rightarrow \infty}\, 0, i.e. \approx 0\quad due\, to\,
 {\vec \kappa}_{+}\approx 0),
\label{V27}
\end{eqnarray}

\noindent to be compared with Eq.(\ref{III4}).

In this way we have obtained the canonical transformation
(\ref{III7})

\begin{equation}
\begin{minipage}[t]{1cm}
\begin{tabular}{|l|} \hline
${\vec \eta}_i$ \\  \hline
 ${\vec \kappa}_i$ \\ \hline
\end{tabular}
\end{minipage} \ {\longrightarrow \hspace{.2cm}} \
\begin{minipage}[t]{2 cm}
\begin{tabular}{|l|l|} \hline
${\vec q}_{+}$   & ${\vec \rho}_{qa}$   \\ \hline ${\vec
\kappa}_{+}$&${\vec \pi}_{qa}$ \\ \hline
\end{tabular}
\end{minipage}
\label{VI28}
\end{equation}

\noindent even if it is not known how to get the inverse canonical transformation.

When we add the gauge fixings ${\vec q}_{+}\approx 0$ for ${\vec \kappa}_{+}
\approx 0$ and we go to Dirac brackets, we get

\begin{eqnarray}
&&{\vec \rho}_{qa} \equiv {\vec \rho}_a,\quad\quad {\vec \pi}_{qa}
\equiv {\vec \pi}_a,\quad\quad {\vec S}_q=\sum_{a=1}^{N-1}{\vec \rho}_{qa}\times
{\vec \pi}_{qa}\, {\buildrel {def}\over =}\,
\sum_{a=1}^{N-1}{\vec \rho}_a\times {\vec \pi}_a,\nonumber \\
 &&{}\nonumber \\
H_{(rel)}&=& \sqrt{\Pi} = H_M(\infty )=\sum_{i=1}^N\sqrt{m_i^2+N
\sum_{ab}^{1..N-1}\gamma_{ai}\gamma_{bi}{\vec \pi}_{qa}\cdot {\vec \pi}_{qb}}
\equiv M_{sys} = H_M.
\label{V29}
\end{eqnarray}

See Appendix C for the case of spinning particles.

\vfill\eject

\section{Relativistic Rotational Kinematics.}

All the results of the previous Sections are needed to get the
separation of the {\it internal} center-of-mass degrees of freedom in
the relativistic theory: this has been accomplished by adding the
gauge fixings ${\vec q}_{+}\approx 0$ to the rest-frame constraints
${\vec \kappa}_{+}\approx 0$ and by going to Dirac brackets. We are
left with the relative canonical variables ${\vec
\rho}_{qa}\equiv {\vec \rho}_a$, ${\vec \pi}_{qa}\equiv {\vec \pi}_a$
and the Hamiltonian

\begin{equation}
H_{(rel)}= H_M(\infty )= \sum_{i=1}^N\sqrt{m_i^2+N
\sum_{ab}^{1..N-1}\gamma_{ai}\gamma_{bi}{\vec \pi}_{qa}\cdot {\vec \pi}_{qb}}
\equiv M_{sys},
\label{VI1}
\end{equation}

\noindent which replaces to the non-relativistic
one $H_{rel,nr}={1\over 2}\sum_{a,b=1}^{N-1}
k^{-1}_{ab}[m_i,\gamma_{ai}]\,\, {\vec \pi}_{qa}\cdot {\vec
\pi}_{qb}$ of Eq.(2.9) of Ref.\cite{iten2} for $c \rightarrow \infty$.

Let us remark that in the Hamiltonian for the relative motions in the
rest frame instant form, each square root identifies a $(N-1)\times
(N-1)$ matrix $K^{-1}_{(i)ab}=N\gamma_{ai}\gamma_{bi}=K^{-1}_{(i)ba}$
\footnote{At the non-relativistic level there is only one such matrix at the
Hamiltonian level, i.e. $k^{-1}_{ab}=\sum_{i=1}^N {1\over {m_i}}
K^{-1}_{(i)ab}$, see Eq.(2.9) of Ref.\cite{iten2}.}. The existence of
relativistic normal Jacobi coordinates would require the simultaneous
diagonalization of these N matrices. But this is impossible because

\beq
[ K^{-1}_{(i)}, K^{-1}_{(j)}
]_{ab}=G_{(ij)ab}=-G_{(ij)ba}=-G_{(ji)ab},
\label{VI2}
\eeq

\noindent with $G_{(ij)ab}=-N[\gamma_{ai}\gamma_{bj}-\gamma_{aj}\gamma_{bi}]$.
There are ${1\over 2}N(N-1)$ matrices $G_{ij}$, each one with ${1\over
2}(N-1)(N-2)$ independent elements. Therefore, the conditions
$G_{(ij)ab}=0$ are ${1\over 4}N(N-1)^2(N-2)$ conditions and we have
only ${1\over 2}(N-1)(N-2)$ free parameters in the $\gamma_{ai}$
\footnote{For N=3, there are 3 conditions and only 1 parameter; for N=4, 18
conditions and 3 parameters.}.

Therefore, it is impossible to diagonalize simultaneously the N
quadratic forms under the square roots: {\it there are no relativistic
normal Jacobi coordinates and no definition of reduced masses and
tensors of inertia}.

To find the analogue of $L_{rel,nr}={1\over 2} \sum_{a,b=1}^{N-1}
k_{ab}[m_i,\gamma_{ai}]\,\, {\dot {\vec \rho}}_a\cdot {\dot {\vec
\rho}}_b$ (Eq.(2.9) of Ref.\cite{iten2}),
we should perform an inverse Legendre transformation.
The first half of Hamilton equations gives

\begin{eqnarray}
{\dot \rho}^r_{qa}\, &{\buildrel \circ \over =}\,& \{
\rho^r_{qa},H_{(rel)}
\}= \sum_{i=1}^N { {N\gamma_{ai} \sum_{b=1}^{N-1} \gamma_{bi}
\pi^r_{qb}}\over {\sqrt{m_i^2+N
\sum_{ab}^{1..N-1}\gamma_{ai}\gamma_{bi}{\vec \pi}_{qa}\cdot {\vec \pi}_{qb}} }},
\nonumber \\
 &&{}\nonumber \\
 &&\Downarrow \nonumber \\
 &&{}\nonumber \\
 {\dot {\vec \rho}}_{qa}\cdot {\dot {\vec \rho}}_{qb}&=&
 \sum_{i,j}^{1..N} { {N\gamma_{ai} \sum_{e=1}^{N-1} \gamma_{ei} {\vec
\pi}_{qe}}\over {\sqrt{m_i^2+N
\sum_{a_1b_1}^{1..N-1}\gamma_{a_1i}\gamma_{b_1i}{\vec \pi}_{qa_1}\cdot
{\vec \pi}_{qb_1}} }}\cdot
{ {N\gamma_{bj} \sum_{f=1}^{N-1} \gamma_{fj} {\vec \pi}_{qf}}\over {
\sqrt{m_j^2+N
\sum_{a_2b_2}^{1..N-1}\gamma_{a_2j}\gamma_{b_2j}{\vec \pi}_{qa_2}\cdot
{\vec \pi}_{qb_2}} }}.\nonumber \\
 &&{}
\label{VI3}
\end{eqnarray}

\noindent To get ${\vec \pi}_{qa}\cdot {\vec \pi}_{qb}$ in terms
of ${\dot {\vec \rho}}_{qa}\cdot {\dot {\vec \rho}}_{qb}$ we should
solve higher order algebraic equations. As already pointed out, this
implies that $L_{rel}({\vec \rho}_{qa},{\dot {\vec
\rho}}_{qa})=\sum_{a=1}^{N-1} {\vec \pi}_{qa}\cdot {\dot {\vec \rho}}_{qa}-H$
is a hyperelliptic function already in the free case. This  in turn
means that, unlike  the non-relativistic case, it is not possible to
define  either an Euclidean or a Riemannian metric on the space of
velocities from the kinetic energy  (see Ref.\cite{iten2}). Therefore
we cannot visualize the Lagrangian dynamics as in the non-relativistic
case. The form of the canonical momenta

\begin{equation}
\pi^r_{qa}={{\partial L_{rel}}\over {\partial {\dot \rho}^r_{qa}}}
=\sum_{b=1}^{N-1} f_{ab}({\dot {\vec \rho}}_{qc}\cdot {\dot {\vec
\rho}}_{qd}) {\dot \rho}^r_{qb},
\label{VI4}
\end{equation}

\noindent can only be given in implicit form.

Moreover, we cannot evaluate the restrictions on the relative
velocities ${\dot {\vec \rho}}_{qa}(\tau )$ resulting from the
existence of the limiting light velocity $c$ \footnote{For the
absolute velocities ${\dot {\vec \eta}}_i(\tau )$ we have $|{\dot
{\vec \eta}}_i(\tau )| \leq c=1$.}.

If we try to follow  the non-relativistic pattern of the {\it static}
orientation-shape bundle approach (see Ref.\cite{iten2}), we
get\footnote{The first line defines the {\it body frame components}
${\check \rho}^r_{qa}$ of the vectors ${\vec \rho}_{qa}$ in this
approach ($\theta^{\alpha}$ are Euler angles). The body frame
components of the relative velocities are the ${\check v}^r_{qa}$ of
the second line, while those of the spin and of the angular velocity
are ${\check S}^r_q$ and ${\check \omega}^r$ respectively.}

\begin{eqnarray}
\rho^r_{qa}&=&R^{rs}(\theta^{\alpha}) {\check \rho}^s_{qa}(q),\nonumber \\
{\dot \rho}^r_{qa}\, &{\buildrel {def} \over =}\,&
R^{rs}(\theta^{\alpha}) {\check v}^s_{qa},\qquad
 {\vec v}_{qa} = {\vec \omega}\times {\vec
\rho}_{qa}+ {{\partial {\vec \rho}_{qa}}\over {\partial q^{\mu}}}
{\dot q}^{\mu},\nonumber \\
 &&{}\nonumber \\
{\dot \rho}^r_{qa}\, &{\buildrel \circ \over =}\,& \{ \rho^r_{qa},
H_{rel}\},\nonumber \\
\pi^r_{qa}&=&\sum_{b=1}^{N-1}f_{ab}({\dot {\vec \rho}}_{qc}\cdot {\dot {\vec
\rho}}_{qd}) {\dot \rho}^r_{qb}=R^{rs}(\theta^{\alpha}) {\check \pi}^s_{qa},\nonumber \\
{\check \pi}^r_{qa}&=&\sum_{b=1}^{N-1}f_{ab}({\check {\vec
v}}_{qc}\cdot {\check {\vec v}}_{qd}) {\check v}^r_{qb},\nonumber \\
 &&{}\nonumber \\
 S_q^r&=&R^{rs}(\theta^{\alpha}) {\check S}_q^s =
\sum_{a=1}^{N-1} [{\vec
\rho}_{qa}\times {\vec \pi}_{qa}]^r,\nonumber \\
{\vec S}_q&=& \sum_{a=1}^{N-1} {\vec
\rho}_{qa}\times {\vec \pi}_{qa}=\nonumber \\
&=&\sum_{ab}^{1...N-1} f_{ab}\Big[ \Big( {\vec \omega}
\times {\vec \rho}_{qc}+{{\partial {\vec \rho}_{qc}}\over
{\partial q^{\mu}}} {\dot q}^{\mu}\Big) \cdot \Big( {\vec
\omega}\times {\vec \rho}_{qd}+ {{\partial {\vec
\rho}_{qd}}\over {\partial q^{\mu}}} {\dot q}^{\mu}\Big)
\Big] \nonumber \\
 &&{\vec \rho}_{qa}\times \Big(
{\vec \omega}\times {\vec \rho}_{qb}+ {{\partial {\vec
\rho}_{qb}}\over {\partial q^{\mu}}} {\dot q}^{\mu}
\Big),\nonumber \\
{\check S}^r_q&=&
\sum_{ab}^{1...N-1} f_{ab}\Big[ {\check I}^{uv}_{(cd)}(q){\check \omega}^u{\check
\omega}^v+{\vec \omega}\cdot ({\vec \rho}_{qc}\times
{{\partial {\vec \rho}_{qd}}\over {\partial q^{\mu}}}+{\vec
\rho}_{qd}\times {{\partial {\vec \rho}_{qc}}\over {\partial
q^{\mu}}} ) {\dot q}^{\mu}+{{\partial {\vec
\rho}_c}\over {\partial q^{\mu}}}\cdot {{\partial {\vec
\rho}_d}\over {\partial q^{\nu}}}{\dot q}^{\mu}{\dot q}^{\nu}\Big]\nonumber \\
&&\Big[ {\check I}^{rs}_{(ab)}(q) {\check
\omega}^s+{\check a}^r_{(ab)\mu}(q){\dot q}^{\mu}\Big] =\nonumber \\
&=&\sum_{ab}^{1...N-1} f_{ab}\Big[ {\check I}^{uv}_{(cd)}(q)\Big(
{\check
\omega}^u+ {\check A}^u_{(cd)\mu}(q){\dot q}^{\mu}\Big) \Big( {\check
\omega}^v+ {\check A}^v_{(cd)\nu}(q){\dot q}^{\nu}\Big) \Big]\nonumber \\
&&{\check I}^{rs}_{(ab)}(q)\Big( {\check
\omega}^s+{\check A}^s_{(ab)\mu}(q){\dot q}^{\mu}\Big) ,\nonumber \\
&&{}\nonumber \\
 {\check I}^{rs}_{(ab)}(q)&=&{\vec
\rho}_{qa}\cdot {\vec \rho}_{qb}\delta^{rs}- {1\over
2}({\check \rho}^r_{qa}{\check
\rho}^s_b+{\check \rho}^r_{qb}{\check \rho}^s_{qa}),\nonumber \\
 {\check {\cal I}}^{rs}({\check
{\vec \omega}},q,m)&=&
\sum_{ab}^{1..N-1} f_{ab}({\vec
v}_{qc}\cdot {\vec v}_{qd})  {\check I}^{rs}_{(ab)}(q),\nonumber
\\
 &&{}\nonumber \\
 {\check a}^u_{(ab)\mu}(q)&=&{1\over 2}\Big[ {\vec
\rho}_{qa}\times {{\partial {\vec \rho}_{qb}}\over {\partial
q^{\mu}}}+{\vec \rho}_{qb}\times {{\partial {\vec
\rho}_{qa}}\over {\partial q^{\mu}}}
\Big] {}^u {\buildrel {def} \over =}\, {\check I}^{uv}_{(ab)}(q)
{\check A}^v_{(ab)\mu}(q),\nonumber \\
 {\check a}^u_{\mu}({\check
{\vec \omega}},q,m)&=&\sum_{ab}^{1..N-1} f_{ab}({\vec v}_{qc}\cdot
{\vec v}_{qd})  {\check a}^u_{(ab)\mu}(q)={\check {\cal
I}}^{uv}({\check {\vec
\omega}},q,m) {\check {\cal A}}^v_{\mu}({\vec \omega},q,m).
\label{VI5}
\end{eqnarray}

We see that {\it there is no more a linear relation} between the body
frame spin and angular velocity. By expanding $f_{ab}(x)$ in a power
series around $x=0$, we get that ${\check S}_q^r$ is an infinite
series with all the powers of the body frame angular velocity. The
lowest term is ${\check S}^r_{q(o)}=\sum_{a,b}^{N-1} f_{ab}(0){\check
I}^{rs}_{(ab)}(q)\Big( {\check \omega}^s+{\check
A}^s_{(ab)\mu}(q){\dot q}^{\mu}\Big)$ with $f_{ab}(0)$ playing the
role of the non-relativistic $k_{ab}$ \footnote{Recall that its
diagonalization defines the Jacobi coordinates and the reduced
masses.}.

Therefore, the tensor of inertia looses a clear identification: only
its building blocks ${\check I}^{rs}_{(ab)}$, existing also in the
non-relativistic theory, appear in the relativistic construction.

The N=2 case with equal masses $m_1=m_2=m$ is the only case in which
we can evaluate the relative Lagrangian. We get

\begin{equation}
L_{rel}(\vec \rho ,{\dot {\vec \rho}})=-m \sqrt{4-{\dot {\vec
\rho}}^2}.
\label{VI6}
\end{equation}

Therefore in this case the only existing relative velocity has the
bound $|{\dot {\vec \rho}}|\leq 2$.

Let us write $\vec \rho = \rho \hat \rho$ with $\rho = |\vec \rho |$
and $\hat \rho ={{\vec \rho}\over {|\vec \rho |}}$. With only one
relative variable the three Euler angles $\theta^{\alpha}$ are
redundand: there are only two independent angles, those identifying
the position of the unit 3-vector $\hat \rho$ on $S^2$. We shall use
the following parametrization (Euler angles $\theta^1=\phi$,
$\theta^2=\theta$, $\theta^3=0$)

\begin{eqnarray}
{\hat \rho}^r &=& R^{rs}(\theta ,\phi ) {\hat \rho}_o^s,\nonumber \\
 &&{\hat \rho}_o=(0,0,1),\quad\quad (reference\,\, unit\,\, 3-vector),
 \nonumber \\
 &&{}\nonumber \\
R^{rs}(\theta ,\phi )&=&R_z(\theta )R_y(\phi )=\left(
\begin{array}{ccc} cos\, \theta cos\, \phi& -sin\, \phi& sin\, \theta
cos\, \phi \\ cos\,
\theta sin\, \phi& cos\, \phi& sin\, \theta sin\, \phi \\
-sin\, \theta& 0& cos\, \theta \end{array} \right) ,\nonumber \\
 &&{}\nonumber \\
 &&\Downarrow \nonumber \\
 &&{}\nonumber \\
{\dot R}^{rs}&=&\left( \begin{array}{ccc} -sin\, \theta cos\, \phi
\dot \theta&
-cos\, \phi \dot \phi& cos\, \theta cos\, \phi \dot \theta -sin\, \theta sin\,
\phi \dot \phi \\
-sin\, \theta sin\, \phi \dot \theta +cos\, \theta cos\, \phi \dot \phi&
-sin\, \phi \dot \phi& cos\, \theta sin\, \phi \dot \theta +sin\, \theta cos\,
\phi \dot \phi \\
-cos\, \theta \dot \theta& 0& -sin\, \theta \dot \theta \end{array} \right) ,
\nonumber \\
R^T\dot R &=& \left( \begin{array}{ccc} 0&-cos\, \theta \dot \phi &
\dot
\theta \\
cos\, \theta \dot \phi& 0& sin\, \theta \dot \phi \\
-\dot \theta& -sin\, \theta \dot \phi& 0 \end{array} \right) .
\label{VI7}
\end{eqnarray}

Following the  orientation-shape bundle approach, we get the following
body frame velocity and angular velocity ($\rho$ is the only shape
variable in this case)

\begin{eqnarray}
{\check v}^r&=&R^{T\, rs}{\dot \rho}^s=\rho (R^T\dot R)^{rs}{\hat
\rho}^s_o+\dot \rho {\hat \rho}_o^r=\rho \epsilon^{ru3}{\check \omega}^u
+\dot \rho {\hat \rho}_o^r=\nonumber \\
 &=&\rho ({\vec \omega}\times {\hat \rho}_o)^r+\dot \rho {\hat
\rho}^r_o,\nonumber \\
  &&{}\nonumber \\
{\vec \omega}&=&({\check \omega}^1=-sin\, \theta \dot \phi ,{\check
\omega}^2=\dot \theta ,0),\nonumber \\
 &&{}\nonumber \\
{\vec v}^2&=&\check I(\rho ){\vec \omega}^2+{\dot
\rho}^2,\nonumber \\
 &&{}\nonumber \\
\check I(\rho )&=&\rho^2.
\label{VI8}
\end{eqnarray}

The non-relativistic inertia tensor of the dipole ${\check I}_{nr}=\mu
\rho^2$ \footnote{$\mu ={{m_1m_2}\over {m_1+m_2}}$ is the reduced mass; see Ref.\cite{iten2}.}
is replaced by $\check I = {\check I}_{nr}/ \mu =\rho^2$. The
Lagrangian in anholonomic variables become

\begin{equation}
{\tilde L}({\vec \omega},\rho ,\dot \rho )=-m\sqrt{4-\check I(\rho
){\vec \omega}^2-{\dot \rho}^2}.
\label{VI9}
\end{equation}

\noindent It is clear that the bound $|{\dot {\vec \rho}}|\leq 2$ put upper
bounds on the kinetic energy of both the rotational and vibrational
motions.

The canonical momenta are

\begin{eqnarray}
{\vec S}&=&{{\partial \tilde L}\over {\partial {\vec
\omega} }}
={{m \check I(\rho ) {\vec \omega}}\over {\sqrt{4-\check I(\rho ){\vec
\omega}^2-{\dot \rho}^2}}},\nonumber \\
\pi &=&{{\partial \tilde L}\over {\partial \dot \rho}}={{m\dot \rho}\over
{\sqrt{4-\check I(\rho ){\vec \omega}^2-{\dot \rho}^2}}}.
\label{VI10}
\end{eqnarray}

The body frame spin  is not linear in the body frame angular velocity
(only approximately for slow rotations). When $|{\dot {\vec
\rho}}|$ varies between $0$ and 2 the momenta vary between $0$ and
$\infty$, namely in phase space there is no bound coming from the
limiting light velocity. This shows  once more that in special
relativity it is convenient to work  in the Hamiltonian framework
avoiding relative and angular velocities.

Since we have $\sqrt{4-\check I(\rho ){\vec \omega}^2-{\dot
\rho}^2}={{2m}\over {\sqrt{m^2+{\check I}^{-1}(\rho ){\vec S}^2+\pi^2}
}}$, the inversion formulas are

\begin{eqnarray}
{\vec \omega}&=&{{ {\vec S}}\over {m\check I(\rho )}}\sqrt{4-
\check I(\rho ){\vec \omega}^2-{\dot
\rho}^2}={{2{\check I}^{-1}(\rho ) {\vec S}}\over {\sqrt{m^2+
{\check I}^{-1}(\rho ){\vec S}^2+\pi^2}}},\nonumber \\
\dot \rho &=&{{\pi}\over m}\sqrt{4-\check I(\rho ){\vec \omega}^2-{\dot
\rho}^2}={{2\pi}\over {\sqrt{m^2+{\check I}^{-1}(\rho ){\vec S}^2+\pi^2}}}.
\label{VI11}
\end{eqnarray}

On the other hand, the Hamiltonian becomes

\begin{equation}
\tilde H=\pi \dot \rho +{\vec S}\cdot {\vec
\omega}-\tilde L= 2\sqrt{m^2+{\check I}^{-1}(\rho ){\vec S}^2+\pi^2}.
\label{VI12}
\end{equation}

Therefore, this special case identifies the following point-canonical
transformation followed by a transformation to an anholonomic basis
like in the non-relativistic framework

\begin{eqnarray}
&&\begin{minipage}[t]{3cm}
\begin{tabular}{|l|} \hline
$\vec \rho$  \\ \hline
 ${\vec \pi}_q$  \\ \hline
\end{tabular}
\end{minipage}
\ {\longrightarrow \hspace{.2cm}}\
\begin{minipage}[t]{3cm}
\begin{tabular}{|ll|l|} \hline
$\theta$ & $\phi$ & $\rho$ \\ \hline
 $\pi_{\theta}$   & $\pi_{\phi}$& $\pi$ \\ \hline
\end{tabular}
\end{minipage}
\ {{\buildrel {non\,\, can.} \over \longrightarrow} \hspace{.2cm}}\
\begin{minipage}[b]{3cm}
\begin{tabular}{|ll|l|} \hline
$\theta$ & $\phi$ & $\rho$ \\ \hline
 ${\check S}^1$ & ${\check S}^2$ & $\pi$  \\
 \hline
 \end{tabular}
 \end{minipage} \nonumber \\
 &&{}\nonumber \\
\rho^r&=&\rho R^{rs}(\theta ,\phi ) {\hat \rho}_o^s=\rho (sin\, \theta
cos\, \phi , sin\, \theta sin\, \phi , cos\, \theta ),\nonumber \\
 &&{}\nonumber \\
\pi^r_q&=&R^{rs}(\theta ,\phi ) {\check \pi}^s_q=R^{rs}(\theta ,\phi ){{m{\check
v}^s}\over {\sqrt{4-{\vec v}^2}}}=R^{rs}(\theta ,\phi ) {{m[\rho {\vec
\omega}\times {\hat
\rho}_o+\dot \rho {\hat \rho}_o]^s}\over {\sqrt{4-\check I(\rho
){\vec \omega}^2-{\dot \rho}^2}}}=\nonumber \\
 &=&R^{rs}(\theta ,\phi )\Big[
\rho {\check I}^{-1}(\rho ){\vec S}\times {\hat \rho}_o+\pi {\hat
\rho}_o \Big] {}^s.
\label{VI13}
\end{eqnarray}

In conclusion, due to the absence of a workable Lagrangian approach,
we are forced to try to define everything at the Hamiltonian level. In
order to get an extension of this results for arbitrary N, we will
abandon the {\it static} orientation-shape bundle approach and we
shall investigate the {\it  canonical spin bases} as in the
non-relativistic case of Ref.\cite{iten2}. In this approach we have to
guess in some way a set of non-point canonical transformations from
the canonical variables ${\vec \rho}_{qa}$, ${\vec \pi}_{qa}$ to a
basis which generalizes the previous result for the N=2 equal mass
case.

The non-relativistic non-Abelian rotational symmetry generating the
Noether constants of motion $\vec S=constant$ is replaced by the {\it
internal} non-Abelian rotational symmetry generating the constants of
motion ${\vec S}_q$ inside the Wigner hyperplane with the rest-frame
conditions ${\vec \kappa}_{+}\approx 0$.

\vfill\eject

\section{Canonical Spin Bases}

In this Section we show that the construction of the {\it canonical
spin bases} with the associated {\it spin frame} and {\it evolving
dynamical body frames} in the relativistic case starting from the
relative canonical variables ${\vec \rho}_{qa}$, ${\vec
\pi}_{qa}$, $a=1,..,N-1$, is identical to that proposed in
Ref.\cite{iten2} for the non-relativistic case. This happens because
the total conserved rest-frame spin is ${\vec S}_q=\sum_{a=1}^{N-1}
{\vec \rho}_{qa}\times {\vec \pi}_{qa} = \sum_{a=1}^{N-1} {\vec
S}_{qa}$ like in the non-relativistic case and because the
construction is based only on the possible {\it spin clusterings}
which can be obtained from the individual ${\vec S}_{qa}$. Only the
Hamiltonian for relative motions is different.

We shall sketch the construction for $N=2$, $N=3$ and $N \geq 4$ by
referring to Ref.\cite{iten2} for the relevant calculations.

\subsection{2-Body Systems.}

Let us start from the case N=2, $i=1,2$

\begin{equation}
\begin{minipage}[t]{1cm}
\begin{tabular}{|l|} \hline
${\vec \eta}_i$ \\  \hline
 ${\vec \kappa}_i$ \\ \hline
\end{tabular}
\end{minipage} \ {\longrightarrow \hspace{.2cm}} \
\begin{minipage}[t]{2 cm}
\begin{tabular}{|l|l|} \hline
${\vec q}_{+}$   & ${\vec \rho}_q$   \\ \hline
${\vec
\kappa}_{+}$&${\vec \pi}_q$ \\ \hline
\end{tabular}
\end{minipage}
\label{VII1}
\end{equation}

After the elimination of the internal center-of-mass degrees of
freedom with the gauge fixings ${\vec q}_{+}\approx 0$, the rest-frame
dynamics of the relative motions is governed by the Hamiltonian

\begin{equation}
H_{(rel)}  = \sqrt{m_1^2+{\vec \pi}_q^2} + \sqrt{m_2^2+{\vec
\pi}_q^2} \equiv  M_{sys}.
\label{VII2}
\end{equation}

The spin is ${\vec S}_q={\vec \rho}_q\times {\vec \pi}_q$
[$S_q=\sqrt{{\vec S}_q^2}$]. Let us define the following decomposition
(the notation $\hat R$ for the unit vector ${\hat \rho}_q$ is used for
comparison with Ref.\cite{lucenti})

\begin{eqnarray}
{\vec \rho}_q&=& \rho_q \hat R,\quad\quad \rho_q=\sqrt{{\vec \rho}_q^2},\quad
\quad \hat R={{{\vec \rho}_q}\over {\rho_q}}={\hat \rho}_q,\quad\quad {\hat R}^2=1,
\nonumber \\
&&{}\nonumber \\ {\vec \pi}_q&=& {\tilde \pi}_q \hat R -{{S_q}\over
{\rho_q}} \hat R\times {\hat S}_q={\tilde \pi}_q{\hat
\rho}_q-{{S_q}\over {\rho_q}}{\hat \rho}_q\times {\hat S}_q,\nonumber \\
&& {\tilde \pi}_q={\vec \pi}_q\cdot \hat R={\vec \pi}_q\cdot {\hat
\rho}_q,\quad\quad {\hat S}_q={{{\vec S}_q}\over {S_q}},\qquad
{\hat S}_q \cdot \hat R =0.
\label{VII3}
\end{eqnarray}

The {\it spin frame} of $R^3$ is given by  ${\hat S}_q$, $\hat R$,
$\hat R\times {\hat S}_q$ with $\{ S^i_q, S^j_q \} = \epsilon^{ijk}
S^k_q$, $\{ {\hat R}^i, {\hat R}^j \} =0$, $\{ {\hat R}^i, S^j_q \} =
\epsilon^{ijk} {\hat R}^k$. The vectors ${\vec S}_q$ and $\hat R$ are
the generators of an E(3) group containing SO(3) as a subgroup.

Let us consider the following canonical transformation adapted to the
{\it spin}

\begin{equation}
\begin{minipage}[t]{1cm}
\begin{tabular}{|l|} \hline
${\vec \rho}_q$ \\ \hline
${\vec \pi}_q$ \\ \hline
\end{tabular}
\end{minipage} \ {\longrightarrow \hspace{.2cm}} \
\begin{minipage}[t]{2 cm}
\begin{tabular}{|ll|l|} \hline
$\alpha$ & $\beta$ & $\rho_q$\\ \hline
 $S_q$   & $S^3_q$ & ${\tilde \pi}_q$  \\ \hline
\end{tabular}
\end{minipage}
\label{VII4}
\end{equation}

\noindent where

\begin{eqnarray}
\alpha &=& tg^{-1} {1\over {S_q}} \Big( {\vec \rho}_q\cdot {\vec \pi}_q-
{{(\rho_q)^2}\over {\rho^3_q}} \pi^3_q\Big) ,\nonumber \\
&&{}\nonumber \\
\beta &=& tg^{-1} {{S^2_q}\over {S^1_q}},\quad\quad sin\, \beta ={{S^2_q}\over
{\sqrt{ (S_q)^2-(S^3_q)^2}}},\quad\quad cos\, \beta
={{S^1_q}\over {\sqrt{ (S_q)^2-(S^3_q)^2}}}.
\label{VII5}
\end{eqnarray}

We have

\begin{eqnarray}
S^1_q&=& \sqrt{ (S_q)^2-(S^3_q)^2} cos\, \beta ,\nonumber \\
 S^2_q&=&\sqrt{ (S_q)^2-(S^3_q)^2} sin\, \beta ,\nonumber \\
 S^3_q&&,
\label{VII6}
\end{eqnarray}

\begin{eqnarray}
{\hat R}^1&=&{\hat \rho}^1_q=sin\, \theta cos\, \varphi = sin\, \beta
sin\, \alpha -{{S^3_q}\over {S_q}} cos\, \beta cos\, \alpha ,\nonumber
\\ {\hat R}^2&=& {\hat \rho}^2_q=sin\, \theta sin\,
\varphi =-cos\, \beta sin\, \alpha - {{S^3_q}
\over {S_q}} sin\, \beta cos\, \alpha ,\nonumber \\
{\hat R}^3&=& {\hat \rho}^3_q=cos\, \theta ={1\over {S_q}} \sqrt{
(S_q)^2-(S^3_q)^2} cos\, \alpha ,\nonumber \\
 &&{}\nonumber \\
({\hat S}_q\times \hat R)^1&=&{\hat S}^2_q{\hat R}^3-{\hat S}^3_q{\hat
R}^2=
 sin\, \beta cos\, \alpha +{{S^3_q}\over {S_q}} cos\, \beta sin\, \alpha ,
 \nonumber \\
({\hat S}_q\times \hat R)^2&=&{\hat S}^3_q{\hat R}^1-{\hat S}^1_q{\hat
R}^3=-cos\, \beta cos\, \alpha +{{S^3_q}\over {S_q}} sin\, \beta sin\,
\alpha ,\nonumber \\
 ({\hat S}_q\times \hat R)^3&=&-{\hat S}^1_q{\hat R}^2-{\hat S}^2_q{\hat R}^1=
 {1\over {S_q}} \sqrt{(S_q)^2-(S^3_q)^2} sin\, \alpha .
\label{VII7}
\end{eqnarray}

Then we have the following inverse canonical transformation

\begin{eqnarray}
{\vec \rho}_q&=&\rho_q \hat R(\alpha ,\beta ,S_q,S^3_q),\nonumber \\
{\vec \pi}_q&=&{\tilde \pi}_q \hat R(\alpha ,\beta
,S_q,S^3_q)-{{S_q}\over {\rho_q}} \hat R(\alpha ,\beta
,S_q,S^3_q)\times {\hat S}_q(\beta ,S_q,S^3_q),\nonumber \\
&&{}\nonumber \\
\Rightarrow&& {\vec \pi}_q^2={\tilde \pi}_q^2+{{S_q^2}\over {\rho_q^2}},\nonumber \\
 &&{}\nonumber \\
 &&{\hat S}_q \times \hat R(\alpha ) = {{\partial \hat R(\alpha )}\over
 {\partial \alpha}}= \hat R(\alpha + {{\pi}\over 2}),\nonumber \\
 &&{}\nonumber \\
 \Rightarrow && \alpha =- tg^{-1}\, {{ ({\hat S}_q \times \hat R)^3}\over
 {[{\hat S}_q \times ( {\hat S}_q \times \hat R)]^3}}.
\label{VII8}
\end{eqnarray}

From the last line of this equation we see that the angle $\alpha$ can
be expressed in terms of ${\hat S}_q$ and $\hat R$.

The conjugate variables $\rho_q$, ${\tilde \pi}_q$ can be called {\it
dynamical shape variables} : they describe the vibration of the
dipole.

Then the rest-frame Hamiltonian for the relative motion becomes

\begin{eqnarray}
H_{(rel)}&=& H_M(\infty)= \sqrt{m_1^2+{\vec
\pi}_q^2}+ \sqrt{m_2^2+{\vec \pi}_q^2}=\nonumber \\
&=& \sqrt{m_1^2+{1\over {\check I}}(S_q)^2+{\tilde
\pi}_q^2}+  \sqrt{m_2^2+{1\over {\check I}}(S_q)^2+{\tilde \pi}_q^2},\nonumber \\
 &&{}\nonumber \\
 \Rightarrow&& \check \omega \, {\buildrel \circ \over =}\, {{\partial
 H_{(rel)}}\over {\partial S_q}}= {{S_q}\over {\check I}} \Big( {1\over
 {\sqrt{m_1^2+{1\over {\check I}}(S_q)^2+{\tilde \pi}_q^2}}}+ {1\over
 {\sqrt{m_2^2+{1\over {\check I}}(S_q)^2+{\tilde \pi}_q^2}}}\Big),\nonumber \\
 {\dot \rho}_q\, &{\buildrel \circ \over =}\,& {{\partial H_{(rel)}}\over
 {\partial {\tilde \pi}_q}}=  {\tilde \pi}_q \Big( {1\over
 {\sqrt{m_1^2+{1\over {\check I}}(S_q)^2+{\tilde \pi}_q^2}}}+ {1\over
 {\sqrt{m_2^2+{1\over {\check I}}(S_q)^2+{\tilde \pi}_q^2}}}\Big),\nonumber \\
 &&{}\nonumber \\
 {\tilde \pi}_q {|}_{{\dot \rho}_q=0} =0,\nonumber \\
 H^{(S)}_{(rel)}&=& H_{(rel)}{|}_{{\dot \rho}_q=0} =
  \sqrt{m_1^2+{1\over {\check I}}(S_q)^2}+  \sqrt{m_2^2+{1\over {\check I}}(S_q)^2},
 \nonumber \\
 H^{(S=0)}_{(rel)}&=& H_{(rel)}{|}_{S_q=0} =
  \sqrt{m_1^2+{\tilde \pi}_q^2}+  \sqrt{m_2^2+{\tilde \pi}_q^2}.
\label{VII9}
\end{eqnarray}

\noindent where $\check I=\rho_q^2={\check I}_{nr}/\mu$ is the
non-relativistic baricentric inertia tensor ${\check I}_{nr}$ of the
dipole divided by the reduced mass $\mu = {{m_1m_2}\over {m_1+m_2}}$.
The quantities $H^{(S)}_{(rel)}$ and $H^{(S=0)}_{(rel)}$ are the
purely rotational and purely vibrational Hamiltonians, respectively.

For equal masses we get formally the same results of the previous
Section but in a different canonical basis, if we make the
identifications $\pi ={\tilde \pi}_q$, $\rho
=\rho_q$ and ${\hat  R}^r=R^{rs}(\theta ,\phi ) {\hat \rho}^s_o$.
Only after a non-point transformation $\alpha$, $S_q$, $\beta$,
$S^3_q$ $\mapsto$ $\theta$, $\pi_{\theta}$, $\phi$, $\pi_{\phi}$ [i.e.
from Eq.(\ref{VII7}) to Eq.(\ref{VII4})], Eqs.(\ref{VII11}) become
Eqs.(\ref{VI13}).

\subsection{3-Body Systems.}

In the case N=3 the range of the indices is $i=1,2,3$, $a=1,2$. The
spin is ${\vec S}_q=\sum_{a=1}^2{\vec \rho}_{qa}\times {\vec
\pi}_{qa}=\sum_{a=1}^2 {\vec S}_{qa}$ after the canonical
transformation which separates the internal center of mass

\begin{equation}
\begin{minipage}[t]{1cm}
\begin{tabular}{|l|} \hline
${\vec \eta}_i$ \\    \hline
 ${\vec \kappa}_i$ \\ \hline
\end{tabular}
\end{minipage} \ {\longrightarrow \hspace{.2cm}} \
\begin{minipage}[t]{2 cm}
\begin{tabular}{|l|l|} \hline
${\vec q}_{+}$   & ${\vec \rho}_{qa}$  \\ \hline
 ${\vec \kappa}_{+}$ & ${\vec \pi}_{qa}$ \\ \hline
\end{tabular}
\end{minipage}
\label{VII10}
\end{equation}

After the gauge fixings ${\vec q}_{+}\approx 0$, the relative motions
in the rest-frame instant form are governed by the Hamiltonian

\begin{equation}
H_{(rel)}=H_M(\infty )=
\sum_{i=1}^3\sqrt{m_i^2+3\sum_{a,b=1}^2
\gamma_{ai}\gamma_{bi} {\vec \pi}_{qa}\cdot {\vec \pi}_{qb}}\equiv M_{sys}.
\label{VII11}
\end{equation}

We shall assume ${\vec S}_q \not= 0$, because the exceptional SO(3)
orbit $S_q=0$ has to be studied separately. This is done by adding
$S_q \approx 0$ as a first class constraint and studying separately
the following three disjoint strata (they have a different number of
first class constraints): a) ${\vec S}_q\approx 0$, but ${\vec
S}_{q1}=-{\vec S}_{q2}\not= 0$; b) ${\vec S}_{qa}\approx 0$, $a=1,2$
(in this case we have ${\vec \rho}_{qa}- k_a {\vec \pi}_{qa} \approx
0$).

For each value of $a=1,2$, we can consider the canonical
transformation (\ref{VII7})
\begin{equation}
\begin{minipage}[t]{1cm}
\begin{tabular}{|l|} \hline
${\vec \rho}_{qa}$ \\ \hline ${\vec \pi}_{qa}$ \\ \hline
\end{tabular}
\end{minipage} \ {\longrightarrow \hspace{.2cm}} \
\begin{minipage}[t]{2 cm}
\begin{tabular}{|ll|l|} \hline
$\alpha_a$ & $\beta_a$ & $\rho_{qa}$\\ \hline
 $S_{qa}$   & $S^3_{qa}$ & ${\tilde \pi}_{qa}$  \\ \hline
\end{tabular}
\end{minipage}
\label{VII12}
\end{equation}

\noindent where

\begin{eqnarray}
\alpha_a &=& tg^{-1} {1\over {S_{qa}}} \Big( {\vec \rho}_{qa}\cdot {\vec \pi}
_{qa}-{{(\rho_{qa})^2}\over {\rho^3_{qa}}} \pi^3_{qa}\Big) ,\nonumber \\
&&{}\nonumber \\
\beta_a &=& tg^{-1} {{S^2_{qa}}\over {S^1_{qa}}},\quad\quad sin\, \beta_a =
{{S^2_{qa}}\over {\sqrt{ (S_{qa})^2-(S^3_{qa})^2}}},\quad\quad cos\,
\beta_a ={{S^1_{qa}}\over {\sqrt{ (S_{qa})^2-(S^3_{qa})^2}}}.
\label{VII13}
\end{eqnarray}

\begin{eqnarray}
{\vec \rho}_{qa}&=& \rho_{qa} {\hat R}_a,\quad\quad \rho_{qa}=\sqrt{{\vec \rho}
_{qa}^2},\quad\quad {\hat R}_a={{{\vec \rho}_{qa}}\over {\rho_{qa}}}={\hat \rho}_{qa},
\quad\quad {\hat R}_a^2=1,\nonumber \\
 &&{}\nonumber \\
 {\vec \pi}_{qa}&=&
{\tilde \pi}_{qa} {\hat R}_a -{{S_{qa}}\over {\rho_{qa}}} {\hat
R}_a\times {\hat S}_{qa},\quad\quad {\tilde
\pi}_{qa}= {\vec \pi}_{qa}\cdot {\hat R}_a.
\label{VII14}
\end{eqnarray}

\begin{eqnarray}
{\vec \rho}_{qa}&=&\rho_{qa} {\hat \rho}_{qa}(\alpha_a ,\beta_a
,S_{qa},S^3_{qa}),\nonumber
\\
{\vec \pi}_{qa}&=&{\tilde \pi}_{qa} {\hat \rho}_{qa}(\alpha_a ,\beta_a
,S_{qa},S^3_{qa})-{{S_{qa}}\over {\rho_{qa}}} {\hat
\rho}_{qa}(\alpha_a ,\beta_a ,S_{qa},S^3_{qa})\times {\hat
S}_{qa}(\beta_a ,S_{qa},S^3_{qa})=\nonumber \\
 &=& {\tilde \pi}_{qa} {\hat R}_a(\alpha_a ,\beta_a
,S_{qa},S^3_{qa})-{{S_{qa}}\over {\rho_{qa}}} {\hat R}_a(\alpha_a
,\beta_a ,S_{qa},S^3_{qa})\times {\hat S}_{qa}(\beta_a
,S_{qa},S^3_{qa}).
\label{VII15}
\end{eqnarray}

We have now {\it two} unit vectors ${\hat R}_a$ and {\it two} E(3)
realizations generated by ${\vec S}_{qa}$, ${\hat R}_a$ respectively
and fixed invariants ${\hat R}^2_a=1$, ${\vec S}_{qa}\cdot {\hat
R}_a=0$ (non-irreducible, type 2\cite{lucenti}).

In order to implement a SO(3) {\it Hamiltonian right action} in
analogy with  the rigid body theory, we must construct an orthonormal
triad or {\it body frame} $\hat N$, $\hat
\chi$, $\hat N\times \hat \chi$. The decomposition
$\vec S = {\check S}^1\hat \chi +{\check S}^2 \hat N\times \hat \chi +
{\check S}^3 \hat N \,\,\, {\buildrel {def}\over =}\,\,\, {\check S}^r
{\hat e}_r$, identifies the SO(3) scalar generators ${\check S}^r$ of
the right action provided they satisfy $\{ {\check S}^r,{\check S}^s
\} =-\epsilon^{rsu} {\check S}^u$. This latter condition
together with the obvious requirement that $\hat N$, $\hat \chi$,
$\hat N\times \hat \chi$ be SO(3) vectors [$\{ {\hat N}^r,S^s
\} =\epsilon^{rsu}{\hat N}^u$, $\{ {\hat \chi}^r,S^s \}
=\epsilon^{rsu}{\hat \chi}^u$, $\{ {\hat N\times \hat \chi}^r,S^s \}
=\epsilon^{rsu}{\hat N\times \hat \chi}^u$] entails the equations
\footnote{With ${\check S}^r=\vec
S\cdot {\hat e}_r$, the conditions $\{ {\check S}^r,{\check S}^s \} =
-\epsilon^{rsu} {\check S}^u$ imply the equations $\vec S\cdot {\hat
e}_r\times {\hat e}_s +S^iS^j \{ {\hat e}^i_r,{\hat e}^j_s\} =
\epsilon_{rsu} S^k{\hat e}_k^u$, hence the quoted result.}
$\{ {\hat N}^r,{\hat N}^s \} = \{ {\hat N}^r,{\hat \chi}^s \} = \{
{\hat \chi}^r,{\hat \chi}^s \} =0$.

To each solution of these equations is associated a couple of
canonical realizations of the E(3) group (type 2, non-irreducible):
one with generators $\vec S$, $\vec N$ and non-fixed invariants
${\check S}^3=\vec S \cdot \hat N$ and $|{\vec N}|$; another with
generators $\vec S$, $\vec \chi$ and non-fixed invariants ${\check
S}^1=\vec S\cdot \hat \chi$ and $|{\vec \chi}|$. These latter contain
the relevant information for constructing the angle $\alpha$ and the
new canonical pair ${\check S}^3$, $\gamma =tg^{-1}{{{\check
S}^2}\over {{\check S}^1}}$ of SO(3) scalars. Since $\{  \alpha ,
{\check S}^3 \} = \{ \alpha , \gamma \} =0$ must hold, it follows that
the vector $\vec N$ necessarily belongs to the $\vec S$-$\hat  R$
plane. The three canonical pairs $S$, $\alpha$, $S^3$, $\beta$,
${\check S}^3$, $\gamma$ will describe the {\it orientational}
variables of our Darboux basis, while $|\vec N|$ and $|\vec \chi |$
will belong to the {\it shape} variables. Alternatively, an
anholonomic basis can be constructed by replacing the above six
variables by ${\check S}^r$ and three uniquely determined Euler angles
$\tilde \alpha$, $\tilde \beta$, $\tilde \gamma$.

In the N=3 case it turns out that a solution of the previous equation
corresponding to a {\it body frame} determined by the 3-body system
configuration only, as in the {\it rigid body} case, is completely
individuated  once two orthonormal vectors $\vec N$ and $\vec \chi$,
functions of the relative coordinates and independent of the momenta,
are found such that $\vec N$ lies in the $\vec S$ - $\hat R$
plane\footnote{Let us remark that any pair of orthonormal vectors
$\vec N$, $\vec \chi$ function only of the relative coordinates can be
used to build a body frame. This freedom is connected to the
possibility of redefining a body frame by using a
configuration-dependent arbitrary rotation, which leaves $\vec N$ in
the $\vec S$-$\hat R$ plane.}. We do not known whether in the case N=3
other solutions of the previous equations exist leading to momentum
dependent body frames\footnote{Anyway, our constructive method
necessarily leads to momentum-dependent solutions of the previous
equations for $N\geq 4$ and therefore to momentum-dependent or {\it
dynamical body frames}. See the following Subsection C.}.

The {\it simplest choice} for the orthonormal vectors $\vec N$ and
$\vec \chi$ functions only of the coordinates is

\begin{eqnarray}
\vec N&=& {1\over 2} ({\hat R}_1+{\hat R}_2)=
{1\over 2} ({\hat \rho}_{q1}+{\hat \rho}_{q2}),\quad\quad \hat
N={{\vec N}\over {|\vec N|}},\qquad |\vec N|=\sqrt{ {{1+{\hat \rho}_{q1}\cdot {\hat
\rho}_{q2}}\over 2} },\nonumber \\
\vec \chi &=&{1\over 2}({\hat R}_1-{\hat R}_2)=
  {1\over 2}({\hat \rho}_{q1}-{\hat \rho}_{q2})
,\quad\quad \hat \chi ={{\vec \chi}\over {|\vec \chi |}},\qquad
  |\vec \chi |=\sqrt{ { {1-{\hat \rho}_{q1}\cdot {\hat
\rho}_{q2} }\over 2} }=\sqrt{1-{\vec N}^2},\nonumber \\
 &&{}\nonumber \\
  \vec N\times \vec \chi &=&-{1\over 2}{\hat
\rho}_{q1}\times {\hat \rho}_{q2},\quad \quad |\vec N\times \vec \chi |=
|\vec N| |\vec \chi |={1\over 2}\sqrt{1-({\hat \rho}_{q1}\cdot {\hat
\rho}_{q2})^2},\nonumber \\
 &&{}\nonumber \\
 &&\vec N\cdot \vec \chi = 0,\qquad \{ N^r,N^s \}=
 \{ \chi^r,\chi^s \} = \{ N^r, \chi^s \} =0,\nonumber \\
 &&{}\nonumber \\
 {\hat R}_1&=&{\hat \rho}_{q1}= \vec N+\vec \chi ,\quad\quad {\hat R}_2=
 {\hat \rho}_{q2}=\vec N-\vec \chi ,\quad\quad {\hat R}_1\cdot {\vec R}_2=
 {\hat \rho}_{q1}\cdot {\hat \rho}_{q2}={\vec N}^2-{\vec \chi}^2.
\label{VII16}
\end{eqnarray}

Likewise, we have for the spins

\begin{eqnarray}
 {\vec S}_q&=& {\vec S}_{q1}+{\vec S}_{q2},\nonumber \\
 {\vec W}_q&=&{\vec S}_{q1}-{\vec S}_{q2},\nonumber \\
 &&{}\nonumber \\
 {\vec S}_{q1}&=&{1\over 2} ({\vec S}_q+{\vec W}_q),\quad\quad {\vec
S}_{q2}= {1\over 2} ({\vec S}_q-{\vec W}_q),\nonumber \\
 &&{}\nonumber \\
 &&\{ W^r_q,W^s_q \} = \epsilon^{rsu} S^u_q.
\label{VII17}
\end{eqnarray}

We  therefore succeeded in constructing an orthonormal triad (the {\it
dynamical body frame}) and two E(3) realizations (non-irreducible,
type 3): one with generators ${\vec S}_q$, $\vec N$ and non-fixed
invariants $|{\vec N}|$ and $\vec S\cdot
\hat N$, the other with generators ${\vec S}_q$ and $\vec \chi$ and
non-fixed invariants $|{\vec \chi}|$ and ${\vec S}_q\cdot {\hat
\chi}$. As said in Ref.\cite{iten2} this is
equivalent to the determination of the
non-conserved generators ${\check S}_q^r$ of a  Hamiltonian {\it right
action} of SO(3): ${\check S}^1_q={\vec S}_q\cdot \hat \chi = {\vec
S}_q\cdot {\hat e}_1$, ${\check S}^2_q={\vec S}_q\cdot \hat N\times
\hat \chi ={\vec S}_q\cdot {\hat e}_2$, ${\check S}^3_q={\vec S}_q\cdot \hat N
={\vec S}_q\cdot {\hat e}_3$.

The realization of the E(3) group with generators ${\vec S}_q$, $\vec
N$ and non-fixed invariants ${\cal I}_1={\vec N}^2$, ${\cal I}_2={\vec
S}_q\cdot \vec N$ leads to the final canonical transformation

\begin{equation}
\begin{minipage}[t]{1cm}
\begin{tabular}{|l|} \hline
${\vec \rho}_{qa}$ \\ \hline
 ${\vec \pi}_{qa}$ \\ \hline
\end{tabular}
\end{minipage}\ {\longrightarrow \hspace{.2cm}} \
\begin{minipage}[t]{4cm}
\begin{tabular}{|ll|ll|l|} \hline
$\alpha_1$ & $\beta_1$ & $\alpha_2$ & $\beta_2$ & $\rho_{qa}$\\ \hline
 $S_{q1}$   & $S^3_{q1}$ & $S_{q2}$ & $S^3_{q2}$ & ${\tilde \pi}_{qa}$\\ \hline
\end{tabular}
\end{minipage}
\ {\longrightarrow \hspace{.2cm}}\
\begin{minipage}[b]{4cm}
\begin{tabular}{|lll|l|l|} \hline
$\alpha$ & $\beta$ & $\gamma$ & $|\vec N|$ & $\rho_{qa}$ \\ \hline
 $S_q={\check S}_q$   & $S^3_q$& ${\check S}_q^3={\vec S}_q\cdot \hat
N$ & $\xi$ & ${\tilde \pi}_{qa}$ \\ \hline
\end{tabular}
\end{minipage}
\label{VII18}
\end{equation}

\noindent where

\begin{eqnarray}
|\vec N|&=& \sqrt{ {{1+{\hat \rho}_{q1}\cdot {\hat \rho}_{q2}}\over 2}
},\nonumber \\
 {\check S}^3_q&=&{\vec S}_q\cdot \hat N={1\over {\sqrt{2}}} \sum_{a=1}^2 {\vec \rho}_{qa}
 \times {\vec \pi}_{qa} \cdot {{ {\hat \rho}_{q1}+{\hat \rho}_{q2}}\over
 { \sqrt{1+{\hat\rho}_{q1}\cdot {\hat \rho}_{q2}} }}\equiv S_q cos\, \psi ,
 \nonumber \\
  && cos\, \psi = {\hat S}_q\cdot \hat N ={{{\check S}^3_q}\over {S_q}},
  \quad\quad sin\, \psi = {1\over {S_q}}
  \sqrt{(S_q)^2-({\check S}^3_q)^2},\nonumber \\
  S_q&=&{\check S}_q= |\sum_{a=1}^2 {\vec \rho}_{qa}\times {\vec \pi}_{qa}|,\nonumber \\
  S^3_q&=&\sum_{a=1}^2 ({\vec \rho}_{qa}\times {\vec \pi}_{qa})^3,\nonumber \\
  &&{}\nonumber \\
\alpha &=& -tg^{-1}\, {{ ({\hat S}_q\times \hat N)^3}\over
{[{\hat S}_q\times ({\hat S}_q\times \hat N)]^3}}=\nonumber \\
  &=&-tg^{-1}\, {{ [{\hat S}_q\times ({\hat \rho}_{q1}+{\hat \rho}_{q2})]^3}\over
{[{\hat S}_q\times ({\hat S}_q\times [{\hat \rho}_{q1}+{\hat
\rho}_{q2}])]^3}},\nonumber \\
\beta &=& tg^{-1}\, {{S^2_q}\over {S^1_q}},\nonumber \\
\gamma &=& tg^{-1}\, {{{\vec S}_q\cdot (\hat N\times \hat \chi )}\over
{{\vec S}_q\cdot \hat \chi}}= tg^{-1}\, {{{\check S}_q^2}\over
{{\check S}_q^1}},\nonumber \\
  \Rightarrow && sin\, \gamma ={{ {\check S}_q^2}\over
  { \sqrt{({\check S}_q)^2-({\check S}_q^3)^2}}},\quad\quad
  cos\, \gamma ={{{\check S}_q^1}\over { \sqrt{({\check S}_q)^2-({\check S}_q^3)^2}}},
  \nonumber \\
 &=& tg^{-1}\, {{\sqrt{2} {\vec S}_q\cdot {\hat \rho}_{q2}\times {\hat \rho}_{q1}}\over
 { \sqrt{1+{\hat \rho}_{q1}\cdot {\hat \rho}_{q2}}\, {\vec S}_q\cdot
 ({\hat \rho}_{q1}-{\hat \rho}_{q2})}},\nonumber \\
\xi &=& {{{\vec W}_q\cdot (\hat N\times \hat \chi )}\over {|\vec \chi |}}=
{{{\vec W}_q\cdot (\hat N\times \hat \chi )}\over {\sqrt{1-{\vec
N}^2}}}\nonumber \\
 &=&{{ \sqrt{2} \sum_{a=1}^2(-)^{a+1}{\vec
\rho}_{qa}\times {\vec
\pi}_{qa}\cdot ({\hat \rho}_{q2}\times {\hat \rho}_{q1})}\over
{[1-{\hat \rho}_{q1}\cdot {\hat \rho}_{q2}]\sqrt{1+{\hat \rho}_{q1}
\cdot {\hat \rho}_{q2}} }}.
\label{VII19}
\end{eqnarray}

For N=3 the {\it dynamical shape variables}, functions  of the
relative coordinates ${\vec \rho}_{qa}$ only, are $|\vec N|$, ${\hat
\rho}_{qa}$, while the conjugate shape momenta are $\xi$, ${\tilde \pi}_{qa}$.

The final array (\ref{VII18}) is nothing else than a {\it scheme
B}\cite{pauri2} of a realization of an E(3) group with generators
${\vec S}_q$, $\vec N$ (non-irreducible, type 3). In particular, the
two canonical pairs $S^3_q$, $\beta$, $S_q$, $\alpha$, constitute the
irreducible kernel of the E(3) {\it scheme A}, whose invariants are
${\check S}^3_q$, $|\vec N|$; $\gamma$ and $\xi$ are the so-called
{\it supplementary variables} conjugated to the invariants; finally,
the two pairs $\rho_{qa}$, ${\tilde \pi}_{qa}$ are so-called {\it
inessential variables}. Let us remark that $S^3_q$, $\beta$, $S_q$,
$\alpha$, $\gamma$, $\xi$, are a local coordinatization of every E(3)
coadjoint orbit with ${\check S}^3_q=const.$, $|\vec N| =const.$ and
fixed values of the inessential variables, present in the 3-body phase
space.

We can now reconstruct ${\vec S}_q$ and define a {\it new} unit vector
$\hat R$ orthogonal to ${\vec S}_q$ by adopting the prescription of
Eq.(\ref{VII10}) as follows

\begin{eqnarray}
{\hat S}^1_q&=&{1\over {S_q}} \sqrt{(S_q)^2-(S^3_q)^2} cos\, \beta
,\nonumber \\
 {\hat S}^2_q&=&{1\over {S_q}} \sqrt{(S_q)^2-(S^3_q)^2}
sin\, \beta ,\nonumber \\
 {\hat S}^3_q&=& {{S^3_q}\over {S_q}},\nonumber \\
 &&{}\nonumber \\
 {\hat R}^1&=&sin\, \beta sin\, \alpha - {{S^3_q}\over {S_q}} cos\,
 \beta cos\, \alpha ,\nonumber \\
 {\hat R}^2&=&-cos\, \beta sin\, \alpha - {{S^3_q}\over {S_q}} sin\,
 \beta cos\, \alpha ,\nonumber \\
 {\hat R}^3&=&{1\over {S_q}} \sqrt{(S_q)^2-(S^3_q)^2} cos\, \alpha ,\nonumber \\
 &&{}\nonumber \\
 && {\hat R}^2=1,\quad\quad {\hat R}\cdot {\vec S}_q=0,\quad\quad \{
{\hat R}^r,{\hat R}^s \} =0,\nonumber \\
 &&{}\nonumber \\
 ({\hat S}_q\times \hat R)^1&=&{\hat S}^2_q{\hat R}^3-{\hat S}^3_q{\hat R}^2=
 sin\, \beta cos\, \alpha +{{S^3_q}\over {S_q}} cos\, \beta sin\, \alpha ,
 \nonumber \\
({\hat S}_q\times \hat R)^2&=&{\hat S}^3_q{\hat R}^1-{\hat S}^1_q{\hat
R}^3=-cos\, \beta cos\, \alpha +{{S^3_q}\over {S_q}} sin\, \beta sin\,
\alpha ,\nonumber \\
 ({\hat S}_q\times \hat R)^3&=&-{\hat S}^1_q{\hat R}^2-{\hat S}^2_q{\hat R}^1=
 {1\over {S_q}} \sqrt{(S_q)^2-(S^3_q)^2} sin\, \alpha ,
\label{VII20}
\end{eqnarray}

The vectors  ${\hat S}_q$, $\hat R$, ${\hat S}_q\times \hat R$ build
up the {\it spin frame} for N=3. The angle $\alpha$ conjugate to $S_q$
is explicitly given by\footnote{ The two expressions of $\alpha$ given
above are consistent with the fact that ${\hat S}_q$, $\hat R$ and
$\hat N$ are coplanar, so that $\hat R$ and $\hat N$ differ by a term
in ${\hat S}_q$.}

\beq
\alpha = -tg^{-1}\, {{ ({\hat S}_q\times \hat N)^3}\over {[{\hat
S}_q\times ({\hat S}_q\times \hat N)]^3}}=- tg^{-1}\, {{({\hat S}_q
\times \hat R)^3}\over {[{\hat S}_q \times ({\hat S}_q
\times \hat R)]^3}}.
\label{VII21}
\eeq

{\it As a consequence of this definition of} $\hat R$, we get the
following expressions for the {\it dynamical body frame} $\hat N$,
$\hat \chi$, $\hat N\times \hat \chi$  in terms of the final canonical
variables

\begin{eqnarray}
 \hat N&=& cos\, \psi {\hat S}_q+sin\, \psi \hat R={{{\check S}^3_q}\over {S_q}}
 {\hat S}_q+{1\over {S_q}}\sqrt{(S_q)^2-({\check S}^3_q)^2}\hat R=\nonumber \\
 &=&\hat N [S_q,\alpha ;S^3_q,\beta ;{\check S}^3_q,\gamma ],\nonumber \\
 &&{}\nonumber \\
 \hat \chi &=&sin\, \psi cos\, \gamma {\hat S}_q-cos\, \psi cos\, \gamma \hat R +sin\,
 \gamma {\hat S}_q\times \hat R=\nonumber \\
  &=& {1\over {S_q}}\sqrt{(S_q)^2-({\check S}^3_q)^2} cos\, \gamma {\hat S}_q-
  {{{\check S}^3_q}\over {S_q}}cos\, \gamma \hat R +sin\, \gamma
  {\hat S}_q\times \hat R=\nonumber \\
  &=& {{{\check S}^1_q}\over {S_q}}\, {\hat S}_q- {{{\check S}^3_q}\over {S_q}}
  {{ {\check S}_q^1\, \hat R + {\check S}_q^2\,  {\hat S}_q\times \hat R}\over
{ \sqrt{({ S}_q)^2-({\check S}_q^3)^2}}}=\nonumber \\
  &=&\hat \chi [S_q,\alpha ;S^3_q,\beta ;{\check S}^3_q,\gamma ],\nonumber \\
 &&{}\nonumber \\
 \hat N\times \hat \chi &=&
sin\, \psi sin\, \gamma {\hat S}_q-cos\, \psi sin\, \gamma \hat R
 -cos\, \gamma  {\hat S}_q\times \hat R=\nonumber \\
  &=& {1\over {S_q}}\sqrt{(S_q)^2-({\check S}^3_q)^2} sin\, \gamma {\hat S}_q-
  {{{\check S}^3_q}\over {S_q}}sin\, \gamma \hat R -cos\, \gamma
  {\hat S}_q\times \hat R=\nonumber \\
 &=& {{{\check S}^2_q}\over {S_q}}\, {\hat S}_q- {{{\check S}^3_q}\over {S_q}}
  {{ {\check S}_q^1\, \hat R - {\check S}_q^2\,  {\hat S}_q\times \hat R}\over
{ \sqrt{({S}_q)^2-({\check S}_q^3)^2}}}=\nonumber \\
  &=&(\hat N\times \hat \chi ) [S_q,\alpha ;S^3_q,\beta ;{\check S}^3_q,\gamma ],
  \nonumber \\
  &&\Downarrow\nonumber \\
  {\hat S}_q &=& sin\, \psi cos\, \gamma \hat \chi +sin\, \psi sin\, \gamma
  \hat N \times \hat \chi + cos\, \psi \hat N\nonumber \\
  &{\buildrel {def} \over =}& {1\over {S_q}} \Big[ {\check S}^1_q \hat \chi +{\check S}^2_q
  \hat N \times \hat \chi +{\check S}^3_q \hat N\Big],\nonumber \\
  &&{}\nonumber \\
  \hat R &=& -cos\, \psi cos\, \gamma \hat \chi -cos\, \psi sin\, \gamma
  \hat N \times \hat \chi +sin\, \psi \hat N,\nonumber \\
  &&{}\nonumber \\
  \hat R \times {\hat S}_q &=& -sin\, \gamma \hat \chi +
  cos\, \gamma \hat N \times \hat \chi .
\label{VII22}
\end{eqnarray}

While $\psi$ is the angle between ${\hat S}_q$ and $\hat N$, $\gamma$
is the angle between the plane $\hat N - \hat \chi$ and the plane
${\hat S}_q - \hat N$. As in the case N=2, $\alpha$ is the angle
between the plane ${\hat S}_q - {\hat f}_3$ and the plane ${\hat S}_q
- \hat R$, while $\beta$ is the angle between the plane ${\hat S}_q
- {\hat f}_3$ and the plane ${\hat f}_3- {\hat f}_1$.

Owing to the results of Appendix C of Ref.\cite{iten2}, which allow to
reexpress $S_{qa}=|{\vec S}_{qa}|$, $S^3_{qa}$, $\beta_a=tg^{-1}
{{S^2_{qa}}\over {S^1_{qa}}}$ and $\alpha_a =-tg^{-1} {{({\hat
S}_{qa}\times {\hat R}_a )^3}\over {({\hat S}_{qa}\times ({\hat
S}_{qa}\times {\hat R}_a))^3}}$ in terms of the final variables, we
can reconstruct the inverse canonical transformation.

The existence  of the {\it spin frame} and of the {\it dynamical body
frame} allows to define two decompositions of the relative variables,
which make explicit the inverse canonical transformation. For the
relative coordinates we get from Eqs. (\ref{VII16}) and Appendix C of
I

\begin{eqnarray}
{\vec \rho}_{qa}&=&\rho_{qa}{\hat R}_a=\rho_{qa}[\vec N+(-)^{a+1} \vec
\chi ]= \rho_{qa}[|\vec N|\hat N+(-)^{a+1}\sqrt{1-{\vec N}^2}\hat \chi ]=
\nonumber \\
 &=& [{\vec \rho}_{qa}\cdot {\hat  S}_q] {\hat S}_q+[{\vec
\rho}_{qa}\cdot \hat R] \hat R +[{\vec \rho}_{qa}\cdot  {\hat S}_q\times \hat R]
 {\hat S}_q\times \hat R =\nonumber \\
 &=&{{\rho_{qa}}\over {S_q}} \Big[ \Big( |\vec N| \, {\check S}^3_q+(-)^{a+1}
 \sqrt{1-{\vec N}^2}\, {\check S}^1_q\Big) {\hat S}_q+\nonumber \\
 &+&\Big( |\vec N| \sqrt{(S_q)^2-({\check S}^3_q)^2}-(-)^{a+1} \sqrt{1-{\vec N}^2}
 {{ {\check S}^1_q{\check S}^3_q}\over {\sqrt{(S_q)^2-({\check S}^3_q)^2}}}\Big) \hat R -
 \nonumber \\
 &-& (-)^{a+1} \sqrt{1-{\vec N}^2} {{{\check S}^2_q}\over {\sqrt{(S_q)^2-({\check S}^3_q)^2}}}
 {\hat S}_q\times \hat R \Big] =\nonumber \\
 &=&{\vec \rho}_{qa}[S_q,\alpha ;S^3_q,\beta ;{\check
S}^3_q,\gamma ; \rho_{qa},|\vec N|].
\label{VII23}
\end{eqnarray}

The results of Appendix C  of Ref.\cite{iten2} give the analogous
formulas for the relative momenta \footnote{See Appendix C of
Ref.\cite{iten2} for the expression of the body frame components of
${\vec \pi}_{qa}$.}

\begin{eqnarray}
 {\vec \pi}_{qa}&=& {\tilde \pi}_{qa} {\hat R}_a+{{S_{qa}}\over {\rho_{qa}}}
{\hat S}_{qa}\times {\hat R}_a=
 {\tilde \pi}_{qa} {\hat \rho}_{qa}+{{S_{qa}}\over {\rho_{qa}}}
{\hat S}_{qa}\times {\hat \rho}_a=\nonumber \\
 &=&[{\vec \pi}_{qa}\cdot \hat N] \hat N +[{\vec \pi}_{qa}\cdot \hat \chi ]
  \hat \chi +[{\vec \pi}_{qa}\cdot \hat N\times \hat \chi ] \hat N\times \hat \chi
  =\nonumber \\
  &&{}\nonumber \\
&=& [{\vec \pi}_{qa}\cdot {\hat S}_q] {\hat S}_q+[{\vec \pi}_{qa}\cdot
\hat R] \hat R +[{\vec \pi}_{qa}\cdot {\hat S}_q\times \hat R]
{\hat S}_q\times \hat R=\nonumber \\
 &=&{1\over {S_q}} \Big[ \Big( ({\vec \pi}_{qa}\cdot \hat N) {\check S}^3_q+({\vec \pi}_{qa}\cdot
 \hat \chi ) {\check S}^1_q +({\vec \pi}_{qa}\cdot \hat N\times \hat \chi )
{\check S}^2_q\Big) {\hat S}_q +\nonumber \\ &+&\Big( ({\vec
\pi}_{qa}\cdot
\hat N) \sqrt{(S_q)^2-({\check S}^3_q)^2} -[({\vec \pi}_{qa}\cdot \hat \chi +
({\vec \pi}_{qa}\cdot \hat N\times \hat \chi )] {{{\check
S}^1_q{\check S}^3_q}\over {\sqrt{(S_q)^2-({\check S}^3_q)^2}}} \Big)
\hat R +\nonumber \\
 &+&[({\vec \pi}_{qa}\cdot \hat \chi ) - ({\vec \pi}_{qa}\cdot \hat N\times
 \hat \chi )] {{{\check S}^2_q{\check S}^3_q}\over {\sqrt{(S_q)^2-({\check S}^3_q)^2}}}
 \hat R \times {\hat S}_q \Big] =\nonumber \\
  &=&{\vec \pi}_{qa}[S_q,\alpha ;S^3_q,\beta ;{\check S}^3_q,\gamma ;
 |\vec N|,\xi ; \rho_{qa},{\tilde \pi}_{qa}].
\label{VII24}
\end{eqnarray}

Finally, the results in Appendix D  allow to perform a sequence of a
canonical transformation to  Euler angles $\tilde \alpha$, $\tilde
\beta$, $\tilde \gamma$ with their conjugate momenta, followed by a
transition to the anholonomic basis used in the orientation-shape
bundle approach\cite{little}

\begin{eqnarray}
&&\begin{minipage}[t]{3cm}
\begin{tabular}{|lll|} \hline
$\alpha$ & $\beta$ & $\gamma$ \\ \hline
 $S_q={\check S}_q$   & $S^3_q$& ${\check S}_q^3$ \\ \hline
\end{tabular}
\end{minipage}
\ {\longrightarrow \hspace{.2cm}}\
\begin{minipage}[t]{3cm}
\begin{tabular}{|lll|} \hline
$\tilde \alpha$ & $\tilde \beta$ & $\tilde \gamma$ \\ \hline
 $p_{\tilde \alpha}$   & $p_{\tilde \beta}$& $p_{\tilde \gamma}$ \\ \hline
\end{tabular}
\end{minipage}
\ {{\buildrel {non\,\, can.} \over \longrightarrow} \hspace{.2cm}}\
\begin{minipage}[b]{3cm}
\begin{tabular}{|lll|} \hline
$\tilde \alpha$ & $\tilde \beta$ & $\tilde \gamma$ \\ \hline
 ${\check S}^1_q$ & ${\check S}^2_q$ & ${\check S}^3_q$  \\
 \hline
 \end{tabular}
 \end{minipage} \nonumber \\
 &&{}\nonumber \\
 S_q&=& {\check S}_q = \sqrt{ ({\check S}^1_q)^2+({\check S}^2_q)^2
 +({\check S}^3_q)^2},\nonumber \\
 S^3_q&=&-sin\, \tilde \beta cos\, \tilde \gamma {\check S}^1_q
 +sin\, \tilde \beta sin\, \tilde \gamma {\check S}^2_q +cos\, \tilde \beta
 {\check S}^3_q,\nonumber \\
 \alpha &=& arctg\, {{ p_{\tilde \beta} tg\, \tilde \beta}\over { {\check S}_q -
 {{p_{\tilde \alpha}p_{\tilde \gamma}}\over {{\check S}_q cos\, \tilde \beta}} }},
 \nonumber \\
 \gamma &=& {{\pi}\over 2} -\tilde \gamma - arctg\, {{ctg\, \tilde \beta p_{\tilde \gamma}-
 {{p_{\tilde \alpha}}\over {sin\, \tilde \beta}} }\over {p_{\tilde \beta}}},
 \nonumber \\
 \beta &=& \tilde \alpha + arctg\, {{ ctg\, \tilde \beta p_{\tilde \alpha} -
 {{p_{\tilde \gamma}}\over {sin\, \tilde \beta}} }\over {p_{\tilde \beta}}} -
 {{\pi}\over 2},
\label{VII25}
\end{eqnarray}

\noindent Here $p_{\tilde \alpha}$, $p_{\tilde \beta}$, $p_{\tilde \gamma}$
are the functions of $\tilde \alpha$, $\tilde \beta$, $\tilde \gamma$,
${\check S}^r_q$, given in Eqs.(\ref{d3}). Eqs.(\ref{d3}),
(\ref{VII25}), (\ref{VII22}) and ${\check S}^2_q={\vec S}_q\cdot \hat
N\times \hat \chi$ lead to the determination of the {\it dynamical
orientation variables} $\tilde \alpha$, $\tilde \beta$, $\tilde
\gamma$ in terms of ${\vec \rho}_{qa}$, ${\vec \pi}_{qa}$. Let us stress that, while in the
orientation-shape bundle approach  the orientation variables
$\theta^{\alpha}$ are gauge variables, the Euler angles $\tilde
\alpha$, $\tilde \beta$, $\tilde \gamma$ are {\it uniquely determined} in
terms of the original variables.

In conclusion the complete transition to the anholonomic basis used in
the {\it static} theory of the orientation-shape bundle is

\begin{equation}
\begin{minipage}[t]{5cm}
\begin{tabular}{|lll|l|l|} \hline
$\alpha$ & $\beta$ & $\gamma$& $|\vec N|$ & $\rho_{qa}$ \\ \hline
 $S_q={\check S}_q$   & $S^3_q$& ${\check S}_q^3$ & $\xi$ & ${\tilde \pi}_{qa}$ \\ \hline
\end{tabular}
\end{minipage}
\ {{\buildrel {non\,\, can.} \over \longrightarrow} \hspace{.2cm}}\
\begin{minipage}[b]{5cm}
\begin{tabular}{|lll|l|l|} \hline
$\tilde \alpha$ & $\tilde \beta$ & $\tilde \gamma$& $|\vec N|$ &
$\rho_{qa}$ \\ \hline
 ${\check S}^1_q$ & ${\check S}^2_q$ & ${\check S}^3_q$ & $\xi$ & ${\tilde \pi}_{qa}$ \\
 \hline
 \end{tabular}
 \end{minipage},
\label{VII26}
\end{equation}

\noindent with the 3 pairs of  conjugate canonical  dynamical shape variables:
$\rho_{qa}$, ${\tilde \pi}_{qa}$, $|\vec N|$, $\xi$.

Eqs.(\ref{VII23}), (\ref{VII26}), (\ref{VII21}) and (\ref{d2}) imply

\bea
\rho^r_{qa} &=& {\cal R}^r{}_s(\tilde \alpha ,\tilde \beta ,\tilde \gamma ) {\check
\rho}^s_{qa}(q),\quad\quad with\nonumber\\
&&{}\nonumber \\
 &&{\check \rho}^1_{qa}(q) = (-)^{a+1} \rho_{qa}
\sqrt{1-{\vec N}^2},\quad\quad {\check \rho}^2_{qa}(q)=0,\quad\quad
{\check \rho}^3_{qa}(q) = \rho_{qa} |\vec N|,\nonumber \\
 &&{}\nonumber \\
 &&and\nonumber \\
 &&{}\nonumber \\
 S^r_q &=&  {\cal R}^r{}_s(\tilde \alpha ,\tilde \beta ,\tilde \gamma ) {\check
 S}^s_q,
\label{VII27}
\eea

\noindent so that the final visualization of our sequence of transformations is

\beq
\begin{minipage}[t]{5cm}
\begin{tabular}{|l|} \hline
${\vec \rho}_{qa}$ \\ \hline
 ${\vec \pi}_{qa}$ \\ \hline
\end{tabular}
\end{minipage}
\ {{\buildrel {non\,\, can.} \over \longrightarrow} \hspace{.2cm}}\
\begin{minipage}[b]{5cm}
\begin{tabular}{|lll|l|} \hline
$\tilde \alpha$ & $\tilde \beta$ & $\tilde \gamma$& $q^{\mu}({\vec
\rho}_{qa})$
\\ \hline
 ${\check S}^1_q$ & ${\check S}^2_q$ & ${\check S}^3_q$ &
 $p_{\mu}({\vec \rho}_{qa},{\vec \pi}_{qa})$ \\ \hline
 \end{tabular}
 \end{minipage}.
 \label{VII28}
 \eeq

Note furthermore that  we get ${\check \rho}^2_{qa} = {\vec
\rho}_{qa}\cdot \hat N\times \hat \chi =0$ by construction and this entails that
using our {\it dynamical body frame} is equivalent to a convention
({\it xxzz gauge}) about the body frame of the type of {\it xxz} and
similar gauges quoted in Ref.\cite{little}.

Finally, we can give the  expression of the  Hamiltonian for the
relative motions \footnote{The   Hamiltonian in the basis
(\ref{VII18}) can be obtained with the following replacements ${\check
S}^1_q=\sqrt{(S_q)^2-({\check S}^3_q)^2} cos\,
\gamma$ and ${\check S}^1_q=\sqrt{(S_q)^2-({\check S}^3_q)^2} sin\,
\gamma$.} in terms of the anholonomic Darboux basis (\ref{VII25}).
By using Eqs. (\ref{e12}) and (\ref{e13}) we get

\begin{eqnarray}
H_{(rel)}&\equiv&M_{sys}= H_M(\infty
)=\sum_{i=1}^3\sqrt{m_i^2+3\sum_{a,b=1}^2
 \gamma_{ai}\gamma_{bi} {\vec \pi}_{qa}\cdot {\vec \pi}_{qb}}=\nonumber \\
 &=&  \sum_{i=1}^3 \Big( m_i^2+{3\over {{\vec
N}^2}}\Big[ {{(\gamma_{1i})^2}\over
{2\rho^2_{q1}}}+{{(\gamma_{2i})^2}\over
{2\rho^2_{q2}}}+{{\gamma_{1i}\gamma_{2i}}\over
{\rho_{q1}\rho_{q2}}}\Big] ({\check S}^1_q)^2+\nonumber \\
 &+& 3\Big[ {{(\gamma_{1i})^2}\over {2\rho^2_{q1}}}+{{(\gamma_{2i})^2}\over
{2\rho^2_{q2}}}+{{\gamma_{1i}\gamma_{2i}(2{\vec N}^2-1)}\over
{\rho_{q1}\rho_{q2}}}\Big] ({\check S}^2_q)^2+\nonumber \\
 &+&{3\over {1-{\vec N}^2}} \Big[
{{(\gamma_{1i})^2}\over {2\rho^2_{q1}}}+{{(\gamma_{2i})^2}\over
{2\rho^2_{q2}}}-{{\gamma_{1i}\gamma_{2i}}\over
{\rho_{q1}\rho_{q2}}}\Big] ({\check S}^3_q)^2+\nonumber \\
 &+&3\sqrt{1-{\vec N}^2} \Big[ \xi ({{(\gamma_{1i})^2}\over {2\rho^2_{q1}}}
 -{{(\gamma_{2i})^2}\over {2\rho^2_{q2}}} ) +4\gamma_{1i}\gamma_{2i}
  |\vec N| \sqrt{1-{\vec N}^2} ({{{\tilde \pi}_{q1}}\over {\rho_{q2}}}-{{{\tilde \pi}_{q2}}
  \over {\rho_{q1}}})\Big] {\check S}_q^2 -\nonumber \\
  &-& {3\over {|\vec N| \sqrt{1-{\vec N}^2}}}
  ({{(\gamma_{1i})^2}\over {\rho^2_{q1}}}-{{(\gamma_{2i})^2}\over
  {\rho^2_{q2}}}) {\check S}^1_q {\check S}^3_q+\nonumber \\
  &+&6(\gamma_{1i})^2 \Big[ {\tilde \pi}^2_{q1}+{{\xi^2 (1-{\vec N}^2)}\over {4\rho^2_{q1}}}\Big] +
  6(\gamma_{2i})^2 \Big[ {\tilde \pi}^2_{q2}+{{\xi^2 (1-{\vec N}^2)}\over {4\rho^2_{q2}}}\Big]
+\nonumber \\
 &+& 12\gamma_{1i}\gamma_{2i} \Big[ (2{\vec N}^2-1) {\tilde \pi}_{q1} {\tilde \pi}_{q2}
 -|\vec N| (1-{\vec N}^2) \xi ({{{\tilde \pi}_{q1}}\over {\rho_{q2}}}+{{{\tilde \pi}_{q2}}\over
 {\rho_{q1}}}) +\nonumber \\
  &+&{{\xi^2 (1-{\vec N}^2)(2{\vec N}^2-1)}\over {4\rho_{q1}\rho_{q2}}}\Big]
 \Big)^{1/2} =\nonumber \\
 &=& \sum_{i=1}^3 H_{(rel) i},
 \label{VII29}
 \end{eqnarray}

\noindent where $q^{\mu}= (\rho_{q1}, \rho_{q2}, |\vec N|)$,
$p_{\mu} = ({\tilde \pi}_{q1}, {\tilde \pi}_{q2}, \xi )$ are the
dynamical shape variables.

By using  the results of Appendix E,  Eq(\ref{e4}), we can put  the
Hamiltonian  in a form reminiscent of the non-relativistic
orientation-shape bundle approach \footnote{The N=3 non-relativistic
Hamiltonian is $H_{rel}={1\over 2}\Big[ ({\check {\cal
I}}^{-1}(q))^{rs} {\check S}^r_q{\check S}^s_q + {\tilde
g}^{\mu\nu}(q) (p_{\mu}-{\vec S}_q\cdot {\vec {\cal A}}_{\mu}(q))
(p_{\nu}-{\vec S}_q\cdot {\vec {\cal A}}_{\nu}(q))\Big]$, with the
quantities ${\check A}^r_{\mu}(q)$, ${\tilde g}^{\mu\nu}(q)$, ${\check
{\cal I}}^{(-1rs}(q)$ evaluated in Appendix E of Ref.\cite{iten2}.
While the purely rotational Hamiltonian (defined by ${\dot q}^{\mu}=0$
implying $p_{\mu}= {\vec S}_q\cdot {\vec {\cal A}}_{\mu}(q)$) is
$H^{(rot)}_{rel}={1\over 2} ({\check {\cal I}}^{-1}(q))^{rs} {\check
S}^r_q{\check S}^s_q$, the purely vibrational Hamiltonian
$H^{(vib)}_{rel}$ is defined in our approach by the requirement
${\check \omega}^r=0$. Since ${\check S}^r_q{|}_{{\check \omega}^s=0}
\not= 0$,  unlike in the static orientation-shape bundle approach we
have $H_{rel} \not= H^{(rot)}_{rel}+H^{(vib)}_{rel}$.}

\begin{eqnarray}
H_{(rel)} &=& \sum_{i=1}^3 H_{(rel) i}=\nonumber \\
 &=& \sum_{i=1}^3
 \sqrt{m_i^2+{\check {\cal T}}_i^{-1 rs}(q){\check S}_q^r{\check S}_q^s+
 {\tilde v}_i^{\mu\nu}(q)\Big( p_{\mu}-{\check {\vec S}}_q\cdot {\check {\vec A}}_{i\mu}(q)\Big)
 \Big( p_{\nu}-{\check {\vec S}}_q\cdot {\check {\vec A}}_{i\nu}(q)\Big)}.
 \label{VII30}
 \end{eqnarray}

A purely rotational (vertical) Hamiltonian is
$H^{(rot)}_{(rel)}=H_{(rel)}{|}_{\dot
q=0}=H_{(rel)}{|}_{p_{\mu}={\check {\vec S}}_q\cdot {\check {\vec
{\cal C}}}_{\mu}({\check {\vec S}}_q,q)}$: since under each square
root there is a different gauge potential ${\check {\vec {\cal
A}}}_{i\mu}(q)$ we get that under each square root there is the mass
term plus a quadratic expression in the body spin but with
coefficients depending on the shape variable and on ${\check {\vec
S}}_q$ itself, since $[p_{\mu}-{\check {\vec S}}_q\cdot {\check {\vec
{\cal A}}}_{i\mu}(q)]{|}_{\dot q=0}={\check {\vec S}}_q\cdot [{\check
{\vec {\cal C}}}_{\mu}({\check {\vec S}}_q,q)-{\check {\vec {\cal
A}}}_{i\mu}(q)]$. Therefore, we have

\begin{eqnarray}
H^{(rot)}_{(rel)}&=&H^{(rot)}_{(rel)}({\check {\vec S}}_q,q)=\nonumber
\\ &=&\sum_{i=1}^N\sqrt{m_i^2+\Big[ {\check {\cal T}}_i^{-1
rs}(q)+{\tilde v}_i^{\mu\nu}(q) \Big( {\check {\cal
C}}^r_{\mu}({\check {\vec S}}_q,q)-{\check {\cal A}}^r_{i\mu}(q)\Big)
\Big( {\check {\cal C}}^s_{\nu}({\check {\vec S}}_q,q)-{\check {\cal
A}}^s_{i\nu}(q)\Big)
\Big] {\check S}_q^r {\check S}_q^s}.\nonumber \\
 &&{}
\label{VII31}
\end{eqnarray}

In the non-relativistic limit (where $p_{\mu}={\check {\vec S}}_q\cdot
{\check {\vec {\cal A}}}_{\mu}(q)$ with ${\check {\vec {\cal
A}}}_{\mu}(q)$ the non-relativistic gauge potential) we get

\begin{eqnarray}
H^{(S)}_{(rel)}\, &\rightarrow_{c \rightarrow \infty}\,&
\sum_{i=1}^Nm_ic^2 + {1\over 2}\sum_{i=1}^N {1\over {m_i}}\Big( lim_{c\rightarrow \infty}
\Big[ {\check {\cal T}}_i^{-1 rs}(q)+\nonumber \\
 &+& {\tilde
v}_i^{\mu\nu}(q) \Big( {\check {\cal C}}^r_{\mu}({\check {\vec
S}}_q,q)-{\check {\cal A}}^r_{i\mu}(q)\Big) \Big( {\check {\cal
C}}^s_{\nu}({\check {\vec S}}_q,q)-{\check {\cal A}}^s_{i\nu}(q)\Big)
\Big] {\check S}_q^r {\check S}_q^s\Big) +O(1/c)=\nonumber \\
 &=&\sum_{i=1}^N m_ic^2 +{1\over 2} {\check {\cal I}}^{-1
rs}(q){\check S}_q^r{\check S}_q^s +O(1/c),
\label{VII32}
\end{eqnarray}

\noindent so that the non-relativistic inverse tensor of inertia is recovered
as

\begin{equation}
{\check {\cal I}}^{-1 rs}(q)=
 lim_{c\rightarrow \infty} \Big( \sum_{i=1}^N {1\over {m_i}}
\Big[ {\check {\cal T}}_i^{-1 rs}(q)+{\tilde
v}_i^{\mu\nu}(q) \Big( {\check {\cal C}}^r_{\mu}({\check {\vec
S}}_q,q)-{\check {\cal A}}^r_{i\mu}(q)\Big) \Big( {\check {\cal
C}}^s_{\nu}({\check {\vec S}}_q,q)-{\check {\cal A}}^s_{i\nu}(q)\Big)
\Big] \Big).
\label{VII33}
\end{equation}

These results together with

\begin{eqnarray}
 H_{(rel)}- \sum_{i=1}^3 m_i
 &\rightarrow_{c \rightarrow \infty}\,& {1\over 2}\sum_{i=1}^N\Big( { { {\check {\cal
 T}}_i^{-1 rs}(q)}\over {m_i}} {\check S}_q^r{\check S}_q^s+\nonumber \\
  &+& {{ {\tilde v}_i^{\mu\nu}(q)}\over
 {m_i}} \Big( p_{\mu}-{\check {\vec S}}_q\cdot {\check {\vec A}}_{i\mu}(q)\Big)
 \Big( p_{\nu}-{\check {\vec S}}_q\cdot {\check {\vec A}}_{i\nu}(q)\Big) \Big) =\nonumber \\
 &=&{1\over 2}\Big( ({\check {\cal I}}^{-1}(q))^{rs}{\check S}_q^r{\check S}_q^s+{\tilde
 g}^{\mu\nu}(q)(p_{\mu}-{\check {\vec S}}_q\cdot {\check {\vec {\cal A}}}_{\mu}(q))
 (p_{\nu}-{\check {\vec S}}_q\cdot {\check {\vec {\cal A}}}_{\nu}(q))\Big).\nonumber \\
  &&{}
\label{VII34}
\end{eqnarray}

\noindent imply that the
 non-relativistic gauge potential and metric are

\begin{eqnarray}
{\check {\vec {\cal A}}}_{\mu}(q)&=& lim_{c\rightarrow \infty} {\check
{\vec {\cal C}}}_{\mu}({\check {\vec S}}_q, q),\nonumber \\
 &&{}\nonumber \\
 {\tilde g}^{\mu\nu}(q)&=& lim_{c\rightarrow \infty}
\sum_{i=1}^N {{{\tilde v}_i^{\mu\nu}(q)}\over {m_i}}.
\label{VII35}
\end{eqnarray}

A purely vibrational Hamiltonian $H_{(rel)}^{(vib)}$ can be defined by
requiring the vanishing of the (now measurable) body frame components
of the angular velocity ${\check \omega}^r=0$. These conditions
transform Eq.(\ref{e9}) in equations for the determination of ${\check
S}^r_q{|}_{{\check \omega}^s=0}$: if we put their solution into
Eq.(\ref{VII30}), we get $H_{(rel)}^{(vib)}$.

On the other hand, the orientation-shape bundle approach privileges
the gauge choice ${\check S}^r_q=0$ (special connection C quoted in
Ref.\cite{iten2}) to define a purely vibrational (C-horizontal)
Hamiltonian

\begin{equation}
H_{(rel)}^{(S=0)}=H_{(rel)} {|}_{{\check {\vec S}}_q=0}=\sum_{i=1}^N
\sqrt{m_i^2+{\tilde v}_i^{\mu\nu}(q)p_{\mu}p_{\nu}},
\label{VII36}
\end{equation}

\noindent with the non-relativistic limit

\begin{eqnarray}
H_{(rel)}^{(S=0)}\, &\rightarrow_{c \rightarrow \infty}\,&
\sum_{i=1}^Nm_ic^2+{1\over 2} \sum_{i=1}^N {{{\tilde v}^{\mu\nu}_i(q)}\over {m_i}}
p_{\mu}p_{\nu}+O(1/c)=\nonumber \\
 &=&\sum_{i=1}^Nm_ic^2+ {1\over 2}
{\tilde g}^{\mu\nu}(q)p_{\mu}p_{\nu}+O(1/c).
\label{VII37}
\end{eqnarray}

However, we cannot use this definition because our canonical
construction is valid only if $S^r_q\not= 0$.

\subsection{N-Body Systems.}

Let us now consider the general case with $N \geq 4$. Instead of
coupling the centers of mass of particle clusters as it is done with
Jacobi coordinates ({\it center-of-mass clusters}), the {\it canonical
spin bases} will be obtained by coupling the spins of the 2-body
subsystems ({\it relative particles}) ${\vec \rho}_{qa}$, ${\vec
\pi}_{qa}$, $a=1,..,N-1$, defined in Eqs.(\ref{III7}), in all possible
ways ({\it spin clusters} from the addition of angular momenta). Let
us stress that we can build a {\it spin basis} with a pattern of {\it
spin clusters} which is completely unrelated to a possible
pre-existing {\it center-of-mass clustering}.

Let us consider the case $N=4$ as a prototype of the general
construction. We have now three relative variables ${\vec \rho}_{q1}$,
${\vec \rho}_{q2}$, ${\vec \rho}_{q3}$ and related momenta ${\vec
\pi}_{q1}$, ${\vec \pi}_{q2}$, ${\vec \pi}_{q3}$. In the following
formulas we use the convention that the subscripts $a,b,c$ mean any
permutation of $1,2,3$.

As in Ref.\cite{iten2}, we define the following sequence of canonical
transformations (we assume $S_q\not= 0$; $S_{qA}\not= 0$, $A=a,b,c$)
corresponding to the {\it spin clustering} pattern $abc
\mapsto (ab) c \mapsto ((ab)c)$ [build first the spin cluster $(ab)$,
then the spin cluster $((ab)c)$]:

\begin{eqnarray}
&&\begin{minipage}[t]{3cm}
\begin{tabular}{|lll|} \hline
${\vec \rho}_{qa}$ & ${\vec \rho}_{qb}$ & ${\vec \rho}_{qc}$\\ ${\vec
\pi}_{qa}$ & ${\vec \pi}_{qb}$ & ${\vec \pi}_{qc}$ \\ \hline
\end{tabular}
\end{minipage}
\ {\longrightarrow \hspace{.2cm}}\      \nonumber \\
\ {\longrightarrow \hspace{.2cm}}\
&&\begin{minipage}[t]{7cm}
\begin{tabular}{|ll|ll|ll|lll|} \hline
$\alpha_a$ & $\beta_a$ & $\alpha_b$& $\beta_b$ & $\alpha_c$ &
$\beta_c$ & $\rho_{qa}$ & $\rho_{qb}$ & $\rho_{qc}$ \\
\hline
 $S_{qa}$   & $S^3_{qa}$& $S_{qb}$ & $S^3_{qb}$ & $S_{qc}$ & $S^3_{qc}$ &
 ${\tilde \pi}_{qa}$ & ${\tilde \pi}_{qb}$ & ${\tilde \pi}_{qc}$ \\ \hline
\end{tabular}
\end{minipage}
\ {\longrightarrow \hspace{.2cm}}\     \nonumber \\
\ { {\buildrel {(ab)c} \over \longrightarrow} \hspace{.2cm}} \
&&\begin{minipage}[t]{12cm}
\begin{tabular}{|ccc|cc|cccc|} \hline
$\alpha_{(ab)}$ & $\beta_{(ab)}$ & $\gamma_{(ab)}$& $\alpha_c$ &
$\beta_c$ & $|{\vec N}_{(ab)}|$ & $\rho_{qa}$ & $\rho_{qb}$ &
$\rho_{qc}$\\
 \hline
 $S_{q(ab)}$ & $S^3_{q(ab)}$   &  ${\check S}_{q(ab)}^3=
 {\vec S}_{q(ab)}\cdot {\hat N}_{(ab)}$ & $S_{qc}$ & $S^3_{qc}$
  & $\xi_{(ab)}$ & ${\tilde \pi}_{qa}$ & ${\tilde \pi}_{qb}$ & ${\tilde \pi}_{qc}$\\ \hline
\end{tabular}
\end{minipage}
\ {\longrightarrow \hspace{.2cm}}\     \nonumber \\
\ {\longrightarrow \hspace{.2cm}}\
&&\begin{minipage}[t]{13cm}
\begin{tabular}{|ccc|cccccc|} \hline
$\alpha_{((ab)c)}$ & $\beta_{((ab)c)}$ & $\gamma_{((ab)c)}$& $|{\vec
N}_{((ab)c)}|$ & $\gamma_{(ab)}$& $|{\vec N}_{(ab)}|$ & $\rho_{qa}$ &
$\rho_{qb}$ & $\rho_{qc}$\\
\hline
 $S_q={\check S}_q$   & $S^3_q$& ${\check S}_q^3={\vec S}_q\cdot {\hat N}_{((ab)c)}$
 & $\xi_{((ab)c)}$ & ${\vec S}_{q(ab)}\cdot {\hat N}_{(ab)}$ &
 $\xi_{(ab)}$ & ${\tilde \pi}_{qa}$ & ${\tilde \pi}_{qb}$ & $ {\tilde \pi}_{qc}$\\ \hline
\end{tabular}
\end{minipage}
\ {\rightarrow \hspace{.2cm}}\     \nonumber \\
\ {{\buildrel {non\, can.} \over \longrightarrow} \hspace{.2cm}}\
&&\begin{minipage}[t]{12cm}
\begin{tabular}{|ccc|cccccc|} \hline
$\tilde \alpha$ & $\tilde \beta$ & $\tilde \gamma$& $|{\vec
N}_{((ab)c)}|$ & $\gamma_{(ab)}$& $|{\vec N}_{(ab)}|$ & $\rho_{qa}$ &
$\rho_{qb}$ & $\rho_{qc}$\\
\hline
 ${\check S}_q^1$   & ${\check S}^2_q$& ${\check S}_q^3$
 & $\xi_{((ab)c)}$ & $\Omega_{(ab)}={\vec S}_{q(ab)}\cdot {\hat N}_{(ab)}$ &
 $\xi_{(ab)}$ & ${\tilde \pi}_{qa}$ & ${\tilde \pi}_{qb}$ & $ {\tilde \pi}_{qc}$\\ \hline
\end{tabular}
\end{minipage}.  \nonumber \\
&&{}
\label{VII38}
\end{eqnarray}

See Appendix F of Ref.\cite{iten2} for the explicit construction of
the canonical transformations.

The first non-point canonical transformation is based on the existence
of the three unit vectors ${\hat R}_A$, $A=a,b,c$, and of three E(3)
groups with fixed values (${\hat R}_A^2=1$, ${\vec S}_A\cdot {\hat
R}_A=0$) of their invariants. One uses Eqs.(\ref{VII13}),
(\ref{VII14}) and (\ref{VII15}).

In the next canonical transformation the  spins of the {\it relative
particles} $a$ and $b$ are coupled to form the spin cluster $(ab)$,
leaving the {\it relative particle} $c$ as a spectator. We use the
definitions ${\vec N}_{(ab)}={1\over 2}({\hat R}_a+{\hat R}_b)$,
${\vec
\chi}_{(ab)}={1\over 2}({\hat R}_a-{\hat R}_b)$, ${\vec
S}_{(ab)}={\vec S}_{qa}+{\vec S}_{qb}$, ${\vec W}_{q(ab)}={\vec
S}_{qa}-{\vec S}_{qb}$. We  get ${\vec N}_{(ab)}\cdot {\vec
\chi}_{(ab)}=0$, $\{ N^r_{(ab)},N^s_{(ab)} \} = \{ N^r_{(ab)},
\chi^s_{(ab)} \} = \{ \chi^r_{(ab)}, \chi^s_{(ab)} \} =0$ and a new
E(3) realization generated by ${\vec S}_{(ab)}$ and ${\vec N}_{(ab)}$,
with non-fixed invariants $|{\vec N}_{(ab)}|$, ${\vec S}_{(ab)}\cdot
{\hat N}_{(ab)}\, {\buildrel {def} \over =}\, \Omega_{(ab)}$. From
Eqs.(\ref{VII23}) we get

\bea
&&{\vec \rho}_{qa} = \rho_{qa} \Big[ |{\vec N}_{(ab)}| {\hat N}_{(ab)}
+ \sqrt{1-{\vec N}^2_{(ab)}} {\hat \chi}_{(ab)}\Big] ,\nonumber \\
&&{\vec \rho}_{qb} = \rho_{qb} \Big[ |{\vec N}_{(ab)}| {\hat N}_{(ab)}
- \sqrt{1-{\vec N}^2_{(ab)}} {\hat \chi}_{(ab)}\Big] ,\nonumber \\
 &&{\vec \rho}_{qc} = \rho_{qc} {\hat R}_c.
 \label{VII39}
 \eea

 \noindent Eq.(\ref{VII19}) defines $\alpha_{(ab)}$ and $\beta_{(ab)}$, so that
 Eq.(\ref{VII20}) defines a unit vector ${\hat R}_{(ab)}$ with ${\vec
 S}_{(ab)}\cdot {\hat R}_{(ab)}=0$, $\{ {\hat R}^r_{(ab)},{\hat R}^s_{(ab)} \} =0$.
This unit vector identifies the {\it spin cluster} $(ab)$ in
 the same way as the unit vectors ${\hat R}_A={\hat {\vec \rho}}_{qA}$ identify
 the {\it relative particles} $A$.

 The next step is the coupling of the {\it spin cluster} $(ab)$
 with unit vector ${\hat R}_{(ab)}$ [described by the canonical variables $\alpha_{(ab)}$,
 $S_{(ab)}$, $\beta_{(ab)}$ $S^3_{(ab)}$] with the {\it relative particle} $c$
 with unit vector ${\hat R}_c$ and  described by $\alpha_c$, $S_{qc}$,
 $\beta_c$, $S^3_{qc}$: this builds the {\it spin cluster} $((ab)c)$.
 Again Eq.(\ref{VII16}) allows to define ${\vec N}_{((ab)c)}=
 {1\over 2}({\hat R}_{(ab)}+{\hat R}_c)$,
 ${\vec \chi}_{((ab)c)}={1\over 2}({\hat R}_{(ab)}-{\hat R}_c)$, ${\vec S}_q
 ={\vec S}_{q((ab)c)}= {\vec S}_{q(ab)}+{\vec S}_{qc}$, ${\vec W}_{q((ab)c)}=
 {\vec S}_{q(ab)}-{\vec S}_{qc}$. Since we have ${\vec N}_{((ab)c)}\cdot {\vec \chi}_{((ab)c)}=0$
 and  $\{ N^r_{((ab)c)},N^s_{((ab)c)} \} = \{ N^r_{((ab)c)}, \chi^s_{((ab)c)} \} =
 \{ \chi^r_{((ab)c)}, \chi^s_{((ab)c)} \} =0$ due to $\{ {\hat R}^r_{(ab)}, {\hat R}_c^s \} =0$,
  a new E(3) group generated by ${\vec S}_q$
 and ${\vec N}_{((ab)c)}$ with non-fixed invariants $|{\vec N}_{((ab)c)}|$,
 ${\vec S}_q\cdot {\vec N}_{((ab)c)}= {\check S}^3_q
 |{\vec N}_{((ab)c)}|$ emerges. Eq.(\ref{VII19}) defines $\alpha_{((ab)c)}$ and
 $\beta_{((ab)c)}$, so that Eq.(\ref{VII20}) allows to identify a final unit
 vector ${\hat R}_{((ab)c)}$ with ${\vec S}_q\cdot {\hat R}_{((ab)c)}=0$ and
 $\{ {\hat R}_{((ab)c)}^r,{\hat R}^s_{((ab)c)} \} =0$.

 In conclusion, when $S_q\not= 0$,
  we find both a {\it dynamical body frame} ${\hat \chi}_{((ab)c)}$,
 ${\hat N}_{((ab)c)}\times {\hat \chi}_{((ab)c)}$, ${\hat N}_{((ab)c)}$, and
 a {\it spin frame} ${\hat S}_q$, ${\hat R}_{((ab)c)}$, ${\hat R}_{((ab)c)}\times {\hat S}_q$
 like in the 3-body case. There is an {\it important difference}, however:
 the orthonormal vectors ${\vec N}_{((ab)c)}$ and
 ${\vec \chi}_{((ab)c)}$ depend on the momenta of the relative particles $a$ and $b$
 through ${\hat R}_{(ab)}$, so that our results do not share any relation with the N=4
 non-trivial SO(3) principal bundle of the orientation-shape bundle approach.

 The final 6 {\it dynamical shape variables} are $q^{\mu} = \{ |{\vec N}_{((ab)c)}|, \gamma_{(ab)},
 |{\vec N}_{(ab)}|, \rho_{qa}, \rho_{qb}, \rho_{qc} \}$. While the last four
 depend only on the original relative coordinates ${\vec \rho}_{qA}$, $A=a,b,c$,
 the first two depend also on the original momenta ${\vec \pi}_{qA}$:
 therefore they are {\it generalized shape variables}.
 By using Appendix D, we obtain

\beq
  \rho^r_{qA}={\cal R}^{rs}(\tilde \alpha ,
 \tilde \beta ,\tilde \gamma )\, {\check \rho}^s_{qA}(q^{\mu},
 p_{\mu}, {\check S}^r_q),\qquad A=a,b,c,
 \label{VII40}
 \eeq

\noindent This means that for N=4 the dynamical body frame components
 ${\check \rho}^r_{qA}$ depend also on the dynamical shape momenta
 and on the dynamical body frame components of the spin. It is clear that
 this result stands outside the orientation-shape bundle approach completely.

As shown in Appendix F of Ref.\cite{iten2}, starting from the
Hamiltonian $H_{rel ((ab)c)}$ in the final variables, we can define a
rotational Hamiltonian $H^{(rot)}_{rel ((ab)c)}$ (for ${\dot
q}^{\mu}=0$, see Eqs.(F18) of Ref.\cite{iten2}) and a vibrational
Hamiltonian $H^{(vib)}_{rel ((ab)c)}$ (vanishing of the physical
dynamical angular velocity ${\check \omega}^r_{((ab)c)}=0$, see
Eqs.(F21) of Ref.\cite{iten2}), but $H_{rel ((ab)c)}$ is not the sum
of these two Hamiltonians. In the rotational Hamiltonian and in the
spin-angular velocity relation we find two {\it inertia-like tensors}
depending only on the dynamical shape variables.

The price to be paid for the existence of 3 global {\it dynamical body
frames} for N=4 is a more complicated form of the Hamiltonian kinetic
energy. On the other hand, {\it dynamical vibrations} and {\it
dynamical angular velocity} are  measurable quantities in each
dynamical body frame.

For N=5 we can repeat the previous construction either with the
sequence of spin clusterings $abcd \mapsto (ab)cd \mapsto ((ab)c)d)
\mapsto (((ab)c)d)$ or with the sequence $abcd \mapsto (ab)(cd) \mapsto
((ab)(cd))$ [$a,b,c,d$ any permutation of 1,2,3,4] as said in the
Introduction. Each {\it spin cluster} $(...)$ will be identified by
the unit vector ${\hat R}_{(...)}$, axis of the {\it spin frame} of
the cluster. All the final {\it dynamical body frames} built  with
this construction will have their axes depending on both the original
configurations and momenta.

This construction is trivially generalized to any N: we have only to
classify all the possible {\it spin clustering patterns}.

Therefore, for $N\geq 4$ our sequence of canonical and non-canonical
transformations leads to the following result to be compared with
Eq.(\ref{VII25}) of the 3-body case

\beq
\begin{minipage}[t]{5cm}
\begin{tabular}{|l|} \hline
${\vec \rho}_{qA}$ \\ \hline
 ${\vec \pi}_{qA}$ \\ \hline
\end{tabular}
\end{minipage}
\ {{\buildrel {non\,\, can.} \over \longrightarrow} \hspace{.2cm}}\
\begin{minipage}[b]{5cm}
\begin{tabular}{|lll|l|} \hline
$\tilde \alpha$ & $\tilde \beta$ & $\tilde \gamma$& $q^{\mu}({\vec
\rho}_{qA},{\vec \pi}_{qA})$
\\ \hline
 ${\check S}^1_q$ & ${\check S}^2_q$ & ${\check S}^3_q$ &
 $p_{\mu}({\vec \rho}_{qA},{\vec \pi}_{qA}$) \\ \hline
 \end{tabular}
 \end{minipage}.
\label{III41}
\eeq

This state of affairs suggests that for $N\geq 4$ and with $S_q\not=
0$, $S_{qA}\not= 0$, $A=a,b,c$, viz. when the standard
(3N-3)-dimensional orientation-shape bundle is not trivial,  the
original (6N-6)-dimensional relative phase space admits the definition
of as many {\it dynamical body frames} as spin canonical bases
\footnote{Recall that  a different Hamiltonian {\it right} SO(3)
action on the relative phase space corresponds to each  of them.},
which are globally defined (apart isolated coordinate singularities)
for the non-singular N-body configurations with ${\vec S}_q\not= 0$
(and with non-zero spin for each spin subcluster).

These {\it dynamical body frames} do not correspond to local cross
sections of the static non-trivial orientation-shape SO(3) principal
bundle and the spin canonical bases do not coincide with the canonical
bases associated with the static theory. Therefore, we do not get
gauge potentials and all the other quantities evaluated in Appendix E
for N=3.

\vfill\eject

\section{The case of Interacting Particles.}

As shown in Ref.\cite{crater} and in its bibliography, the
action-at-a-distance interactions inside the Wigner hyperplane may be
introduced either under the square roots (scalar and vector
potentials) or outside (scalar potential like the Coulomb one)
appearing in the free Hamiltonian (\ref{II19}) or (\ref{VI1}).

In the rest-frame instant form the most general Hamiltonian with only
action-at-a-distance interactions is

\beq
H = \sum_{i=1}^N \sqrt{ m_i^2+U_i+[{\vec k}_i-{\vec V}_i]^2} + V,
\label{VIII1}
\eeq

\noindent with $U=U({\vec \kappa}_k, {\vec \eta}_h-{\vec \eta}_k)$,
${\vec V}_i={\vec V}_i({\vec k}_{j\not= i}, {\vec \eta}_i-{\vec
\eta}_{j\not= i})$, $V=V_o(|{\vec \eta}_i-{\vec \eta}_j|)+V^{'}({\vec k}_i,
{\vec \eta}_i-{\vec \eta}_j)$.

In the rest frame the Hamiltonian for the relative motion becomes

\beq
H_{(rel)} =
\sum_{i=1}^N\sqrt{m_i^2+{\tilde U}_i+[\sqrt{n}\sum_{a=1}^{N-1}\gamma_{ai}{\vec \pi}_{qa}-
{\tilde {\vec V}}_i]^2} +\tilde V,
\label{VIII2}
\eeq

\noindent with

\bea
{\tilde U}_i&=&U([\sqrt{n}\sum_{a=1}^{N-1}\gamma_{ak}{\vec \pi}_{qa},
{1\over{\sqrt{N}}}\sum_{a=1}^{N-1}(\gamma_{ah}-\gamma_{ak}){\vec
\rho}_{qa}),\nonumber \\
 {\tilde {\vec V}}_i&=&{\vec V}_i([\sqrt{n}\sum_{a=1}^{N-1}
\gamma_{aj\not= i}{\vec \pi}_{qa},{1\over{\sqrt{N}}}\sum_{a=1}^{N-1}
(\gamma_{ai}-\gamma_{aj\not= i}){\vec \rho}_{qa}),\nonumber \\
 \tilde V &=& V_o(|{1\over
{\sqrt{N}}}\sum_{a=1}^{N-1}(\gamma_{ai}-\gamma_{aj}){\vec
\rho}_{qa}|) + V^{'}([\sqrt{n}\sum_{a=1}^{N-1}\gamma_{ai}{\vec \pi}_{qa},
{1\over{\sqrt{N}}}\sum_{a=1}^{N-1}(\gamma_{ai}-\gamma_{aj}){\vec
\rho}_{qa}).
\label{VIII3}
\eea

The prices for the existence of 3 possible global {\it dynamical body
frames} for N=4 are:

i) a more complicated form of the Hamiltonian kinetic energy but with
a definition of measurable {\it dynamical vibrations} and {\it
dynamical angular velocity} in each dynamical body frame;

ii) the fact that a potential $V({\vec \eta}_{ij}\cdot {\vec
\eta}_{hk}))$ with ${\vec \eta}_{ij}={\vec \eta}_i-{\vec \eta}_j$
becomes dependent also on the shape momenta, since we have

\beq
V({\vec
\eta}_{ij}\cdot {\vec \eta}_{hk})= V[{1\over N} \sum_{a,b}^{1,..,N-1}
(\Gamma_{ai}-\Gamma_{aj})(\Gamma_{bh}-\Gamma_{bk}) {\vec
\rho}_{qa}\cdot {\vec \rho}_{qb}].
\label{VIII4}
\eeq

\noindent For N=4, due to Eq.(\ref{e6}), in the pattern $((ab)c)$ we have

\beq
V={\tilde V}_{((ab)c)}[\rho_{qa}, \rho_{qb}, \rho_{qc}, |{\vec
N}_{((ab)c)}|, \gamma_{(ab)}, |{\vec N}_{(ab)}|;\, \xi_{((ab)c)},
\Omega_{(ab)};\, {\check S}^r_q].
\label{VIII5}
\eeq

For more general potentials $V({\vec \eta}_{ij}\cdot {\vec \eta}_{hk},
{\vec \kappa}_i\cdot {\vec
\eta}_{hk}, {\vec \kappa}_i\cdot {\vec \kappa}_j)$, like the non-relativistic
limit of the relativistic Darwin potential of Ref.\cite{crater}, more
complicated expressions are obtained.

\vfill\eject

\section{Conclusions.}

In this paper we have explored the relativistic kinematics of a system
of N scalar positive-energy particles. In the framework of the
rest-frame instant form of dynamics it is possible to find the
relativistic extension of the Abelian translational and non-Abelian
rotational symmetries whose associated Noether constants of motion are
fundamental for the study of isolated systems. In the relativistic
case the rest-frame description on the Wigner hyperplanes allows to
clarify all the problems by virtue of a doubling of all the concepts:
they can be either {\it external} (namely observed by an arbitrary
inertial Lorentz frame) or {\it internal} (namely observed by an
inertial observer at rest inside the Wigner hyperplane).
Correspondingly two realizations of the Poincar\'e algebra are
naturally defined.

After a clarification of the possible {\it external} and {\it
internal} definitions of relativistic center of mass, we have shown
that it is possible to define a family of canonical transformations
for the definition of canonical relative variables. The Hamiltonian in
the rest frame can be expressed in terms of these variables. It turns
out that, due to the presence of the square roots in the Hamiltonian,
the non-relativistic concepts of {\it Jacobi normal relative
coordinates}, {\it reduced masses} and {\it barycentric tensor of
inertia} cannot be extended to the relativistic formulation.

On the other hand, the rest-frame description with the Wigner
hyperplanes allows to use the non-relativistic formalism developed in
Ref.\cite{iten2} for the study of the rotational kinematics, since it
it is independent of Jacobi coordinates. Therefore, we can extend the
concepts of {\it canonical spin bases}, {\it spin frames} and {\it
dynamical body frames} to the relativistic level and we find again
that, due to the non-Abelian nature of rotations,  a global separation
of rotations from vibrations is {\it not} possible.

In a future paper\cite{iten3} we will conclude the study of
relativistic kinematics by defining Dixon's multipoles\cite{dixon} for
the relativistic N-body problem in the rest-frame instant form of
dynamics. It will be shown that, in this framework, we can recover
concepts like the {\it tensor of inertia} by using the quadrupole
moment.

The final task should be the extension of all these results to
relativistic extended ({\it continua}) isolated systems.

\vfill\eject

\appendix

\section{Parametrized Minkowski Theories.}

In this Appendix we review the main aspects of parametrized Minkowski
theories and of the canonical reduction of gauge systems, following
Refs.\cite{lus,crater,india}, where a complete treatment of N scalar
charged positive energy particles plus the electromagnetic field is
given.

The starting point was Dirac's\cite{dirac} reformulation of classical
field theory on spacelike hypersurfaces foliating Minkowski spacetime
$M^4$. The foliation is defined by an embedding $R\times
\Sigma \rightarrow M^4$, $(\tau ,\vec
\sigma ) \mapsto z^{\mu}(\tau ,\vec \sigma )$,
with $\Sigma$ an abstract 3-surface diffeomorphic to $R^3$
 In this way one gets a parametrized field theory with a covariant
3+1 splitting of flat spacetime, which is already in a form suited to
the transition to general relativity in its ADM canonical formulation.

The price to be paid is that one has to add the embeddings
$z^{\mu}(\tau ,\vec \sigma )$ identifying the points of the spacelike
hypersurface $\Sigma_{\tau}$ \footnote{The only ones carrying Lorentz
indices; the scalar parameter $\tau$ labels the leaves of the
foliation and $\vec \sigma$ are curvilinear coordinates on
$\Sigma_{\tau}$.} as new configuration variables and then to redefine
the fields on $\Sigma_{\tau}$ in such a way they know the whole
hypersurface $\Sigma_{\tau}$ of $\tau$-simultaneity \footnote{For a
Klein-Gordon field $\phi (x)$, this new field is $\tilde \phi (\tau
,\vec \sigma )=\phi (z(\tau ,\vec
\sigma ))$: it contains the nonlocal information about the embedding
and the associated notion of {\it equal time}.}.

Then one rewrites the Lagrangian of the given isolated system in the
form required by the coupling to an external gravitational field,
makes the 3+1 splitting of Minkowski spacetime and replaces all the
fields of the system by the new fields on $\Sigma_{\tau}$
\footnote{These are Lorentz scalars, having only surface indices.}.
Instead of considering the 4-metric as describing a gravitational
field \footnote{Namely as an independent field like in metric gravity,
where one adds the Hilbert action to the action for the matter
fields.}, here one replaces the 4-metric with the  induced metric $g_{
AB}[z]
=z^{\mu}_{A}\eta_{\mu\nu}z^{\nu}_{B}$ on
$\Sigma_{\tau}$, a functional of $z^{\mu}$, and considers the
embedding coordinates $z^{\mu}(\tau ,\vec \sigma )$ as independent
fields. We use the notation $\sigma^{A}=(\tau ,\sigma^{\check r})$ of
Refs.\cite{lus,crater}. The $z^{\mu}_{A}(\sigma )=
\partial z^{\mu}(\sigma )/\partial \sigma^{A}$ are flat cotetrad fields on Minkowski
spacetime \footnote{I.e. ${}^4\eta^{\mu\nu}=z^{\mu}_A\, {}^4g^{AB}\,
z^{\nu}_B$ with ${}^4g^{AB}$ the inverse of ${}^4g_{AB}$.} with the
$z^{\mu}_r$'s tangent to $\Sigma_{\tau}$\footnote{Note that in metric
gravity the $z^{\mu}_{A}\not=\partial z^{\mu}/\partial
\sigma^{A}$ are not cotetrad fields since no holonomic coordinates
$z^{\mu}(\sigma )$  exist.}.

The evolution vector is given by $z^{\mu}_{\tau}=N_{[z](flat)}l^{\mu}+
N^{\check r}_{[z](flat)}z^{\mu}_{\check r}$, where $l^{\mu}(\tau ,\vec
\sigma )$ is the normal to $\Sigma_{\tau}$ in $z^{\mu}(\tau ,\vec
\sigma )$ and

\bea
N_{[z](flat)}(\tau ,\vec \sigma
)&=&\sqrt{{}^4g_{\tau\tau}-{}^3\gamma^{\check r\check s}\, {}^4g_{\tau
\check r}\, {}^4g_{\tau \check s}}=\sqrt{{}^4g
/{}^3\gamma},\nonumber \\
 N_{[z](flat) \check r}(\tau ,\vec \sigma )&=&{}^3g
_{\check r\check s}(\tau ,\vec \sigma )N^{\check s}_{[z](flat)}(\tau ,\vec
\sigma )={}^4g_{\tau \check r},
\label{a1}
\eea

\noindent are the flat lapse and shift functions
defined through the metric like in metric gravity (here ${}^3g^{\check
r\check u}\, {}^4g_{\check u\check s}=\delta^{\check r}_{\check s}$);
however, in Minkowski spacetime they are functionals of $z^{\mu}(\tau
,\vec \sigma )$ instead of being independent variables.   See Appendix
B for notations on spacelike hypersurfaces.

{}From this Lagrangian for the isolated system \footnote{See for
instance Eq.(\ref{II3}) for the case of N free scalar particles.}, we
have that: i) the possible constraints of the system  are Lorentz
scalars; ii) four primary first class constraints are added which
imply the independence of the description from the choice of the
foliation with spacelike hypersufaces:

\beq
{\cal H}_{\mu}(\tau ,\vec
\sigma )=\rho_{\mu}(\tau ,\vec \sigma )-l_{\mu}(\tau ,\vec \sigma )
T_{system}^{\tau\tau}(\tau ,\vec \sigma )-z_{\check r \mu}(\tau ,\vec
\sigma )T_{system}^{\tau \check r}(\tau ,\vec \sigma ) \approx 0.
\label{a2}
\eeq

\noindent
Here $T_{system}^{\tau\tau}(\tau ,\vec \sigma )$, $T_{system}^{\tau
\check r}(\tau ,\vec \sigma )$, are the components of the
energy-momentum tensor in the holonomic coordinate system on
$\Sigma_{\tau}$ corresponding to the energy- and momentum-density of
the isolated system. These four constraints satisfy an Abelian Poisson
algebra being solved in 4-momenta $\rho_{\mu}(\tau ,\vec \sigma )$
conjugate to the embedding variables $z^{\mu}(\tau ,\vec \sigma )$:
$\quad \lbrace {\cal H}_{\mu}(\tau ,\vec
\sigma ), {\cal H}_{\nu}(\tau ,{\vec \sigma}^{'}) \rbrace
=0$.

We see  that the embedding fields $z^{\mu}(\tau ,\vec \sigma )$ are
the {\it gauge} variables associated with this kind of general
covariance.

The Dirac Hamiltonian is

\beq
H_D = H_{(c)} + \int d^3\sigma
\lambda^{\mu}(\tau ,\vec \sigma ){\cal H}_{\mu} (\tau ,\vec
\sigma )+ (\mbox{\it system-dependent\, primary\, constraints}),
\label{a3}
\eeq

\noindent
with $\lambda^{\mu}(\tau ,\vec \sigma )$ arbitrary Dirac multipliers
\footnote{$H_{(c)}$ is the canonical part: it is either zero or weakly
vanishing due to system-dependent secondary constraints.}. By using
${}^4\eta^{\mu\nu}=[l^{\mu}l^{\nu}- z^{\mu}_{\check r}\, {}^3g^{\check
r\check s} z^{\nu}_{\check s}](\tau ,
\vec \sigma )$ we can write

\bea
\lambda_{\mu}(\tau ,\vec
\sigma ){\cal H}^{\mu}(\tau ,\vec \sigma )&=&[(\lambda_{\mu}l^{\mu})
(l_{\nu}{\cal H}^{\nu})-(\lambda_{\mu}z^{\mu}_{\check r})({}^3g
^{\check r\check s} z_{\check s \nu}{\cal H}^{\nu})](\tau ,\vec \sigma )\,
{\buildrel  {def} \over =}\nonumber \\
 &{\buildrel {def} \over =}&\, N_{(flat)}(\tau ,\vec
\sigma ) (l_{\mu}{\cal H}^{\mu})(\tau ,\vec \sigma )
-N_{(flat) \check r}(\tau ,\vec \sigma ) ({}^3g^{\check r\check s} z_{\check
s \nu}{\cal H}^{\nu})(\tau ,\vec \sigma ),
\label{a4}
\eea

\noindent with the (non-holonomic form of
the) constraints ${\tilde {\cal H}}(\tau ,\vec \sigma )=(l_{\mu}{\cal
H}^{\mu})(\tau ,\vec
\sigma )\approx 0$, ${\tilde {\cal H}}_{\check r}(\tau ,\vec \sigma )=(z_{\check r \mu} {\cal
H}^{\mu})(\tau ,\vec \sigma )\approx 0$, satisfying the universal
Dirac algebra

\begin{eqnarray}
\lbrace {\tilde {\cal H}}_r(\tau ,\vec \sigma ),{\tilde {\cal H}}_s
(\tau ,{\vec \sigma}^{'})\rbrace &=& {\tilde {\cal H}}_r(\tau ,{\vec
\sigma}^{'} )\, {{\partial \delta^3(\vec \sigma ,{\vec \sigma}^{'})}\over
{\partial \sigma^s}} + {\tilde {\cal H}}_s(\tau ,\vec \sigma )
{{\partial
\delta^3(\vec \sigma ,{\vec \sigma}^{'})}\over {\partial \sigma^r}},
\nonumber \\
\lbrace {\tilde {\cal H}}(\tau ,\vec \sigma ),{\tilde {\cal H}}_r(\tau ,
{\vec \sigma}^{'})\rbrace &=& {\tilde {\cal H}}(\tau ,\vec \sigma )
{{\partial \delta^3(\vec \sigma ,{\vec \sigma}^{'})}\over {\partial \sigma^r}},
\nonumber \\
\lbrace {\tilde {\cal H}}(\tau ,\vec \sigma ),{\tilde {\cal H}}(\tau ,{\vec
\sigma}^{'})\rbrace &=&[{}^3g^{rs}(\tau ,\vec \sigma ) {\tilde {\cal H}}_s
(\tau ,\vec \sigma )+\nonumber \\ &+&{}^3g^{rs}(\tau ,{\vec
\sigma}^{'}){\tilde {\cal H}}_s(\tau ,{\vec \sigma}^{'})]{{\partial
\delta^3(\vec \sigma ,{\vec
\sigma}^{'})}\over {\partial \sigma^r}}.
\label{a5}
\end{eqnarray}

In this way we have defined new  flat lapse and shift functions

\begin{eqnarray}
N_{(flat)}(\tau ,\vec \sigma )&=& \lambda_{\mu}(\tau ,\vec \sigma )
l^{\mu}(\tau ,\vec \sigma ),\nonumber \\
 N_{(flat) \check r}(\tau
,\vec \sigma )&=& \lambda_{\mu}(\tau ,\vec \sigma ) z^{\mu}_{\check
r}(\tau ,\vec \sigma ).
\label{a6}
\end{eqnarray}

\noindent which have the same content of the arbitrary Dirac multipliers
$\lambda_{\mu}(\tau ,\vec \sigma )$, namely they multiply primary
first class constraints satisfying the Dirac algebra. In Minkowski
spacetime they are quite distinct from the previous lapse and shift
functions $N_{[z](flat)}$, $N_{[z](flat) \check r}$, defined starting
from the metric. Only by  using the Hamilton equations
$z^{\mu}_{\tau}(\tau ,\vec \sigma )\, {\buildrel \circ \over =}\, \{
z^{\mu}(\tau ,\vec \sigma ), H_D \} = \lambda^{\mu}(\tau ,\vec
\sigma )$ we get $N_{[z](flat)}(\tau ,\vec \sigma )\, {\buildrel \circ \over =}\,
N_{(flat)}(\tau ,\vec \sigma )$, $N_{[z](flat)\check r}(\tau ,\vec
\sigma )\, {\buildrel \circ \over =}\, N_{(flat)\check r}(\tau ,\vec \sigma )$.

Therefore, when we consider arbitrary 3+1 splittings of spacetime with
arbitrary spacelike hypersurfaces, the descriptions of metric gravity
plus matter and  the parametrized Minkowski description of the same
matter do not seem to follow the same pattern. The situation changes
however if the allowed 3+1 splittings of spacetime in ADM metric
gravity are restricted to have the leaves approaching Minkowski
spacelike hyperplanes at spatial infinity and if parametrized
Minkowski theories are restricted to spacelike hyperplanes.

The restriction of parametrized Minkowski theories to flat hyperplanes
in Minkowski spacetime  is done by adding the gauge-fixings\cite{lus}

\beq
z^{\mu}(\tau ,\vec \sigma )- x^{\mu}_s(\tau )-b^{\mu}_{\check r}(\tau
)\sigma^{\check r} \approx 0.
\label{a7}
\eeq

\noindent Here $x^{\mu}_s(\tau )$ denotes a point on the hyperplane
$\Sigma_{\tau}$ chosen as an arbitrary origin; the $b^{\mu}_{\check
r}(\tau )$'s form an orthonormal triad at $x^{\mu}_s(\tau )$ and the
$\tau$-independent normal to the family of spacelike hyperplanes is
$l^{\mu}=b^{\mu}_{\tau}=\epsilon^{\mu}{}
_{\alpha\beta\gamma}b^{\alpha}_{\check 1}(\tau )b^{\beta}
_{\check 2}(\tau )b^{\gamma}_{\check 3}(\tau )$. Each hyperplane is
described by 10 configuration variables, $x^{\mu}_s(\tau )$ and the 6
independent degrees of freedom contained in the triad $b^{\mu
}_{\check r} (\tau )$, and by the 10 conjugate momenta: $p^{\mu}_s$
and 6 variables hidden in a spin tensor $S^{\mu\nu}_s$\cite{lus}. With
these 20 canonical variables it is possible to build 10 Poincar\'e
generators ${\bar p}^{\mu}_s
=p^{\mu}_s$, ${\bar J}^{\mu\nu}_s=x^{\mu}_sp^{\nu}_s-x^{\nu}
_sp^{\mu}_s+S^{\mu\nu}_s$.

After the restriction to spacelike hyperplanes, the piece $\int
d^3\sigma \lambda^{\mu}(\tau ,\vec \sigma ) {\cal H}_{\mu}(\tau ,\vec
\sigma )$ of the Dirac Hamiltonian is reduced to

\beq
{\tilde \lambda}^{\mu}(\tau ){\tilde {\cal H}}_{\mu}(\tau )
-{1\over 2}{\tilde \lambda}^{\mu\nu}(\tau ){\tilde {\cal H}}_{\mu\nu}(\tau ),
\label{a8}
\eeq

\noindent because the time constancy of the gauge-fixings $z^{\mu}
(\tau ,\vec \sigma )-x^{\mu}_s(\tau )-b^{\mu}_{\check r}(\tau )\sigma
^{\check r}\approx 0$ implies\hfill\break ($\, \, \, \dot {}$ means $d/d\tau$)

\bea
\lambda_{\mu}(\tau ,\vec \sigma )&=&{\tilde
\lambda}_{\mu}(\tau )+{\tilde \lambda}_{\mu\nu}(\tau )b^{\nu}
_{\check r}\sigma^{\check r},\nonumber \\
 && with \nonumber \\
 &&{\tilde \lambda}^{\mu}(\tau )=-{\dot x}
^{\mu}_s(\tau ),\quad\quad {\tilde \lambda}^{\mu\nu}(\tau )=-{\tilde \lambda}
^{\nu\mu}(\tau )={1\over 2}\sum_{\check r}[{\dot b}^{\mu}_{\check r}
b^{\nu}_{\check r}-b^{\mu}_{\check r}{\dot b}^{\nu}_{\check r}](\tau
).
\label{a9}
\eea

\noindent  Since at this stage we have
$z^{\mu}_{\check r}(\tau ,\vec \sigma )\approx b^{\mu}_{\check r}(\tau
)$, we get

\bea
z^{\mu}_{\tau}(\tau ,\vec \sigma )&\approx& N_{[z](flat)}(\tau ,\vec
\sigma )l^{\mu}(\tau ,\vec \sigma )+N^{\check r}
_{[z](flat)}(\tau ,\vec \sigma )b^{\mu}_{\check r}(\tau ,\vec
\sigma )\approx \nonumber \\
 &\approx& {\dot x}^{\mu}_s(\tau )+{\dot b}^{\mu}_{\check r}(\tau )
\sigma^{\check r}=-{\tilde \lambda}^{\mu}(\tau )-{\tilde
\lambda}^{\mu\nu}(\tau )b_{\check r \nu}(\tau )\sigma^{\check r}.
\label{a10}
\eea

\noindent
Only now we get the coincidence of the two definitions of flat lapse
and shift functions independently from the equations of motion, i.e.

\beq
N_{[z](flat)}(\tau ,\vec \sigma )\approx N_{(flat)}(\tau ,\vec \sigma
),\quad\quad N_{[z](flat)
\check r}(\tau ,\vec \sigma )\approx N_{(flat)\check r}(\tau ,\vec \sigma ).
\label{a11}
\eeq

The description on arbitrary foliations with spacelike hyperplanes is
independent from the choice of the foliation, due to the remaining 10
first class constraints

\bea
{\tilde {\cal H}}^{\mu}(\tau )&=&\int d^3\sigma {\cal H}^{\mu} (\tau
,\vec \sigma )=p^{\mu}_s - p^{\mu}_{sys}=p^{\mu}_s-\nonumber
\\
 &-&[total\, momentum\,
of\, the\, system\, inside\, the\, hyperplane]^{\mu}\approx
0,\nonumber \\
 &&{}\nonumber \\
 {\tilde {\cal H}}^{\mu\nu}(\tau )&=&
 b^{\mu}_{\check r}(\tau ) \int d^3\sigma \, \sigma
^{\check r}{\cal H}^{\nu}(\tau ,\vec \sigma )-b^{\nu}_{\check r}(\tau )
\int d^3\sigma \, \sigma^{\check r} {\cal H}^{\mu}(\tau ,\vec \sigma )=
S_s^{\mu\nu} - S_{sys}^{\mu\nu}=\nonumber \\
 &=&S^{\mu\nu}_s-[intrinsic\, angular\, momentum\, of\, the\, system\nonumber \\
 &&inside\, the\, hyperplane]^{\mu\nu}=S^{\mu\nu}_s -\nonumber \\
 &-&(b^{\mu}_{\check r}(\tau )l^{\nu}-b^{\nu}_{\check r}(\tau )l^{\mu}
)[boost\, part\, of\, system's\, angular\, momentum]^{\tau \check
r}-\nonumber  \\
 &-&(b^{\mu}_{\check r}(\tau )b^{\nu}_{\check
s}(\tau )-b^{\nu}_{\check r}(\tau )b^{\mu}_{\check s}(\tau ))[spin\,
part\, of\, system's\, angular\, momentum]^{\check r\check s}\approx
0.\nonumber \\
 &&{}
\label{a12}
\eea

Therefore, on spacelike hyperplanes in Minkowski spacetime we have

\begin{eqnarray}
N_{(flat)}(\tau ,\vec \sigma )&=&\lambda_{\mu}(\tau ,\vec \sigma
)l^{\mu} (\tau ,\vec \sigma ) \mapsto \nonumber \\
 &&{}\nonumber \\
 &\mapsto&
N_{(flat)}(\tau ,\vec \sigma )=N_{[z](flat)}(\tau ,\vec \sigma )=
\nonumber \\
&&=-{\tilde \lambda}
_{\mu}(\tau )l^{\mu}-l^{\mu}{\tilde \lambda}_{\mu\nu}(\tau )b
^{\nu}_{\check s}(\tau ) \sigma^{\check s}=-\lambda (\tau )-
 {1\over 2} \lambda_{\tau \check s}(\tau ) \sigma^{\check s},\nonumber \\
 N_{(flat)\,
\check r}(\tau ,\vec
\sigma )&=&\lambda_{\mu}(\tau ,\vec
 \sigma )z^{\mu}_{\check r} (\tau ,\vec \sigma ) \mapsto \nonumber \\
 &&{}\nonumber \\
 &\mapsto& N_{(flat )}(\tau ,\vec \sigma )=N_{[z](flat)\check r}
(\tau ,\vec \sigma )=\nonumber \\
 &&=-{\tilde \lambda}
_{\mu}(\tau )b^{\mu}_{\check r}(\tau )-b^{\mu}_{\check r}(\tau ){\tilde
\lambda}_{(\mu )(\nu )}(\tau ) b^{\nu}_{\check s}(\tau ) \sigma^{\check s}=
-\lambda_{\check r}(\tau )-{1\over 2}\lambda_{\check r\check s}(\tau ) \sigma^{\check s},
 \nonumber \\
 &&{}\nonumber \\
 &&\lambda_A(\tau )={\tilde \lambda}_{\mu}(\tau )b^{\mu}
_A(\tau ),\quad\quad {\tilde \lambda}_{\mu}(\tau )=b^A
_{\mu}(\tau )\lambda_A(\tau ),\nonumber \\
&&\lambda_{AB}(\tau )={\tilde \lambda}_{\mu\nu}(\tau )
[b^{\mu}_{A}b^{\nu}_{B}-b^{\nu}_{A}b^{\mu}_{B}](\tau )=2[{\tilde
\lambda}_{\mu\nu } b^{\mu}_{A}b^{\nu}_{B}](\tau
),\nonumber \\
 &&{\tilde \lambda}_{\mu\nu}(\tau )={1\over
4}[b^A_{\mu}b^B_{\nu}-b^B_{\mu}b^A_{\nu}](\tau ) \lambda_{AB}(\tau
)=\nonumber \\
 &&={1\over 2} [b^A_{\mu}b^B_{\nu} \lambda_{AB}](\tau ).
\label{a13}
\end{eqnarray}

This is the main difference of the present approach from the treatment
of parametrized Minkowski theories given in standard references:
there, no configuration action is defined but only a phase space
action, in which people  use,  wrongly, $N_{[z](flat)}$,
$N_{[z](flat)\check r}$ instead of $N_{(flat)}$, $N_{(flat)\check r}$
not only on spacelike hyperplanes but also on arbitrary spacelike
hypersurfaces.

At this stage the embedding canonical variables $z^{\mu}(\tau ,\vec
\sigma )$, $\rho_{\mu}(\tau ,\vec \sigma )$ are reduced to:\hfill\break
\hfill\break
i) $x^{\mu}_s(\tau ), p^{\mu}_s$ with $\{ x^{\mu}_s,p^{\nu}_s\}
=-{}^4\eta^{\mu\nu}$, parametrizing the arbitrary origin of the coordinates on the
family of spacelike hyperplanes. The four constraints ${\cal
H}^{\mu}(\tau )
\approx p_s^{\mu} -p_{sys}^{\mu}\approx0$  mean that
$p_s^{\mu}$ is determined by the 4-momentum of the isolated system.
\hfill\break
ii) $b^{\mu}_A(\tau )$ \footnote{With the $b^{\mu}_r(\tau )$'s being
three orthogonal spacelike unit vectors generating the fixed
$\tau$-independent timelike unit normal $b^{\mu}_{\tau}=l^{\mu}$ to
the hyperplanes.} and $S^{\mu\nu}_s=-S^{\nu\mu}_s$, with the
orthonormality constraints $b^{\mu}_A\, {}^4\eta_{\mu\nu}
b^{\nu}_B={}^4\eta_{AB}$. The non-vanishing Dirac brackets enforcing
the orthonormality constraints \cite{hanson,lus} for the $b^{\mu}_A$'s
are

\bea
\{ b^{\rho}_A, S^{\mu\nu}_s \}&=&{}^4\eta^{\rho\mu} b^{\nu}_A
-{}^4\eta^{\rho\nu} b^{\mu}_A,\nonumber \\
 \{ S^{\mu\nu}_s,S^{\alpha\beta}_s \} &=& C^{\mu\nu\alpha\beta}
 _{\gamma\delta} S^{\gamma\delta}_s,
\label{a14}
\eea

\noindent with $C^{\mu\nu\alpha\beta}_{\gamma\delta}$
the structure constants of the Lorentz algebra.

Then one has that $p^{\mu}_s$, $J^{\mu\nu}_s=x
^{\mu}_sp^{\nu}_s-x^{\nu}_sp^{\mu}_s+S^{\mu\nu}_s$, satisfy
the algebra of the Poincar\'e group, with $S^{\mu\nu}_s$ playing the
role of the spin tensor. The other six constraints $ {\cal
H}^{\mu\nu}(\tau )\approx S^{\mu\nu}
_s-S^{\mu\nu}_{sys}\approx 0$ mean that
$S_s^{\mu\nu}$ coincides with the spin tensor of the isolated system.

For the velocity of the origin $x^{\mu}_s(\tau )$ we get

\begin{eqnarray}
{\dot x}_s^{\mu}(\tau ) &{\buildrel \circ \over =}\,& \{
x^{\mu}_s(\tau ), H_D \} ={\tilde \lambda}_{\nu}(\tau ) \{
x^{\mu}_s(\tau ), {\cal H}^{\mu}(\tau ) \} =\nonumber \\
 &=&-{\tilde \lambda}^{\mu}(\tau
)=[u^{\mu}(p_s)u^{\nu}(p_s)-\epsilon^{\mu}_r(u(p_s))\epsilon^{\nu}_r(u(p_s))]
{\dot x}_{s \nu}(\tau )=\nonumber \\
&=&- u^{\mu}(p_s) \lambda (\tau )+
\epsilon^{\mu}_r(u(p_s)) \lambda_r(\tau ),\nonumber \\
{\dot x}^2_s(\tau )&=& \lambda^2(\tau )-{\vec \lambda}^2(\tau ) >
0,\quad\quad {\dot x}_s\cdot u(p_s)=-\lambda (\tau ),\nonumber \\
U^{\mu}_s(\tau )&=& {{ {\dot x}^{\mu}_s(\tau )}\over {
\sqrt{\epsilon {\dot x}^2
_s(\tau )} }}={{-\lambda (\tau )u^{\mu}(p_s)+\lambda_r(\tau )\epsilon^{\mu}_r
(u(p_s))}\over {\sqrt{\lambda^2(\tau )-{\vec \lambda}^2(\tau )} }},\nonumber \\
&&{}\nonumber \\
\Rightarrow&& x^{\mu}_s(\tau )=x^{\mu}_s(0)-u^{\mu}(p_s)\int_0^{\tau}d\tau_1
\lambda (\tau_1)+\epsilon^{\mu}_r(u(p_s))\int_0^{\tau}d\tau_1\lambda_r(\tau_1).
\label{a15}
\end{eqnarray}

Let us remark that, for each configuration of an isolated system with
timelike total 4-momentum there is a privileged family of hyperplanes
(the {\it Wigner hyperplanes} orthogonal to $p^{\mu}_s$, existing when
$\epsilon p^2_s > 0$) corresponding to the {\it intrinsic rest-frame}
of the isolated system. If we choose these hyperplanes with suitable
gauge fixings for the constraints ${\tilde {\cal H}}^{\mu\nu}(\tau
)\approx 0$ \cite{lus}, we are left with the four constraints ${\cal
H}^{\mu}(\tau )\approx 0$, which can be rewritten as

\bea
\epsilon_s &=&
\sqrt{\epsilon p^2_s} \approx [invariant\, mass\, of\, the\,
isolated\, system\, under\, investigation]= M_{sys}; \nonumber \\
 &&{}\nonumber \\
 {\vec p}_{sys}&=&[3-momentum\, of\, the\, isolated\, system\, inside\, the\,
Wigner\, hyperplane]\approx 0,\nonumber \\
 &&{}\nonumber \\
 H_D &=& H_{(c)}+{\tilde \lambda}^{\mu}(\tau ) {\tilde {\cal H}}_{\mu}(\tau )+
 (system-dependent\, primary\, constraints)=\nonumber \\
 &=&H_{(c)}+\lambda (\tau )[\epsilon_s -M_{sys}]-
 \vec \lambda (\tau ) \cdot {\vec p}_{sys}+\nonumber \\
 &+&(system-dependent\, primary\, constraints).
\label{a16}
\eea

There is no more a restriction on $p_s^{\mu}$, because
$u^{\mu}_s(p_s)=p^{\mu}_s/\sqrt{\epsilon p^2_s}$ gives the orientation
of the Wigner hyperplanes containing the isolated system, with respect
to an arbitrary given external observer.

In this special gauge, after having gone to Dirac brackets we have
$b^{\mu}_A\equiv L^{\mu}{}_A(p_s,{\buildrel
\circ \over p}_s)$ (the standard Wigner boost for timelike Poincar\'e orbits),
$S_s^{\mu\nu}\equiv S_{sys}^{\mu\nu}$, ${\tilde
\lambda}_{\mu\nu}(\tau )\equiv 0$.
The origin $x^{\mu}_s(\tau )$ does not belong any more to the
canonical basis for these Dirac brackets and is replaced by the
non-covariant canonical variable\cite{lus} ${\tilde x}^{\mu}_s(\tau )=
x^{\mu}_s(\tau )-{1\over {\epsilon_s(p^o_s+\epsilon_s)}}\Big[
p_{s\nu}S_s^{\nu\mu}+\epsilon_s(S^{o\mu}_s-S^{o\nu}_s{{p_{s\nu}p_s^{\mu}}\over
{\epsilon^2_s}}) \Big]$.

In general, we have the problem that in the gauges where ${\tilde
\lambda}_{\mu\nu}(\tau )$ or ${\tilde \lambda}_{AB}(\tau )$ are
different from zero, the foliations with leaves $\Sigma_{\tau}$
associated to arbitrary 3+1 splittings of Minkowski spacetime are
geometrically {\it ill-defined} at spatial infinity so that the
variational principle describing the isolated system could make sense
only for those 3+1 splittings having these part of the Dirac's
multipliers vanishing. The problem is that, since on hyperplanes
${\dot l}^{\mu}=0$ and $l^{\mu}\, b_{\check r\mu}(\tau )=0$ imply
$l^{\mu} {\dot b}_{\check r\mu}(\tau )=0$, Eqs.(\ref{a10}) implies
$\lambda_{\tau \check r}(\tau )=0$ (i.e. only three ${\tilde
\lambda}_{\mu\nu}(\tau )$ are independent) on spacelike hyperplane,
because otherwise Lorentz boosts could generate crossing of the
foliation leaves. To avoid incosistencies this suggests to make the
reduction from arbitrary spacelike hypersurfaces either directly to
the Wigner hyperplanes or to spacelike hypersurfaces approaching
Wigner hyperplanes asymptotically
\footnote{Asymptotically, we must fix the gauge freedom generated by
the spin part of Lorentz boosts, see Eq.(\ref{a9}); how this can be
done before the restriction to spacelike hyperplanes has still to be
studied.}.

Till now, therefore, the 3+1 splittings of Minkowski spacetime whose
leaves are Wigner hyperplanes are the only ones for which the
foliation is well defined at spatial infinity: both the induced proper
time interval and shift functions are finite there.

One obtains in this way a new kind of instant form of the dynamics,
the  {\it Wigner-covariant 1-time rest-frame instant
form}\cite{lus,india}.  For any isolated system all the variables
become Wigner covariant except for the {\it external} canonical center
of mass ${\tilde x}^{\mu}_s$, which looses even Lorentz covariance.
This does not matter, however, since it is a completely decoupled
variable. This is the special relativistic generalization of the
non-relativistic separation of the center of mass from the relative
motion [$H={{ {\vec P}^2}\over {2M}}+H_{rel}$]. The role of the center
of mass is taken by the Wigner hyperplane, identified by a point
${\tilde x}^{\mu} (\tau )$ and its normal $p^{\mu}_s$.

\vfill\eject

\section{Notations on spacelike hypersurfaces.}

Let us first review some preliminary results from Refs.\cite{lus}
needed in the description of physical systems on spacelike hypersurfaces.

Let $\lbrace \Sigma_{\tau}\rbrace$ be a one-parameter family of spacelike
hypersurfaces foliating Minkowski spacetime $M^4$ with 4-metric $\eta_{\mu\nu}
=\epsilon (+---)$, $\epsilon =\pm$ \footnote{$\epsilon =+1$ is the particle physics
convention; $\epsilon =-1$ the general relativity one.} and giving a
3+1 decomposition of it. At fixed $\tau$, let $z^{\mu}(\tau ,\vec
\sigma )$ be the coordinates of the points on $\Sigma
_{\tau }$ in $M^4$, $\lbrace \vec \sigma \rbrace$ a system of coordinates on
$\Sigma_{\tau}$. If $\sigma^{\check A}=(\sigma^{\tau}=\tau ;\vec \sigma
=\lbrace \sigma^{\check r}\rbrace)$ \footnote{The notation ${\check A}=(\tau ,
{\check r})$ with ${\check r}=1,2,3$ will be used; note that ${\check A}=
\tau$ and ${\check A}={\check r}=1,2,3$ are Lorentz-scalar indices.} and
$\partial_{\check A}=\partial /\partial \sigma^{\check A}$, one can
define the cotetrads

\begin{equation}
z^{\mu}_{\check A}(\tau ,\vec \sigma )=\partial_{\check A}z^{\mu}(\tau ,\vec
\sigma ),\quad\quad
\partial_{\check B}z^{\mu}_{\check A}-\partial_{\check A}z^{\mu}_{\check B}=0,
\label {b1}
\end{equation}

\noindent so that the metric on $\Sigma_{\tau}$ is

\begin{eqnarray}
&&g_{{\check A}{\check B}}(\tau ,\vec \sigma )=z^{\mu}_{\check A}(\tau ,\vec
\sigma )\eta_{\mu\nu}z^{\nu}_{\check B}(\tau ,\vec \sigma ),\quad\quad
\epsilon g_{\tau\tau}(\tau ,\vec \sigma ) > 0,\nonumber \\
&&g(\tau ,\vec \sigma )=-det\, ||\, g_{{\check A}{\check B}}(\tau ,\vec
\sigma )\, || ={(det\, ||\, z^{\mu}_{\check A}(\tau ,\vec \sigma )\, ||)}^2,
\nonumber \\
&&\gamma (\tau ,\vec \sigma )=-det\, ||\, g_{{\check r}{\check s}}(\tau ,\vec
\sigma )\, ||=det\, ||{}^3g_{\check r\check s}(\tau ,\vec \sigma )||,
\label{b2}
\end{eqnarray}

\noindent where $g_{\check r\check s}=-\epsilon \, {}^3g_{\check r\check s}$
with ${}^3g_{\check r\check s}$ having positive signature $(+++)$.

If $\gamma^{{\check r}{\check s}}(\tau ,\vec \sigma )=-\epsilon \,
{}^3g^{\check r\check s}$ is the inverse of the
3-metric $g_{{\check r}{\check s}}(\tau ,\vec \sigma )$ [$\gamma^{{\check r}
{\check u}}(\tau ,\vec \sigma )g_{{\check u}{\check s}}(\tau ,\vec
\sigma )=\delta^{\check r}_{\check s}$], the inverse $g^{{\check A}{\check B}}
(\tau ,\vec \sigma )$ of $g_{{\check A}{\check B}}(\tau ,\vec \sigma )$
[$g^{{\check A}{\check C}}(\tau ,\vec \sigma )g_{{\check c}{\check b}}(\tau ,
\vec \sigma )=\delta^{\check A}_{\check B}$] is given by

\begin{eqnarray}
&&g^{\tau\tau}(\tau ,\vec \sigma )={{\gamma (\tau ,\vec \sigma )}\over
{g(\tau ,\vec \sigma )}},\nonumber \\
&&g^{\tau {\check r}}(\tau ,\vec \sigma )=-[{{\gamma}\over g} g_{\tau {\check
u}}\gamma^{{\check u}{\check r}}](\tau ,\vec \sigma )=\epsilon
[{{\gamma}\over g} g_{\tau \check u}\, {}^3g^{\check u\check r}](\tau ,\vec
\sigma ),\nonumber \\
&&g^{{\check r}{\check s}}(\tau ,\vec \sigma )=\gamma^{{\check r}{\check s}}
(\tau ,\vec \sigma )+[{{\gamma}\over g}g_{\tau {\check u}}g_{\tau {\check v}}
\gamma^{{\check u}{\check r}}\gamma^{{\check v}{\check s}}](\tau ,\vec \sigma )
=\nonumber \\
&=&-\epsilon \, {}^3g^{\check r\check s}(\tau ,\vec \sigma )+[{{\gamma}\over g}
g_{\tau \check u}g_{\tau \check v}\, {}^3g{\check u\check r}\, {}^3g^{\check v
\check s}](\tau ,\vec \sigma ),
\label{b3}
\end{eqnarray}

\noindent so that $1=g^{\tau {\check C}}(\tau ,\vec \sigma )g_{{\check C}\tau}
(\tau ,\vec \sigma )$ is equivalent to

\begin{equation}
{{g(\tau ,\vec \sigma )}\over {\gamma (\tau ,\vec \sigma )}}=g_{\tau\tau}
(\tau ,\vec \sigma )-\gamma^{{\check r}{\check s}}(\tau ,\vec \sigma )
g_{\tau {\check r}}(\tau ,\vec \sigma )g_{\tau {\check s}}(\tau ,\vec \sigma ).
\label{b4}
\end{equation}

We have

\begin{equation}
z^{\mu}_{\tau}(\tau ,\vec \sigma )=(\sqrt{ {g\over {\gamma}} }l^{\mu}+
g_{\tau {\check r}}\gamma^{{\check r}{\check s}}z^{\mu}_{\check s})(\tau ,
\vec \sigma ),
\label{b5}
\end{equation}

\noindent and

\begin{eqnarray}
\eta^{\mu\nu}&=&z^{\mu}_{\check A}(\tau ,\vec \sigma )g^{{\check A}{\check B}}
(\tau ,\vec \sigma )z^{\nu}_{\check B}(\tau ,\vec \sigma )=\nonumber \\
&=&(l^{\mu}l^{\nu}+z^{\mu}_{\check r}\gamma^{{\check r}{\check s}}
z^{\nu}_{\check s})(\tau ,\vec \sigma ),
\label{b6}
\end{eqnarray}

\noindent where

\begin{eqnarray}
l^{\mu}(\tau ,\vec \sigma )&=&({1\over {\sqrt{\gamma}} }\epsilon^{\mu}{}_{\alpha
\beta\gamma}z^{\alpha}_{\check 1}z^{\beta}_{\check 2}z^{\gamma}_{\check 3})
(\tau ,\vec \sigma ),\nonumber \\
&&l^2(\tau ,\vec \sigma )=1,\quad\quad l_{\mu}(\tau ,\vec \sigma )z^{\mu}
_{\check r}(\tau ,\vec \sigma )=0,
\label{b7}
\end{eqnarray}

\noindent is the unit (future pointing) normal to $\Sigma_{\tau}$ at
$z^{\mu}(\tau ,\vec \sigma )$.

For the volume element in Minkowski spacetime we have

\begin{eqnarray}
d^4z&=&z^{\mu}_{\tau}(\tau ,\vec \sigma )d\tau d^3\Sigma_{\mu}=d\tau [z^{\mu}
_{\tau}(\tau ,\vec \sigma )l_{\mu}(\tau ,\vec \sigma )]\sqrt{\gamma
(\tau ,\vec \sigma )}d^3\sigma=\nonumber \\
&=&\sqrt{g(\tau ,\vec \sigma )} d\tau d^3\sigma.
\label{b8}
\end{eqnarray}

Let us remark that, according to the geometric approach of
Ref.\cite{kuchar},one can write

\beq
z^{\mu}_{\tau}(\tau ,\vec \sigma )=N(\tau ,
\vec \sigma )l^{\mu}(\tau ,\vec \sigma )+N^{\check r}(\tau ,\vec \sigma )
z^{\mu}_{\check r}(\tau ,\vec \sigma ),
 \label{b9}
 \eeq

\noindent
where $N=\sqrt{g/\gamma}=\sqrt{g_{\tau\tau}
-\gamma^{{\check r}{\check s}}g_{\tau{\check r}}g_{\tau{\check s}}}
=\sqrt{g_{\tau\tau}+\epsilon
{}^3g^{{\check r}{\check s}}g_{\tau{\check r}}g_{\tau{\check s}}}$
and $N^{\check r}=g_{\tau \check s}\gamma^{\check s\check r}=-\epsilon
g_{\tau \check s}\, {}^3g^{\check s\check r}$ \hfill\break
\hfill\break
are the standard lapse and shift functions $N_{[z](flat)}$, $N^{\check r}
_{[z](flat)}$ of Appendix A, so that

\bea
g_{\tau \tau}&=&\epsilon N^2+g_{\check r\check s}N^{\check r}N^{\check
s}=\epsilon [N^2-{}^3g_{\check r\check s}N^{\check r}N^{\check s}],
\nonumber \\
 g_{\tau \check r}&=&g_{\check r\check s}N^{\check
s}=-\epsilon \, {}^3g_{\check r\check s}N^{\check s},\nonumber \\
 g^{\tau \tau}&=&\epsilon N^{-2}, \nonumber \\
 g^{\tau \check r}&=&-\epsilon N^{\check r}/N^2, \nonumber \\
 g^{\check r\check s}&=&\gamma^{\check r\check s}+\epsilon {{N^{\check r}N^{\check s}}\over
{N^2}}=-\epsilon [{}^3g^{\check r\check s}- {{N^{\check r}N^{\check
s}}\over {N^2}}],\nonumber \\
 &&{}\nonumber \\
 &&{{\partial}\over {\partial z^{\mu}_{\tau}}}=l_{\mu}\, {{\partial}\over
{\partial N}}+z_{{\check s}\mu}\gamma^{{\check s}{\check r}}
{{\partial}\over {\partial N^{\check r}}}=l_{\mu}\, {{\partial}\over
{\partial N}}-\epsilon z_{{\check s}\mu}\, {}^3g^{{\check s}{\check
r}} {{\partial}\over {\partial N^{\check r}}}, \nonumber \\
 &&d^4z=N\sqrt{\gamma}d\tau d^3\sigma.
 \label{b10}
 \eea

The rest frame form of a timelike four-vector $p^{\mu}$ is $\stackrel
{\circ}{p}{}^{\mu}=\eta \sqrt{\epsilon p^2} (1;\vec 0)= \eta^{\mu
o}\eta
\sqrt{\epsilon p^2}$,
$\stackrel{\circ}{p}{}^2=p^2$, where $\eta =sign\, p^o$.
The standard Wigner boost transforming $\stackrel{\circ}{p}{}^{\mu}$ into
$p^{\mu}$ is

\begin{eqnarray}
L^{\mu}{}_{\nu}(p,\stackrel{\circ}{p})&=&\epsilon^{\mu}_{\nu}(u(p))=
\nonumber \\
&=&\eta^{\mu}_{\nu}+2{ {p^{\mu}{\stackrel{\circ}{p}}_{\nu}}\over {\epsilon
p^2}}-
{ {(p^{\mu}+{\stackrel{\circ}{p}}^{\mu})(p_{\nu}+{\stackrel{\circ}{p}}_{\nu})}
\over {p\cdot \stackrel{\circ}{p} +\epsilon p^2} }=\nonumber \\
&=&\eta^{\mu}_{\nu}+2u^{\mu}(p)u_{\nu}(\stackrel{\circ}{p})-{ {(u^{\mu}(p)+
u^{\mu}(\stackrel{\circ}{p}))(u_{\nu}(p)+u_{\nu}(\stackrel{\circ}{p}))}
\over {1+u^o(p)} },\nonumber \\
&&{} \nonumber \\
\nu =0 &&\epsilon^{\mu}_o(u(p))=u^{\mu}(p)=p^{\mu}/\eta
\sqrt{\epsilon p^2}, \nonumber \\
\nu =r &&\epsilon^{\mu}_r(u(p))=(-u_r(p); \delta^i_r-{ {u^i(p)u_r(p)}\over
{1+u^o(p)} }).
\label{b11}
\end{eqnarray}

The inverse of $L^{\mu}{}_{\nu}(p,\stackrel{\circ}{p})$ is $L^{\mu}{}_{\nu}
(\stackrel{\circ}{p},p)$, the standard boost to the rest frame, defined by

\begin{equation}
L^{\mu}{}_{\nu}(\stackrel{\circ}{p},p)=L_{\nu}{}^{\mu}(p,\stackrel{\circ}{p})=
L^{\mu}{}_{\nu}(p,\stackrel{\circ}{p}){|}_{\vec p\rightarrow -\vec p}.
\label{b12}
\end{equation}

Therefore, we can define the following cotetrads and tetrads
\footnote{The $\epsilon^{\mu}_r(u(p))$'s are also called polarization
vectors; the indices r, s will be used for A=1,2,3 and $\bar o$ for
$A=o$.}

\begin{eqnarray}
&&\epsilon^{\mu}_A(u(p))=L^{\mu}{}_A(p,\stackrel{\circ}{p}),\nonumber \\
&&\epsilon^A_{\mu}(u(p))=L^A{}_{\mu}(\stackrel{\circ}{p},p)=\eta^{AB}\eta
_{\mu\nu}\epsilon^{\nu}_B(u(p)),\nonumber \\
&&{} \nonumber \\
&&\epsilon^{\bar o}_{\mu}(u(p))=\eta_{\mu\nu}\epsilon^{\nu}_o(u(p))=u_{\mu}(p),
\nonumber \\
&&\epsilon^r_{\mu}(u(p))=-\delta^{rs}\eta_{\mu\nu}\epsilon^{\nu}_r(u(p))=
(\delta^{rs}u_s(p);\delta^r_j-\delta^{rs}\delta_{jh}{{u^h(p)u_s(p)}\over
{1+u^o(p)} }),\nonumber \\
&&\epsilon^A_o(u(p))=u_A(p),
\label{b13}
\end{eqnarray}

\noindent which satisfy

\begin{eqnarray}
&&\epsilon^A_{\mu}(u(p))\epsilon^{\nu}_A(u(p))=\eta^{\mu}_{\nu},\nonumber \\
&&\epsilon^A_{\mu}(u(p))\epsilon^{\mu}_B(u(p))=\eta^A_B,\nonumber \\
&&\eta^{\mu\nu}=\epsilon^{\mu}_A(u(p))\eta^{AB}\epsilon^{\nu}_B(u(p))=u^{\mu}
(p)u^{\nu}(p)-\sum_{r=1}^3\epsilon^{\mu}_r(u(p))\epsilon^{\nu}_r(u(p)),
\nonumber \\
&&\eta_{AB}=\epsilon^{\mu}_A(u(p))\eta_{\mu\nu}\epsilon^{\nu}_B(u(p)),\nonumber
\\
&&p_{\alpha}{{\partial}\over {\partial p_{\alpha}} }\epsilon^{\mu}_A(u(p))=
p_{\alpha}{{\partial}\over {\partial p_{\alpha}} }\epsilon^A_{\mu}(u(p))
=0.
\label{b14}
\end{eqnarray}

The Wigner rotation corresponding to the Lorentz transformation $\Lambda$ is

\begin{eqnarray}
R^{\mu}{}_{\nu}(\Lambda ,p)&=&{[L(\stackrel{\circ}{p},p)\Lambda^{-1}L(\Lambda
p,\stackrel{\circ}{p})]}^{\mu}{}_{\nu}=\left(
\begin{array}{cc}
1 & 0 \\
0 & R^i{}_j(\Lambda ,p)
\end{array}
\right) ,\nonumber \\
{} && {}\nonumber \\
R^i{}_j(\Lambda ,p)&=&{(\Lambda^{-1})}^i{}_j-{ {(\Lambda^{-1})^i{}_op_{\beta}
(\Lambda^{-1})^{\beta}{}_j}\over {p_{\rho}(\Lambda^{-1})^{\rho}{}_o+\eta
\sqrt{\epsilon p^2}} }-\nonumber \\
&-&{{p^i}\over {p^o+\eta \sqrt{\epsilon p^2}} }[(\Lambda^{-1})^o{}_j-
{ {((\Lambda^{-1})^o
{}_o-1)p_{\beta}(\Lambda^{-1})^{\beta}{}_j}\over {p_{\rho}(\Lambda^{-1})^{\rho}
{}_o+\eta \sqrt{\epsilon p^2}} }].
\label{b15}
\end{eqnarray}

The polarization vectors transform under the
Poincar\'e transformations $(a,\Lambda )$ in the following way

\begin{equation}
\epsilon^{\mu}_r(u(\Lambda p))=(R^{-1})_r{}^s\, \Lambda^{\mu}{}_{\nu}\,
\epsilon^{\nu}_s(u(p)).
\label{b16}
\end{equation}

\vfill\eject

\section{The Gartenhaus-Schwartz Transformation for Spinning Particles.}

In Ref.\cite{biga} there is the rest-frame instant form description of
a system of N spinning positive-energy particles with the intrinsic
spin described by Grassmann variables.

On the Wigner hyperplane the N spinning particles are described by a
canonical basis containing the center-of-mass variables ${\tilde
x}^{\mu}_s$, $p^{\mu}_s$, the pairs ${\vec \eta}_i$, ${\vec
\kappa}_i$, $i=1,..,N$, of Eq.(\ref{II16}) and three Grassmann variables
for each spin, $\xi^r_i \equiv
\epsilon_{\mu}^r(u(p_s))\xi^{\mu}_i$ \cite{biga} satisfying $\{ \xi^r_i,
\xi^s_j \} = =-i \delta^{rs}\delta_{ij}$ and having vanishing Poisson bracket
with all the other variables.

The rest-frame {\it external} realization of the Poincar\'e algebra is
built  in analogy to Eq.(\ref{II14}) but with a modified spin tensor
 ${\tilde S}_s^{\mu\nu}$

\begin{eqnarray}
{\vec {\bar S}}_s&=&\sum_{i=1}^N \Big( {\vec \eta}_i \times {\vec \kappa}_i +
{\vec {\bar S}}_{i\xi}\Big) ,\nonumber \\
&&{}\nonumber \\
{\bar S}^r_{i\xi} &=& -{i\over 2} \epsilon^{ruv}\xi^u_i\xi^v_i,\quad\quad
\{ {\bar S}^r_{i\xi}, {\bar S}^s_{j\xi} \} =\delta_{ij}\epsilon^{rsu}{\bar S}^u
_{i\xi}.
\label{c1}
\end{eqnarray}

In absence of interactions Eqs.(\ref{II7}), (\ref{II8}), (\ref{II16})
remain valid.

By using the expression of ${\bar S}^{\mu\nu}_s$ on the Wigner
hyperplane given in Ref.\cite{biga} and the methodology of
Ref.\cite{pauri}, the {\it internal} realization (\ref{II15}) of the
Poincar\'e algebra becomes

\begin{eqnarray}
H_M&=& M_{sys} =\sum_{i=1}^N \sqrt{m^2_i+{\vec \kappa}_i^2}=
\sum_{i=1}^N H_i,
\quad\quad H_i= \sqrt{m_i^2+{\vec \kappa}_i^2},\nonumber \\
{\vec \kappa}_{+}&=&\sum_{i=1}^N{\vec \kappa}_i\quad (\approx 0),\nonumber \\
\vec J&=&{\vec {\bar S}}_s=\sum_{i=1}^N \Big( {\vec \eta}_i \times {\vec
\kappa}_i + {\vec {\bar S}}_{i\xi}\Big) ={\vec J}_B+{\vec J}_S,\quad\quad
{\vec J}_S=\sum_{i=1}^N{\vec {\bar S}}_{i\xi},\nonumber \\
\vec K&=&-\sum_{i=1}^N{\vec \eta}_i H_i+\sum_{i=1}^N{{ {\vec {\bar S}}_{i\xi}
\times {\vec \kappa}_i}\over {m_i+H_i}}={\vec K}_B+{\vec K}_s,\quad\quad
{\vec K}_B=-\sum_{i=1}^N{\vec \eta}_i H_i.
\label{c2}
\end{eqnarray}

Following Ref.\cite{pauri}, if we put

\begin{eqnarray}
\vec K&=&-H_M{\vec q}_{+} +{{ {\vec S}_q\times {\vec \kappa}_{+}}\over
{\sqrt{\Pi}+H_M}},\quad\quad \Pi =H^2_M-{\vec \kappa}_{+}^2,\nonumber \\
&&{}\nonumber \\
{\vec S}_q&=&\vec J-{\vec q}_{+} \times {\vec \kappa}_{+},
\label{c3}
\end{eqnarray}

\noindent we get consistently the expression of the {\it internal} canonical
center of mass ${\vec q}_{+}$ given in Eq.(\ref{IV4}), with ${\vec
S}_q$ the associated spin vector.

As in Section VI, let us apply the Gartenhaus-Schwartz transformation
to go from the internal canonical basis ${\vec \eta}_i$, ${\vec
\kappa}_i$, ${\vec \xi}_i$ (with ${\vec \kappa}_{+}\approx 0$)
to the center-of-mass basis ${\vec q}
_{+}$, ${\vec \kappa}_{+}$, ${\vec \rho}_{qa}$, ${\vec \pi}_{qa}$, ${\vec \xi}
_{qi}$ (again with ${\vec \kappa}_{+}\approx 0$) with ${\vec {\bar S}}_{i\xi}
\mapsto {\vec S}_{qi\xi}$. By using Eqs.(\ref{V1}) and (\ref{V4}) with
$\vec K={\vec K}_B+{\vec K}_S$ of Eq.(\ref{c2}), we can find the
differential equations for the $\alpha$-dependence of the various
quantities.

Since $\{ {\vec \pi}_a,{\vec K}_S \} =0$, ${\vec \pi}_{qa}=lim_{\alpha
\rightarrow \infty}\,\, {\vec \pi}_a(\alpha )$ is the same as in the spinless
case; the ${\vec \pi}_{qa}$'s are given in Eqs.(\ref{V13}). As in the
spinless case $\Pi$ is an invariant and we have $\sqrt{\Pi}=H_M(\infty
)=H_{(rel)}=\sum
_{i=1}^NH_i(\infty )$ with $H_i(\infty )=(H_MH_i-{\vec \kappa}_{+}\cdot {\vec
\kappa}_i)/ \sqrt{\Pi}$. Moreover, ${\vec n}_{+}={\vec \kappa}_{+}/
|{\vec \kappa}_{+}|$ is invariant.

For the spin variables ${\vec {\bar S}}_{i\xi}$ we get

\begin{eqnarray}
{{d{\bar S}^r_{i\xi}(\alpha )}\over {d\alpha}}&=& \{ {\bar S}^r_{i\xi}(\alpha ),
{\vec q}_{+}(\alpha )\cdot {\vec \kappa}_{+}(\alpha ) \} = \{ {{\vec K(\alpha )
\cdot {\vec \kappa}_{+}(\alpha )}\over {H_M(\alpha )}},{\bar S}^r_{i\xi} \} =
\nonumber \\
&=&{{\kappa^s_{+}(\alpha )}\over {H_M(\alpha )}} \{ K^s_S(\alpha ), {\bar S}^r
_{i\xi}(\alpha ) \} = {{ [ \Big({\vec \kappa}_{+}(\alpha )\times {\vec \kappa}
_i(\alpha )\Big) \times {\vec {\bar S}}_{i\xi}(\alpha ) ]^r}\over {H_M(\alpha )
(m_i+H_i(\alpha ))}},\nonumber \\
&&{}\nonumber \\
\Rightarrow&& {{d{\vec {\bar S}}_{i\xi}(\alpha )}\over {d\theta (\alpha )}}=
{{ \Big( {\vec n}_{+}\times {\vec \kappa}_i(\alpha )\Big) \times {\vec {\bar
S}}_{i\xi}(\alpha )}\over {m_i+H_i(\alpha )}}.
\label{c4}
\end{eqnarray}

This equation coincides with Eq.(3.10) of Ref.\cite{osborne}. By using
Eq.(3.11) of Ref.\cite{osborne}, its integration provides {\it Thomas
precession} of the spin variable ${\vec {\bar S}}_{i\xi}$ about an
axis ${\vec \kappa}_i\times {\vec n}_{+}$ in the instantaneous
center-of-mass frame\cite{ritus}:

\begin{eqnarray}
{\vec {\bar S}}_{i\xi}(\alpha )&=&cos\, \gamma (\alpha ) {\vec {\bar S}}_{i\xi}
+[1-cos\, \gamma (\alpha )] ({\vec v}_i\cdot {\vec {\bar S}}_{i\xi})
{\vec v}_i-sin\, \gamma (\alpha ) {\vec v}_i \times {\vec {\bar S}}
_{i\xi},\nonumber \\
&&{}\nonumber \\
{\vec v}_i&=& {{ {\vec \kappa}_i\times {\vec n}_{+}}\over {|{\vec \kappa}_i
\times {\vec n}_{+}|}},\nonumber \\
tg\, {{\gamma (\alpha )}\over 2}&=&{{1-cos\, \gamma (\alpha )}\over {sin\,
\gamma (\alpha )}}={{ |{\vec \kappa}_i\times {\vec n}_{+}|}\over {(m_i+H_i)
ctgh\, {{\theta (\alpha )}\over 2}-{\vec \kappa}_i\cdot {\vec n}_{+}}}
\nonumber \\
&&{\rightarrow}_{\alpha \rightarrow \infty}\,\, {{|{\vec \kappa}_i\times {\vec
\kappa}_{+}|}\over {(m_i+H_i)(H_M+\sqrt{\Pi})-{\vec \kappa}_i\cdot {\vec
\kappa}_{+}}}\approx 0,\quad [tg\, {{\gamma (\infty )}\over 2}={{H_M+
\sqrt{\Pi}}\over {|{\vec \kappa}_{+}|}}],\nonumber \\
sin\, \gamma (\alpha )&=&2 {{(m_i+H_i)ctgh\, {{\theta (\alpha )}\over 2}-
{\vec \kappa}_i\cdot {\vec n}_{+}}\over {|{\vec \kappa}_i\times {\vec n}_{+}|^2+
[(m_i+H_i)ctgh\, {{\theta (\alpha )}\over 2}-{\vec \kappa}_i\cdot {\vec n}_{+}]
^2}} |{\vec \kappa}_i\times {\vec n}_{+}|,\nonumber \\
cos\, \gamma (\alpha )&=&1-2{{|{\vec \kappa}_i\times {\vec n}_{+}|^2}\over
{|{\vec \kappa}_i\times {\vec n}_{+}|^2+[(m_i+H_i)ctgh\,
{{\theta (\alpha )}\over 2}-{\vec \kappa}_i\cdot {\vec n}_{+}]^2}}.
\label{c5}
\end{eqnarray}

Therefore, we obtain

\begin{eqnarray}
{\vec S}_{qi\xi}&=& lim_{\alpha \rightarrow \infty}\,\, {\vec {\bar S}}
_{i\xi}(\alpha )=\Big[ 1-{{|{\vec \kappa}_i\times {\vec \kappa}_{+}|^2}\over
{(m_i+H_i)(m_i+H_i(\infty ))(H_M+\sqrt{\Pi})\sqrt{\Pi} }}\Big] {\vec {\bar S}}
_{i\xi}+\nonumber \\
&+&{{ {\vec \kappa}_i\times {\vec \kappa}_{+}\cdot {\vec {\bar S}}_{i\xi}
{\vec \kappa}_i\times {\vec \kappa}_{+} }\over {(m_i+H_i)(m_i+H_i(\infty
))(H_M+\sqrt{\Pi})\sqrt{\Pi} }}-\nonumber \\
&-& {{(m_i+H_i)(H_M+\sqrt{\Pi})-{\vec \kappa}_i\cdot {\vec \kappa}_{+}}\over
{(m_i+H_i)(m_i+H_i(\infty ))(H_M+\sqrt{\Pi})\sqrt{\Pi} }} ({\vec \kappa}_i
\times {\vec \kappa}_{+})\times {\vec {\bar S}}_{i\xi} \approx {\vec {\bar S}}
_{i\xi}.
\label{c6}
\end{eqnarray}

For the Grassmann variables ${\vec \xi}_i$, we get the same
differential equation

\begin{eqnarray}
{{d\xi^r_i(\alpha )}\over {d\alpha}}&=& \{ \xi^r_i(\alpha ), {\vec q}
_{+}(\alpha )\cdot {\vec \kappa}_{+}(\alpha ) \} = {{[\Big( {\vec \kappa}
_{+}(\alpha )\times {\vec \kappa}_i(\alpha )\Big) \times {\vec \xi}_i]^r}\over
{H_M(\alpha )(m_i+H_i(\alpha ))}},\nonumber \\
&&{}\nonumber \\
\Rightarrow&& {{d{\vec \xi}_i(\alpha )}\over {d\theta (\alpha )}}= {{\Big(
{\vec n}_{+}\times {\vec \kappa}_i(\alpha )\Big) \times {\vec \xi}_i(\alpha )
}\over {m_i+H_i(\alpha )}},\nonumber \\
&&\Downarrow \nonumber \\
{\vec \xi}_{qi}&=&lim_{\alpha \rightarrow \infty }\,\, {\vec \xi}_i(\alpha )=
\Big[ 1-{{|{\vec \kappa}_i\times {\vec \kappa}_{+}|^2}\over
{(m_i+H_i)(m_i+H_i(\infty ))(H_M+\sqrt{\Pi})\sqrt{\Pi} }}\Big] {\vec \xi}
_i+\nonumber \\
&+&{{ {\vec \kappa}_i\times {\vec \kappa}_{+}\cdot {\vec \xi}_i
{\vec \kappa}_i\times {\vec \kappa}_{+} }\over {(m_i+H_i)(m_i+H_i(\infty
))(H_M+\sqrt{\Pi})\sqrt{\Pi} }}-\nonumber \\
&-& {{(m_i+H_i)(H_M+\sqrt{\Pi})-{\vec \kappa}_i\cdot {\vec \kappa}_{+}}\over
{(m_i+H_i)(m_i+H_i(\infty ))(H_M+\sqrt{\Pi})\sqrt{\Pi} }} ({\vec \kappa}_i
\times {\vec \kappa}_{+})\times {\vec \xi}_i \approx {\vec \xi}
_i.
\label{c7}
\end{eqnarray}

Let us remark that, now, as one  can easily check, besides the
invariants $I^{(1)}_i$ and $I^{(2)}$ of Eqs. (\ref{V14}) of the
spinless case (we have $I^{(2)}=I_B^{(2)}+I_S^{(2)}$ since $\vec
K={\vec K}_B+{\vec K}_S$) there are also the following invariants

\begin{eqnarray}
I^{(3)}_i&=&({\vec \kappa}_i\times {\vec n}_{+}) \cdot {\vec {\bar S}}_{i\xi},
\nonumber \\
&&{}\nonumber \\
I^{(2)}&=&I^{(2)}_B+I^{(2)}_S,\quad\quad I^{(2)}_S={{|{\vec \kappa}_{+}|}\over
{H_M}}\sum_{i=1}^N {{I^{(3)}_i}\over {m_i+H_i}}.
\label{c8}
\end{eqnarray}

Let us now consider the position vectors. Like in the spinless case
the preliminary calculations for Eq.(\ref{V19}), now give

\begin{eqnarray}
{d\over {d\alpha}} {\vec n}_{+}\cdot {\vec \eta}_i(\alpha )&=&
-{{{\vec n}_{+}\cdot {\vec \eta}_i(\alpha )}\over {H_i(\alpha )}}
{{dH_i(\alpha )}\over {d\alpha}}- {{I^{(1)}_i I^{(2)}}\over {H_i(\alpha )}}
{{dJ^{(1)}(\alpha )}\over {d\alpha}}-\nonumber \\
&-&{1\over {H_i(\alpha )}} {{dH_i(\alpha )}\over {d\alpha}} {{I^{(3)}_i}\over
{(m_i+H_i(\alpha ))^2}}.
\label{c9}
\end{eqnarray}

These equations have the solution

\begin{eqnarray}
{\vec n}_{+}\cdot {\vec \eta}_i(\alpha )&=&
{{H_i}\over {H_i(\alpha )}} {\vec n}_{+}\cdot {\vec \eta}_i -{{I^{(2)}}
\over {|{\vec \kappa}_{+}|}} ( e^{\alpha}- {{H_i}\over {H_i(\alpha )}})+
\nonumber \\
&+&{{I^{(3)}_i}\over {H_i(\alpha )}} \Big( {1\over {m_i+H_i(\alpha )}}-
{1\over {m_i+H_i}}\Big).
\label{c10}
\end{eqnarray}

For ${\vec \eta}_i(\alpha )$ we have

\begin{eqnarray}
{{d{\vec \eta}_i(\alpha )}\over {d\alpha}}&=& \{ {\vec \eta}_i(\alpha ),
{\vec \kappa}_{+}(\alpha )\cdot {\vec q}_{+}(\alpha ) \} =-n^s_{+} {{\partial}
\over{\partial {\vec k}_i(\alpha )}} {{|{\vec \kappa}_{+}(\alpha )|
K^s(\alpha )}\over {H_M(\alpha )}}=\nonumber \\
&=& {\vec n}_{+}\cdot {\vec \eta}_i(\alpha ) {{|{\vec \kappa}_{+}(\alpha )|
{\vec k}_i(\alpha )}\over {H_i(\alpha ) H_M(\alpha )}}+\nonumber \\
&+& {{ \sum_{j=1}^N H_j(\alpha ) {\vec n}_{+}\cdot {\vec \eta}_j(\alpha )}\over
{H_M(\alpha )}}
\Big[ {{|{\vec \kappa}_{+}(\alpha )| {\vec k}_i(\alpha )}\over {H_i(\alpha )
H_M(\alpha )}}-{\vec n}_{+}\Big]+\nonumber \\
&+&{{{\vec n}_{+}\cdot \Big( {\vec {\bar S}}_{i\xi}(\alpha )\times {\vec
\kappa}_i(\alpha )\Big) |{\vec \kappa}_{+}(\alpha )| {\vec \kappa}_i(\alpha )}
\over { H_M(\alpha )H_i(\alpha )(m_i+H_i(\alpha ))^2}} -{{|{\vec \kappa}
_{+}(\alpha )| {\vec n}_{+}\times {\vec {\bar S}}_{i\xi}(\alpha )}\over
{H_M(\alpha )(m_i+H_i(\alpha ))}}=\nonumber \\
&=&{{H_i|{\vec \kappa}_{+}(\alpha )| {\vec \kappa}_i(\alpha )}\over
{H_i^2(\alpha )H_M(\alpha )}} \Big[ {\vec n}_{+}\cdot {\vec \eta}_i+{{I^{(2)}}
\over {|{\vec \kappa}_{+}|}} \Big]-{{{\vec n}_{+}\cdot \vec K(\alpha )
{\vec n}_{+}}\over {H_M(\alpha )}}+\nonumber \\
&+&{{{\vec \kappa}_i(\alpha )|{\vec \kappa}_{+}(\alpha )|}\over {H_i(\alpha )
H_M(\alpha )}} {{{\vec n}_{+}\cdot \Big( {\vec {\bar S}}_{i\xi}(\alpha )\times
{\vec \kappa}_i(\alpha )\Big)}\over {(m_i+H_i(\alpha ))^2}}+\nonumber \\
&+&{{I^{(3)}_i}\over {H_i(\alpha )}} \Big[ {1\over {m_i+H_i(\alpha )}}-{1\over
{m_i+H_i}}\Big] -{{|{\vec \kappa}_{+}(\alpha )| {\vec n}_{+}\times {\vec {\bar
S}}_{i\xi}(\alpha )}\over {H_M(\alpha )(m_i+H_i(\alpha ))}}.
\label{c11}
\end{eqnarray}

The equations for ${\vec \rho}_a(\alpha )=\sqrt{N} \sum_{i=1}^N\gamma
_{ai}{\vec \eta}_i(\alpha )$ are [see Eq.(3.21) of Ref.\cite{osborne}]

\begin{eqnarray}
{{d{\vec \rho}^a(\alpha )}\over {d\alpha}}&=& \sqrt{N}\sum_{i=1}^N\gamma_{ai}
{{d{\vec \eta}_i(\alpha )}\over {d\alpha}}=\nonumber \\
&=&\sqrt{N}\sum_{i=1}^N\gamma_{ai} {{H_i |{\vec \kappa}_{+}(\alpha )| {\vec
\kappa}_i(\alpha )}\over {H^2_i(\alpha )}} ({\vec n}_{+}\cdot {\vec \eta}_i +
{{I^{(2)}}\over {|{\vec \kappa}_{+}|}})+\nonumber \\
&+&\sqrt{N}\sum_{i=1}^N\gamma_{ai} {{{\vec \kappa}_i(\alpha )}\over
{H_i(\alpha )}} \Big[ {{|{\vec \kappa}_{+}(\alpha )|}\over {H_M(\alpha )}}
{{{\vec n}_{+}\cdot \Big( {\vec {\bar S}}_{i\xi}\times {\vec \kappa}_i(\alpha )
\Big) }\over {(m_i+H_i(\alpha ))^2}}+\nonumber \\
&+&{{I^{(3)}_i |{\vec \kappa}_{+}(\alpha )|}\over {H_i H_M(\alpha )}} ({1\over
{m_i+H_i(\alpha )}}-{1\over {m_i+H_i}})\Big] +\nonumber \\
&+&\sqrt{N}\sum_{i=1}^N \gamma_{ai} {{ |{\vec \kappa}_{+}(\alpha )| {\vec n}_{+}
\times {\vec {\bar S}}_{i\xi}(\alpha )}\over {H_M(\alpha ) (m_i+H_i(\alpha ))}}.
\label{c12}
\end{eqnarray}

By using the results contained in Ref.\cite{osborne} this equation can be
integrated with the final result

\begin{eqnarray}
{\vec \rho}_a(\alpha )&=&{\vec \rho}_a -\sum_{i,j=1}^N \sum_{b=1}^{N-1} \gamma
_{aj}(\gamma_{bi}-\gamma_{bj}){{H_iH_j}\over {H_M}} {\vec J}^{(2)}_j(\alpha )
{\vec n}_{+}\cdot {\vec \rho}_b+\nonumber \\
&+&\sqrt{N} \sum_{i=1}^N{{I^{(3)}_i}\over {H_M(m_i+H_i)}}\sum_{j=1}^N\gamma
_{aj} {{{\vec k}_j(\alpha )}\over {H_j(\alpha )}} sh\, \theta (\alpha )+
\nonumber \\
&+&\sqrt{N}\sum_{i=1}^N {{\gamma_{ai}}\over {(m_i+H_i(\alpha ))(m_i+H_i)}}
\Big[ {{I^{(3)}_i ({\vec \kappa}_i-|{\vec n}_{+}\cdot {\vec \kappa}_i| {\vec
n}_{+} sh\, \theta (\alpha )}\over {H_i(\alpha )}}+\nonumber \\
&+&{{I^{(3)}_i [H_i-H_i(\alpha )]{\vec n}_{+}}\over {H_i(\alpha )}}+[ch\,
\theta (\alpha )-1] {\vec n}_{+}\cdot {\vec {\bar S}}_{i\xi} {\vec \kappa}_i
\times {\vec n}_{+} -\nonumber \\
&-&sh\, \theta (\alpha ) {{(m_i+H_i)[ch\, \theta (\alpha )+1]-{\vec n}_{+}\cdot
{\vec k}_i sh\, \theta (\alpha )}\over {ch\, \theta (\alpha ) +1}} \Big] .
\label{c13}
\end{eqnarray}

Then we get

\begin{eqnarray}
{\vec \rho}_{qa}&=& lim_{\alpha \rightarrow \infty}\,\, {\vec \rho}_a(\alpha )=
{\vec \rho}_a-\nonumber \\
&-&\sum_{i,j=1}^N \sum_{b=1}^{N-1} \gamma
_{aj}(\gamma_{bi}-\gamma_{bj}){{H_i}\over {H_M}} \Big[ {{|{\vec \kappa}_{+}|
{\vec \kappa}_j(\infty )}\over {H_j(\infty ) \sqrt{\Pi}}}+({{H_M}\over
{\sqrt{\Pi}}}-1) {\vec n}_{+}\Big] {\vec n}_{+}\cdot {\vec \rho}_b
+\nonumber \\
&+&\sqrt{N} \sum_{i=1}^N {{I^{(3)}_i}\over {H_M(m_i+H_i)}}\sum_{j=1}^N\gamma
_{aj}{{|{\vec \kappa}_{+}| {\vec \kappa}_j(\infty )}\over {H_j(\infty )
\sqrt{\Pi}}}+\nonumber \\
&+&\sqrt{N} \sum_{i=1}^N {{\gamma_{ai}}\over {(m_i+H_i)(m_i+H_i(\infty ))}}
\Big[ {{|{\vec \kappa}_{+}| I^{(3)}_i}\over {H_i(\infty )\sqrt{\Pi}}} ({\vec
\kappa}_i-{\vec n}_{+}\cdot {\vec \kappa}_i {\vec n}_{+})+\nonumber \\
&+&{{I^{(3)}_i(H_i-H_i(\infty ))}\over {H_i(\infty )}} +{{(H_M-\sqrt{\Pi})
{\vec n}_{+}\cdot {\vec {\bar S}}_{i\xi}}\over {\sqrt{\Pi}}} {\vec \kappa}_i
\times {\vec n}_{+}-\nonumber \\
&-&{{|{\vec \kappa}_{+}|}\over {\sqrt{\Pi}}} {{(m_i+H_i)(H_M+\sqrt{\Pi})-
|{\vec \kappa}_{+}| {\vec n}_{+}\cdot {\vec \kappa}_i}\over {H_M-\sqrt{\Pi}}}
{\vec n}_{+} \times {\vec {\bar S}}_{i\xi} \Big] \approx {\vec \rho}_a.
\label{c14}
\end{eqnarray}

In the same way as in the spinless case we obtain

\begin{eqnarray}
{\vec J}_S(\alpha )&=&\sum_{i=1}^N \Big[ {\vec \eta}_i(\alpha )\times {\vec
\kappa}_i(\alpha ) +{\vec {\bar S}}_{i\xi}(\alpha )\Big] \nonumber \\
&&{}\nonumber \\ &&\rightarrow_{\alpha \rightarrow \infty}\,\, {\vec
S}_q=\sum_{a=1}^{N-1} {\vec
\rho}_{qa}\times {\vec \pi}_{qa}+ \sum_{i=1}^N {\vec S}_{qi\xi}.
\label{c15}
\end{eqnarray}

\vfill\eject

\section{Euler Angles.}

Let us denote by $\tilde \alpha$, $\tilde \beta$, $\tilde
\gamma$ the Euler angles chosen as orientation variables $\theta^{\alpha}$.

Let ${\hat f}_1=\hat i$, ${\hat f}_2=\hat j$, ${\hat f}_3=\hat k$ be
the unit 3-vectors along the axes of the space frame and ${\hat
e}_1=\hat \chi$, ${\hat e}_2=\hat N\times \hat \chi$, ${\hat e}_3=\hat
N$, the unit 3-vectors along the axes of a {\it body frame}. Then we
have

\begin{eqnarray}
{\vec S}_q&=& S^r_q {\hat f}_r =R^{rs}(\tilde \alpha, \tilde \beta,
\tilde \gamma ) {\check S}^s_q {\hat f}_r={\check S}^s_q {\hat e}_s,
\nonumber \\
 &&{\hat e}_s= (R^T)^{sr}(\tilde \alpha, \tilde \beta, \tilde \gamma )
 {\hat f}_r={\cal R}_s{}^r(\tilde \alpha, \tilde \beta,
\tilde \gamma ) {\hat f}_r.
\label{d1}
\end{eqnarray}

There are two main conventions for the definition of the Euler angles
$\tilde \alpha$, $\tilde \beta$, $\tilde \gamma$.

A) The {\it y-convention} (see Refs.\cite{c1} (Appendix B) and
\cite{c2}):\hfill\break
 i) perform a first rotation of an angle $\tilde \alpha$
around ${\hat f}_3$ [${\hat f}_1 \mapsto {\hat e}^{'}_1=cos\, \tilde
\alpha {\hat f}_1 +sin\, \tilde \alpha {\hat f}_2$, ${\hat f}_2
\mapsto {\hat e}^{'}_2=-sin\, \tilde \alpha {\hat f}_1+cos\, \tilde \alpha {\hat f}_2$,
${\hat f}_3 \mapsto {\hat e}^{'}_3={\hat f}_3$];\hfill\break
 ii) perform a second rotation of an angle $\tilde \beta$ around ${\hat e}^{'}_2$
 [${\hat e}^{'}_1 \mapsto {\hat e}^{"}_1=cos\, \tilde \beta {\hat e}^{'}_1
 -sin\, \tilde \beta {\hat e}^{'}_3$, ${\hat e}^{'}_2 \mapsto {\hat e}^{"}_2={\hat e}^{'}_2$,
 ${\hat e}^{'}_3 \mapsto {\hat e}^{"}_3=sin\, \tilde \beta {\hat e}^{'}_1+
 cos\, \tilde \beta {\hat e}^{'}_3$];\hfill\break
 iii) perform a third rotation of an angle $\tilde \gamma$ around ${\hat e}^{"}_3$
 [${\hat e}^{"}_1 \mapsto {\hat e}_1=cos\, \tilde \gamma {\hat e}^{"}_1+
 sin\, \tilde \gamma {\hat e}^{"}_2$, ${\hat e}^{"}_2 \mapsto {\hat e}_2=
 -sin\, \tilde \gamma {\hat e}^{"}_1+cos\, \tilde \gamma {\hat e}^{"}_2$].
 In this way one gets

\begin{eqnarray}
&&\left( \begin{array}{c} \hat \chi \\ \hat N\times \hat \chi \\ \hat
N \end{array} \right) \equiv \left( \begin{array}{c} {\hat e}_1 \\
{\hat e}_2
\\ {\hat e}_3 \end{array} \right) ={\cal R}(\tilde \alpha ,\tilde
\beta ,\tilde \gamma ) \left( \begin{array}{c} {\hat f}_1 \\ {\hat f}_2 \\
{\hat f}_3 \end{array} \right) ,\nonumber \\
 &&{}\nonumber \\
 &&{\cal R}_r{}^s(\tilde \alpha ,\tilde \beta ,\tilde \gamma )=R^{T rs}(\tilde \alpha ,\tilde
\beta ,\tilde \gamma )=\nonumber \\
 &=&\left( \begin{array}{ccc} cos\, \tilde \gamma cos\, \tilde \beta
cos\, \tilde \alpha -sin\, \tilde \gamma sin\, \tilde \alpha & cos\,
\tilde \gamma cos\, \tilde \beta sin\, \tilde \alpha +sin\, \tilde \gamma cos\, \tilde \alpha &
-cos\, \tilde \gamma sin\, \tilde \beta \\
-(sin\, \tilde \gamma cos\, \tilde \beta cos\, \tilde \alpha +cos\, \tilde \gamma
sin\, \tilde \alpha & -sin\, \tilde \gamma cos\, \tilde \beta sin\,
\tilde \alpha +cos\, \tilde \gamma cos\, \tilde \alpha & sin\, \tilde \gamma
sin\, \tilde \beta \\ sin\, \tilde \beta cos\, \tilde \alpha & sin\,
\tilde \beta sin\, \tilde \alpha & cos\, \tilde \beta \end{array} \right) ,\nonumber \\
 &&{}\nonumber \\
 &&with\nonumber \\
 &&{}\nonumber \\
 tg\, \tilde \alpha &=& {{ {\hat N}^2}\over {{\hat N}^1}},\nonumber \\
 cos\, \tilde \beta &=& {\hat N}^3,\nonumber \\
 tg\, \tilde \gamma &=& - {{ {\hat \chi}^3}\over {(\hat N\times \hat \chi )^3}}.
\label{d2}
\end{eqnarray}

Since   $\hat N$ and $\hat \chi$ are functions of ${\vec \rho}_{qa}$
only, see Eq.(\ref{VII16}), it follows $\{ \tilde \alpha ,\tilde
\beta \} = \{ \tilde \beta ,\tilde \gamma \} = \{ \tilde \gamma , \tilde
\alpha \} =0$.

 B) The {\it x-convention} (see Refs.\cite{c3}, \cite{c1} (in the text) and
 \cite{pauri1}):
 the Euler angles $\theta$, $\varphi$ and $\psi$ are: i) $\theta =\tilde \beta$;
 ii) $cos\, \varphi =-sin\, \tilde \alpha$, $sin\, \varphi =cos\, \tilde \alpha$;
 iii) $cos\, \psi =sin\, \tilde \gamma$, $sin\, \psi =-cos\, \tilde \gamma$.

 We  use the {\it y-convention}. Following Ref.\cite{pauri1}, let us introduce
 the canonical momenta $p_{\tilde \alpha}$, $p_{\tilde \beta}$, $p_{\tilde
 \gamma}$ conjugated to $\tilde \alpha$, $\tilde \beta$, $\tilde \gamma$:
 $\{ \tilde \alpha ,p_{\tilde \alpha} \} = \{ \tilde \beta ,p_{\tilde
 \beta} \} = \{ \tilde \gamma ,p_{\tilde \gamma} \} =1$ (note that this Darboux chart
 does not exist globally). Then, the results of Ref.\cite{pauri1} imply

 \begin{eqnarray}
S^1_q&=& -sin\, \tilde \alpha p_{\tilde \beta} +{{cos\, \tilde
\alpha}\over {sin\, \tilde \beta}} p_{\tilde \gamma} -cos\, \tilde \alpha ctg\,
\tilde \beta p_{\tilde \alpha},\nonumber \\
 S^2_q&=& cos\, \tilde \alpha p_{\tilde \beta} +{{sin\, \tilde \alpha}\over
 {sin\, \tilde \beta}} p_{\tilde \gamma} - sin\, \tilde \alpha
 ctg\, \tilde \beta p_{\tilde \alpha},\nonumber \\
 S^3_q&=& p_{\tilde \alpha},\nonumber \\
 &&{}\nonumber \\
 {\check S}^1_q&=& sin\, \tilde \gamma p_{\tilde \beta} -{{cos\, \tilde \gamma}\over
 {sin\, \tilde \beta}} p_{\tilde \alpha} +cos\, \tilde \gamma
 ctg\, \tilde \beta p_{\tilde \gamma},\nonumber \\
 {\check S}^2_q&=& cos\, \tilde \gamma p_{\tilde \beta} +
 {{sin\, \tilde \gamma}\over {sin\, \tilde \beta}} p_{\tilde \alpha}
 -sin\, \tilde \gamma ctg\, \tilde \beta p_{\tilde \gamma},\nonumber \\
 {\check S}^3_q&=& p_{\tilde \gamma},\nonumber \\
 &&\Downarrow \nonumber \\
 p_{\tilde \alpha} &=& S^3_q =-sin\, \tilde \beta cos\, \tilde \gamma {\check S}^1_q
 +sin\, \tilde \beta sin\, \tilde \gamma {\check S}^2_q +cos\, \tilde \beta
 {\check S}^3_q,\nonumber \\
 p_{\tilde \beta} &=& -sin\, \tilde \alpha S^1_q +cos\, \tilde \alpha S^2_q =
  sin\, \tilde \gamma {\check S}^1_q -cos\, \tilde \gamma {\check S}^2_q,\nonumber \\
  p_{\tilde \gamma}&=& {\check S}^3_q = cos\, \tilde \alpha sin\, \tilde \beta S^1_q
  +sin\, \tilde \alpha sin\, \tilde \beta S^2_q + cos\, \tilde \beta S^3_q.
 \label{d3}
 \end{eqnarray}

\vfill\eject

\section{The 3-Body Case.}

 Let us try to rewrite the 3-body Hamiltonian (\ref{VII29}) in a form reminiscent
 of the non-relativistic Eq.(3.31) of Ref.\cite{iten2}, since this is the form used
 in the static orientation-shape bundle approach\cite{little}. We shall use
 the notation $q^{\mu}$ for the 3 generalized shape variables ($\rho_{qa}$,
 $|\vec N|$), and $p_{\mu}$ for the conjugate momenta (${\tilde \pi}_{qa}$,
 $\xi$).

In the non-relativistic orientation-shape bundle approach\cite{little}
one adopts the condition ${\vec S}_q=0$\footnote{It corresponds to the
choice of a special connection C on the SO(3) principal bundle
determined by the Euclidean metric in the non-relativistic kinetic
energy.} for the definition of {\it C-horizontal} (corresponding to a
gauge convention on the definition of {\it vibrations}); on the other
hand, the intrinsic concept of {\it vertical} (corresponding to pure
{\it rotational} motion) is defined by the condition of vanishing
shape velocities ${\dot q}^{\mu}=0$. If one would naively follow the
non-relativistic formulation, the following decomposition of the set
$({\vec S}_q,\, p_{\mu})$ into a vertical $()_v$ part and a
C-horizontal $()_{Ch}$ part

\begin{eqnarray}
 ({\vec S}_q,\,\, p_{\mu})&=&
 ({\vec S}_q,\,\, p_{\mu}{|}_{\dot q=0}\, {\buildrel {def} \over =}\,
 {\vec S}_q\cdot {\vec {\cal C}}_{\mu}
 ({\vec S}_q,q,m_i,\gamma_{ai}) )_v+
 \nonumber \\
 &+&(\vec 0,\quad p_{\mu}{|}_{{\vec S}_q=0}\,
 {\buildrel {def} \over =}\, p_{\mu}-{\vec S}_q\cdot
 {\vec {\cal C}}_{\mu}({\vec S}_q, q, m_i,\gamma_{ai}) )_{Ch},
\label{e1}
\end{eqnarray}

\noindent would be expected. This separation would identify the gauge potential ${\vec {\cal
C}}_{\mu}({\vec S}_q, q, m_i,\gamma_{ai})$, which could also be
spin-dependent.

But in  the relativistic case, since $H_{(rel)}=\sum_{i=1}^3 H_{(rel)
i}$ with each term being a square root, the shape velocities ${\dot
q}^{\mu}$, evaluated by means of the first half of Hamilton equations,
have to be written as the sum of 3 terms ${\dot q}^{\mu}_i$,
$i=1,2,3,$

\begin{eqnarray}
  {\dot q}^{\mu}\, &{\buildrel \circ \over =}\,& {{\partial H_{(rel)}}\over {\partial
  p_{\mu}}}=\sum_{i=1}^3 {1\over {2 H_{(rel)i}}}{{\partial H^2_{(rel)i}}\over
  {\partial p_{\mu}}}\, {\buildrel {def} \over =}\, \sum_{i=1}^3 {\dot q}_i^{\mu},\quad
  {\dot q}^{\mu}_i={1\over {2 H_{(rel)i}}}{{\partial H^2_{(rel)i}}\over
  {\partial p_{\mu}}},\nonumber \\
 &&{}\nonumber \\
 {\dot \rho}_{q1}&{\buildrel \circ \over =}& {{\partial H_{(rel)}}\over
 {\partial {\tilde \pi}_{q1}}}=
 \sum_{i=1}^3{1\over {2 H_{(rel)i}}}{{\partial H^2_{(rel)i}}\over
  {\partial {\tilde \pi}_{q1}}}\,
  {\buildrel {def} \over =}\, \sum_{i=1}^3{\dot \rho}_{q1\, i},\nonumber \\
 {\dot \rho}_{q2}&{\buildrel \circ \over =}& {{\partial H_{(rel)}}\over
 {\partial {\tilde \pi}_{q2}}}=\sum_{i=1}^3{1\over {2 H_{(rel)i}}}
 {{\partial H^2_{(rel)i}}\over
  {\partial {\tilde \pi}_{q2}}}\,
  {\buildrel {def} \over =}\, \sum_{i=1}^3{\dot \rho}_{q2\, i},\nonumber \\
 {\dot {|\vec N|}}&{\buildrel \circ \over =}& {{\partial H_{(rel)}}\over
 {\partial \xi}}=\sum_{i=1}^3{1\over {2 H_{(rel)i}}}{{\partial H^2_{(rel)i}}\over
  {\partial \xi}}\,
  {\buildrel {def} \over =}\, \sum_{i=1}^3{\dot {|\vec N|}}_i.
\label{e2}
\end{eqnarray}

Therefore, the presence of the 3 square roots $H_{(rel)i}$
\footnote{Remnants of the positive energy branch of the mass-shell
conditions $p_i^2=m_i^2$, which are characteristic of Lorentz
signature.}with $H_{(rel)}=\sum_{i=1}^3H_{(rel) i}$ allows to
introduce 3 concepts of {\it $i$-vertical} [${\dot q}^{\mu}_i=0$]. As
a consequence, now 3 concepts of {\it $Ch-i$-horizontal} (one for each
particle) can be introduced, each one defining a decomposition of the
type:

\begin{eqnarray}
 ({\vec S}_q,\,\, p_{\mu}){|}_i&=& ({\vec S}_q,\,\,
p_{\mu}{|}_{{\dot q}_i=0}\, {\buildrel {def} \over =}\,
 {\vec S}_q\cdot {\vec {\cal A}}_{i\mu}(q) )_{v-i}+
 \nonumber \\
 &+&(\vec 0,\,\, p_{\mu}{|}_{{\vec S}_q=0}\,
 {\buildrel {def} \over =}\, p_{\mu}-{\vec S}_q\cdot
 {\vec {\cal A}}_{i\mu}(q) )_{Ch-i}
\label{e3}
\end{eqnarray}

This implies the presence of 3 different particle gauge potentials
${\vec {\cal A}}_{i\mu}(q)$ (one for each particle) to be  contrasted
with the global but spin-dependent gauge potential ${\vec {\cal
C}}_{\mu}( {\vec S}_q , q,m_i,\gamma_{ai})$ appearing in the vertical
component of the momenta still given by the first part of
Eq.(\ref{e1}).

Therefore in the dynamical body frame  approach\cite{iten2}  we could
introduce 3 concepts of {\it i-dynamical vibrations}: since, as we
shall see, even the angular velocity has the form $\vec \omega =
\sum_{i=1}^3{\vec \omega}_i$, we could require to have ${\check
\omega}^r_i=0$ separately.

On the other hand, since in this approach the angular velocity is a
measurable quantity, the  global  {\it dynamical vibrations} are
defined by the requirement ${\check \omega}^r=0$.

This suggests to write the Hamiltonian for relative motions in the
form

\begin{eqnarray}
H_{(rel)} &=& \sum_{i=1}^3 H_{(rel) i}=\nonumber \\
 &=& \sum_{i=1}^3
 \sqrt{m_i^2+{\check {\cal T}}_i^{-1 rs}(q){\check S}_q^r{\check S}_q^s+
 {\tilde v}_i^{\mu\nu}(q)\Big( p_{\mu}-{\vec S}_q\cdot {\vec A}_{i\mu}(q)\Big)
 \Big( p_{\nu}-{\vec S}_q\cdot {\vec A}_{i\nu}(q)\Big)},
 \label{e4}
 \end{eqnarray}

\noindent with Eq.(\ref{VII31}) giving its purely rotational content.

It is clear that the generalized shape coordinates are not normal
coordinates for the Hamiltonian. Now there are  3 inverse metrics
${\tilde v}_i^{\mu\nu}(q)$. There is no concept of inertia tensor and
of reduced masses. Instead, there are 3 mass-independent particle
tensors ${\check {\cal T}}_i^{-1 rs}(q)$ replacing the inverse of the
non-relativistic  inertia tensor ${\check {\cal I}}^{-1 rs}(q,m)$ of
Eq.(F18) of Ref.\cite{iten2}.

Let us see how it is possible to find ${\vec {\cal C}}({\vec
S}_q,q,m_i,\gamma_{ai})$, ${\vec {\cal A}}_{i\mu}(q)$, ${\tilde
v}^{\mu\nu}_i(q)$, ${\check {\cal T}}^{-1 rs}_i(q)$ starting from our
choice of variables.

The 3 equations (\ref{e2}) can be inverted to get $p_{\mu}$ in terms
of $q^{\mu}$, ${\dot q}^{\mu}$, ${\check  S}^r_q$, $m_i$,
$\gamma_{ai}$: this is as difficult as finding the Lagrangian for the
relative motion. Then, by definition we have  ${\vec S}_q\cdot {\vec
{\cal C}}_{\mu}({\vec S}_q, q) = p_{\mu}{|}_{\dot q=0}$, namely

\begin{eqnarray}
 {\vec S}_q\cdot {\vec {\cal C}}_{q1}({\vec
S}_q, q, m_i, \gamma_{ai})
 &=& {\tilde \pi}_{q1} {|}_{\dot q=0},\nonumber \\
  {\vec S}_q\cdot {\vec
{\cal C}}_{q2}({\vec S}_q, q, m_i,
\gamma_{ai}) &=& {\tilde \pi}_{q2} {|}_{\dot q=0},\nonumber \\
 {\vec S}_q\cdot {\vec {\cal
C}}_{\xi}({\vec S}_q, q, m_i,
\gamma_{ai}) &=& \xi {|}_{\dot q=0}.
\label{e5}
\end{eqnarray}

From Eqs.(\ref{e2}) we have the following form for the components
${\dot q}^{\mu}_i$

\begin{eqnarray}
  {\dot \rho}_{q1\, i} &=&{1\over {2 H_{(rel)i}}}
  \Big[ 2(\gamma_{1i})^2 {\tilde \pi}_{q1}+2\gamma_{1i}\gamma_{2i}
 \Big[ (2{\vec N}^2-1) {\tilde \pi}_{q2} +|\vec N|(1-{\vec N}^2){{{\check S}^2_q-\xi}\over
 {\rho_{q2}}}\Big],\nonumber \\
  {\dot \rho}_{q2\, i} &=&{1\over {2 H_{(rel)i}}}
  \Big[ 2(\gamma_{2i})^2 {\tilde \pi}_{q2}+2\gamma_{1i}\gamma_{2i}
 \Big[ (2{\vec N}^2-1) {\tilde \pi}_{q1} -|\vec N|(1-{\vec N}^2){{{\check S}^2_q+\xi}\over
 {\rho_{q2}}}\Big], \nonumber \\
  {\dot {|\vec N|}}_i&=&{1\over {2 H_{(rel)i}}}
  \Big[ (1-{\vec N}^2)\Big( {{(\gamma_{1i})^2}\over {2\rho^2_{q1}}}
 +{{(\gamma_{2i})^2}\over {2\rho^2_{q2}}}\Big) \xi -
 2\gamma_{1i}\gamma_{2i}|\vec N| (1-{\vec N}^2) \Big( {{{\tilde \pi}_{q1}}\over {\rho_{q2}}}+
 {{{\tilde \pi}_{q2}}\over {\rho_{q1}}}\Big) +\nonumber \\
 &+&2\sqrt{1-{\vec N}^2} \Big( {{(\gamma_{1i})^2}\over {4\rho^2_{q1}}}-
 {{(\gamma_{2i})^2}\over {4\rho^2_{q2}}}\Big) {\check S}^2_q\Big].
\label{e6}
\end{eqnarray}

Since ${\dot q}_i^{\mu}=0$ implies  ${{\partial H^2_{(rel)i}}\over
{\partial p_{\mu}}}=0$, we get ${\vec S}_q\cdot {\vec {\cal
A}}_{i\mu}(q) = p_{\mu}{|}_{{\dot q}_i=0}$. Then, from these equations
we can  find ${\vec {\cal A}}_{i\mu}(q)$. Using the shape variables of
our canonical basis we find

\begin{eqnarray}
 {\check {\cal A}}^1_{i\mu}(q)&=& {\check {\cal A}}^3_{i\mu}(q)=0,\nonumber \\
 &&{}\nonumber \\
 {\check {\cal A}}^2_{i \xi}(q)&=&{{ [{{{\vec N}^2(1-{\vec N}^2)}\over {1-(2{\vec N}^2-1)^2}}
 -{1\over 4}\sqrt{1-{\vec N}^2}]({{\gamma^2_{1i}}\over {\rho_{q1}^2}}-{{\gamma_{2i}^2}\over
 {\rho_{q2}^2}})}\over { {1\over 4}({{\gamma_{1i}^2}\over {\rho_{q1}^2}}+{{\gamma_{2i}^2}\over
 {\rho_{q2}^2}})+{{(1-{\vec N}^2)(2{\vec N}^2-1)\gamma_{1i}\gamma_{2i}}\over
 {2\rho_{q1}\rho_{q2}}}-{{{\vec N}^2}\over {1-(2{\vec N}^2-1)^2}}({{\gamma_{1i}^2}\over
 {\rho_{q1}^2}}+{{\gamma_{2i}^2}\over {\rho_{q2}^2}}-{{2(2{\vec N}^2-1)\gamma_{1i}\gamma_{2i}}
 \over {\rho_{q1}\rho_{q2}}})}},\nonumber \\
 {\check {\cal A}}^2_{i \rho_1}(q)&=& -{{|\vec N| (1-{\vec N}^2)}\over
 {\gamma_{1i}[1-(2{\vec N}^2-1)]}} \Big[ {{\gamma_{2i}}\over {\rho_{q2}}}+{{(2{\vec
 N}^2-1)\gamma_{1i}}\over {\rho_{q1}}}+({{\gamma_{2i}}\over {\rho_{q2}}}-
 {{(2{\vec N}^2-1)\gamma_{1i}}\over {\rho_{q1}}}) {\check {\cal A}}^2_{i\xi}(q)\Big],
 \nonumber \\
 {\check {\cal A}}^2_{i \rho_2}(q)&=& +{{|\vec N| (1-{\vec N}^2)}\over
 {\gamma_{2i}[1-(2{\vec N}^2-1)]}}\Big[ {{\gamma_{1i}}\over {\rho_{q1}}}+{{(2{\vec
 N}^2-1)\gamma_{2i}}\over {\rho_{q2}}}+({{\gamma_{1i}}\over {\rho_{q1}}}-
 {{(2{\vec N}^2-1)\gamma_{2i}}\over {\rho_{q2}}}) {\check {\cal A}}^2_{i\xi}(q)\Big].
 \label{e7}
 \end{eqnarray}

The term in $H_{(rel)i}$ quadratic in the $p_{\mu}'s$ identifies the
 ${\tilde v}^{\mu\nu}_i(q)$ so that

\begin{equation}
 {\tilde v}_i^{\mu\nu}=\left( \begin{array}{lll}
 \gamma_{1i}^2 & (2{\vec N}^2-1)\gamma_{1i}\gamma_{2i} &
 -{{|\vec N| (1-{\vec N}^2)}\over {\rho_{q2}}}\gamma_{1i}\gamma_{2i} \\
 (2{\vec N}^2-1)\gamma_{1i}\gamma_{2i} & \gamma_{2i}^2 &
 -{{|\vec N| (1-{\vec N}^2)}\over {\rho_{q1}}}\gamma_{1i}\gamma_{2i} \\
 -{{|\vec N| (1-{\vec N}^2)}\over {\rho_{q2}}}\gamma_{1i}\gamma_{2i} &
 -{{|\vec N| (1-{\vec N}^2)}\over {\rho_{q1}}}\gamma_{1i}\gamma_{2i} &
 {{1-{\vec N}^2}\over 2}({{\gamma_{1i}^2}\over {\rho_{q1}^2}}+
 {{\gamma_{2i}^2}\over {\rho_{q2}^2}})+{{(1-{\vec N}^2)(2{\vec N}^2-1)\gamma_{1i}
 \gamma_{2i}}\over {2\rho_{q1}\rho_{q2}}} \end{array} \right) .
 \label{e8}
 \end{equation}

The body frame angular velocity results

\begin{eqnarray}
 {\check \omega}^r\, &{\buildrel \circ \over =}\,& {{\partial H_{(rel)}}\over
 {\partial {\check S}_q^r}}=\sum_{i=1}^N {1\over {2 H_{(rel)i}}} {{\partial H^2_{(rel)i}}
 \over {\partial {\check S}^r_q}}=\sum_{i=1}^N{\check \omega}^r_i,\nonumber \\
 {\check \omega}^r_i &=& {1\over {H_{(rel)i}}}\Big[ {\check {\cal T}}_i^{-1 rs}(q)
 {\check S}_q^s-{\check {\cal A}}^r_{i\mu}(q){\tilde v}_i^{\mu\nu}(q)\Big(
 p_{\nu}-{\check {\vec S}}_q\cdot {\check {\vec {\cal A}}}_{i\nu}(q)\Big) \Big],
 \nonumber \\
 {\check \omega}^1_i {|}_{{\check {\vec S}}_q=0}&=&{1\over {2 H_{(rel)i}}}
 \Big[ {2\over {{\vec N}^2}} \Big( {{(\gamma_{1i})^2}\over {4\rho^2_{q1}}}+{{(\gamma_{2i})^2}\over
 {4\rho^2_{q2}}}+{{\gamma_{1i}\gamma_{2i}}\over {2\rho_{q1}\rho_{q2}}}\Big)
 {\check S}^1_q-\nonumber \\
 &-&{2\over {|\vec N| \sqrt{1-{\vec N}^2}}}\Big( {{(\gamma_{1i})^2}\over {4\rho^2_{q1}}}-
 {{(\gamma_{2i})^2}\over {4\rho^2_{q2}}}\Big) {\check S}^3_q\Big],\nonumber \\
 {\check \omega}^2_i {|}_{{\check {\vec S}}_q=0}&=&{1\over {2H_{(rel)i}}}
 \Big[ 2 \Big( {{(\gamma_{1i})^2}\over {4\rho^2_{q1}}}+{{(\gamma_{2i})^2}\over
 {4\rho^2_{q2}}}+{{\gamma_{1i}\gamma_{2i}(2{\vec N}^2-1)}\over {2\rho_{q1}\rho_{q2}}}\Big)
 {\check S}^2_q+\nonumber \\
 &+&\sqrt{1-{\vec N}^2} \Big[ 2\Big( {{(\gamma_{1i})^2}\over {4\rho^2_{q1}}}-
 {{(\gamma_{2i})^2}\over {4\rho^2_{q2}}}\Big) \xi +2\gamma_{1i}\gamma_{2i} |\vec N|
 \sqrt{1-{\vec N}^2} \Big( {{{\tilde \pi}_{q1}}\over {\rho_{q2}}}-
 {{{\tilde \pi}_{q2}}\over {\rho_{q1}}}\Big) \Big] \Big],\nonumber \\
 {\check \omega}^3_i {|}_{{\check {\vec S}}_q=0}&=&{1\over {2H_{(rel)i}}}
 \Big[ {2\over {\sqrt{1-{\vec N}^2}}} \Big( {{(\gamma_{1i})^2}\over {4\rho^2_{q1}}}+{{(\gamma_{2i})^2}\over
 {4\rho^2_{q2}}}-{{\gamma_{1i}\gamma_{2i}}\over {2\rho_{q1}\rho_{q2}}}\Big)
 {\check S}^3_q-\nonumber \\
 &-&{2\over {|\vec N| \sqrt{1-{\vec N}^2}}}
 \Big( {{(\gamma_{1i})^2}\over {4\rho^2_{q1}}}-{{(\gamma_{2i})^2}\over {4\rho^2_{q2}}}\Big)
 {\check S}^1_q,\nonumber \\
 or &&{\check \omega}^r_i {|}_{{\check {\vec S}}_q=0}=-{1\over {2H_{(rel)i}}}{|}_{{\check
 {\vec S}}_q=0}  {\check {\cal A}}_{i \mu}^r(q) {\tilde v}_i^{\mu\nu}(q)p_{\nu}.
 \label{e9}
 \end{eqnarray}

This allows to determine the functions ${\check {\cal T}}_i^{-1
rs}(q)$ once ${\vec {\cal A}}_{i\mu}(q)$ and ${\tilde
v}_i^{\mu\nu}(q)$ are known. From the equalities

 \begin{eqnarray}
 &&H_{(rel)i} {\check \omega}^1_i={\check {\cal T}}_i^{-1 1s}(q){\check S}_q^s,\nonumber \\
 &&H_{(rel)i} {\check \omega}^2_i={\check {\cal T}}_i^{-1 2s}(q){\check S}_q^s -
 {\check {\cal A}}^2_{i\mu}(q){\tilde v}_i^{\mu\nu}(q)\Big( p_{\nu}-{\check
 S}_q^2{\check {\cal A}}^2_{i\nu}(q)\Big) ,\nonumber \\
 &&H_{(rel)i}{\check \omega}^3_i={\check {\cal T}}_i^{-1 3s}(q){\check S}_q^s,
 \label{e10}
 \end{eqnarray}

 \noindent we find the following non-zero components

\begin{eqnarray}
 {\check {\cal T}}_i^{-1 11}(q)&=&
 {1\over {{\vec N}^2}} \Big( {{(\gamma_{1i})^2}\over
{4\rho^2_{q1}}}+{{(\gamma_{2i})^2}\over
 {4\rho^2_{q2}}}+{{\gamma_{1i}\gamma_{2i}}\over {2\rho_{q1}\rho_{q2}}}\Big)
 ,\nonumber \\
{\check {\cal T}}_i^{-1 13}(q)&=& -{1\over {|\vec N| \sqrt{1-{\vec
N}^2}}}\Big( {{(\gamma_{1i})^2}\over {4\rho^2_{q1}}}-
 {{(\gamma_{2i})^2}\over {4\rho^2_{q2}}}\Big) ,\nonumber \\
 {\check {\cal T}}_i^{-1 22}(q)&=&{1\over 2} \Big[ {{\gamma_{1i}^2}\over {\rho_{q1}^2}}+
 {{\gamma_{2i}^2}\over {\rho_{q2}^2}}+{{(2{\vec N}^2-1)\gamma_{1i}\gamma_{2i}}\over
 {\rho_{q1}\rho_{q2}}}\Big] -\nonumber \\
 &-&{{|\vec N| (1-{\vec N}^2)\gamma_{1i}\gamma_{2i}}\over {\rho_{q1}\rho_{q2}}}\Big(
 \rho_{q1}{\check {\cal A}}^2_{i\rho_1}(q)-\rho_{q2}{\check {\cal A}}^2_{i\rho_2}(q)\Big)
 -{{\sqrt{1-{\vec N}^2}}\over 2} \Big( {{\gamma_{1i}^2}\over {\rho_{q1}^2}}-
 {{\gamma_{2i}^2}\over {\rho_{q2}^2}}\Big) {\check {\cal A}}^2_{i\xi}(q),
 \nonumber \\
 {\check {\cal T}}_i^{-1 33}(q)&=&  {1\over {\sqrt{1-{\vec N}^2}}}
 \Big( {{(\gamma_{1i})^2}\over {4\rho^2_{q1}}}+{{(\gamma_{2i})^2}\over
 {4\rho^2_{q2}}}-{{\gamma_{1i}\gamma_{2i}}\over {2\rho_{q1}\rho_{q2}}}\Big).
 \label{e11}
 \end{eqnarray}

Finally, let us recall the following results of Appendix C of
Ref.\cite{iten2}

\begin{eqnarray}
 {\vec \pi}^2_{q1}&=&{\tilde \pi}^2_{q1}+{1\over {4\rho^2_{q1}}} \Big[
 \xi^2(1-{\vec N}^2)+ ({\check S}_q^2)^2+{1\over { {\vec N}^2}} ({\check S}^1_q)^2
+\nonumber \\
 &+&{{({\check S}_q^3)^2}\over {1-{\vec N}^2}}+2 (\xi \sqrt{1-{\vec N}^2} {\check S}_q^2 -
 {{{\check S}_q^1{\check S}_q^3}\over {|\vec N| \sqrt{1-{\vec N}^2}}}  )
  \Big],\nonumber \\
 {\vec \pi}_{q2}^2&=&{\tilde \pi}^2_{q2}+{1\over {4\rho^2_{q2}}} \Big[
 \xi^2(1-{\vec N}^2)+ ({\check S}_q^2)^2+{1\over { {\vec N}^2}} ({\check S}^1_q)^2
+\nonumber \\
 &+&{{({\check S}_q^3)^2}\over {1-{\vec N}^2}}-2 (\xi \sqrt{1-{\vec N}^2} {\check S}_q^2 -
 {{{\check S}_q^1{\check S}_q^3}\over {|\vec N| \sqrt{1-{\vec N}^2}}}  )
  \Big],\nonumber \\
 {\vec \pi}_{q1}\cdot {\vec \pi}_{q2}&=& (2{\vec N}^2-1) {\tilde \pi}_{q1}{\tilde \pi}_{q2}
 +\nonumber \\
  &+&|\vec N| \sqrt{1-{\vec N}^2}\Big[ \Big( {{{\tilde \pi}_{q1}}\over
 {\rho_{q2}}}-{{{\tilde \pi}_{q2}}\over {\rho_{q1}}}\Big)
 {\check S}_q^2 - \Big({{{\tilde \pi}_{q1}}\over {\rho_{q2}}}+
 {{{\tilde \pi}_{q2}}\over {\rho_{q1}}}\Big) \xi \sqrt{1-{\vec N}^2}\Big]
+\nonumber \\
 &+&{1\over {4\rho_{q1}\rho_{q2}}} \Big[ (2{\vec N}^2-1)
 ({\check S}_q^2)^2+{1\over { {\vec N}^2}} ({\check S}^1_q)^2
-{{({\check S}_q^3)^2}\over {1-{\vec N}^2}}+\nonumber \\
&+& (1-{\vec N}^2)(2{\vec N}^2-1) \xi^2 \Big] ,
\label{e12}
\end{eqnarray}

\begin{eqnarray}
 {\vec \pi}^2_{q1}&=&{\tilde \pi}^2_{q1}+{1\over {4\rho^2_{q1}}} \Big[
 \xi^2(1-{\vec N}^2)+ ({\check S}_q^2)^2+{1\over { {\vec N}^2}} ({\check S}^1_q)^2
+\nonumber \\
 &+&{{({\check S}_q^3)^2}\over {1-{\vec N}^2}}+2 (\xi \sqrt{1-{\vec N}^2} {\check S}_q^2 -
 {{{\check S}_q^1{\check S}_q^3}\over {|\vec N| \sqrt{1-{\vec N}^2}}}  )
  \Big],\nonumber \\
 {\vec \pi}_{q2}^2&=&{\tilde \pi}^2_{q2}+{1\over {4\rho^2_{q2}}} \Big[
 \xi^2(1-{\vec N}^2)+ ({\check S}_q^2)^2+{1\over { {\vec N}^2}} ({\check S}^1_q)^2
+\nonumber \\
 &+&{{({\check S}_q^3)^2}\over {1-{\vec N}^2}}-2 (\xi \sqrt{1-{\vec N}^2} {\check S}_q^2 -
 {{{\check S}_q^1{\check S}_q^3}\over {|\vec N| \sqrt{1-{\vec N}^2}}}  )
  \Big],\nonumber \\
 {\vec \pi}_{q1}\cdot {\vec \pi}_{q2}&=& (2{\vec N}^2-1) {\tilde \pi}_{q1}{\tilde \pi}_{q2}
 +\nonumber \\
  &+&|\vec N| \sqrt{1-{\vec N}^2}\Big[ \Big( {{{\tilde \pi}_{q1}}\over
 {\rho_{q2}}}-{{{\tilde \pi}_{q2}}\over {\rho_{q1}}}\Big)
 {\check S}_q^2 - \Big({{{\tilde \pi}_{q1}}\over {\rho_{q2}}}+
 {{{\tilde \pi}_{q2}}\over {\rho_{q1}}}\Big) \xi \sqrt{1-{\vec N}^2}\Big]
+\nonumber \\
 &+&{1\over {4\rho_{q1}\rho_{q2}}} \Big[ (2{\vec N}^2-1)
 ({\check S}_q^2)^2+{1\over { {\vec N}^2}} ({\check S}^1_q)^2
-{{({\check S}_q^3)^2}\over {1-{\vec N}^2}}+\nonumber \\
 &+&\xi^2 (1-{\vec N}^2)(2{\vec N}^2-1) \Big] .
\label{e13}
\end{eqnarray}

\vfill\eject

\end{document}